\documentclass[nofootinbib,twocolumn]{revtex4-1}

\usepackage{amsmath}
\usepackage{amsfonts}
\usepackage{amssymb}
\usepackage{graphicx}

\usepackage{color}
\usepackage{enumitem}

\usepackage{verbatim}
\usepackage{simplemargins}

\newcommand{\tar}{\mathrm{t}}
\newcommand{\pro}{\mathrm{p}}

\newcommand{\dd}{\mathrm{d}}

\newcommand{\n}{\hat{\mathbf{n}}}

\newcommand{\fix}{\mathrm{fix}}
\newcommand{\cm}{\mathrm{cm}}
\newcommand{\lab}{\mathrm{lab}}

\newcommand{\N}{\mathcal{N}}
\newcommand{\D}{\mathcal{D}}

\newcommand{\X}{\chi}
\newcommand{\R}{\varrho}
\newcommand{\V}{V}
\newcommand{\K}{\kappa}
\newcommand{\LL}{\mathcal{L}}
\newcommand{\e}{\mathrm{e}}
\newcommand{\Z}{\mathcal{Z}}
\newcommand{\ZZ}{\mathbb{Z}}  
\newcommand{\ro}{\xi}  
\newcommand{\Rho}{\mathrm{P}}
\newcommand{\bnd}{\zeta}
\newcommand{\vtx}{\mathrm{Z}}
\newcommand{\tht}{\tilde{\theta}}
\newcommand{\cpz}{\mathsf{Z}}
\newcommand{\cpr}{\mathsf{R}}

\newcommand{\atan}{\mathrm{arctan}}
\newcommand{\acot}{\mathrm{arccot}}

\newcommand{\ifif}{\mathrm{if}}
\newcommand{\sgn}{\mathrm{sgn}}

\newcommand{\pull}{\vspace*{-5.5mm}}
\newcommand{\pullc}{\vspace*{-3.5mm}}

\newcommand{\pulleqa}{\vspace*{-1.5mm}}
\newcommand{\pulleqb}{\vspace*{-1.5mm}}

\setlist[itemize]{leftmargin=*} 

\settopmargin{1.6cm}
\setbottommargin{1.9cm}

\usepackage[labeled,resetlabels]{multibib}
 
\begin{document}

\thispagestyle{empty}

\onecolumngrid
\begin{center}

\textbf{\large A shadow of the repulsive Rutherford scattering in the laboratory frame}\\[.5cm]

Petar \v{Z}ugec$^{1*}$ and Dario Rudec$^1$\\[.1cm]
{\small
{\itshape
$^1$Department of Physics, Faculty of Science, University of Zagreb, Zagreb, Croatia\\}
$^*$Electronic address: pzugec@phy.hr\\[.5cm]
}

\begin{minipage}{400pt}
\small
The paper explores the Rutherford scattering shadow in an entire class of comoving frames---inertial frames moving along the initial projectile direction---of which the laboratory frame, where the target is initially at rest, is a representative example. The paper is a continuation of the previous work addressing the scattering shadow in the fixed-target and the center-of-mass frame. It is shown that the transition from these frames is technically quite involved, due to the scattering shadow forming at an infinite distance from an initial target position. The central procedure involves solving the $5^\mathrm{th}$ degree polynomial as a part of an associated extremization procedure. The shadow existence itself is subject to certain conditions, dependent on a given comoving frame. A new and unexpected phenomenon is found within a certain set of comoving frames, including the laboratory frame itself. It consists of a phase transition between an entirely smooth type of shadow and the one characterized by a formation of a sharp edge.\\
\end{minipage}

\end{center}


\twocolumngrid

\renewcommand{\theequation}{\arabic{equation}}

\section{Introduction}
\label{intro}

The famous historical experiments by Geiger and Marsden \cite{geiger1,geiger2,geiger3}, dealing with the scattering of $\alpha$-particles by thin metal foils, are appropriately considered not only as the turning points in physics, but also as the turning points of a rare kind in the development of the modern civilization. These experiments have, in a most direct manner and for the first time in a history of mankind, allowed Rutherford to reveal the inner structure of the atom \cite{rutherford}, thus discovering the existence of the atomic nucleus and ushering the age of nuclear technology. Aside from the obvious social benefits of having discovered a novel and usable source of energy, these achievements have inadvertently had an unprecedented influence upon our understanding of the entire universe---from a very start of its existence (the primordial nucleosynthesis) to the origin of life on Earth (the stellar nucleosynthesis from the Sun, as a source of all life-sustaining energy on Earth).

In a recent work \cite{zugec} a remarkable feature of this scattering---nowadays known as the Rutherford scattering and understood to be the scattering the electric charges due to the Coulomb interaction---was analyzed in some detail. This feature consist in the \textit{repulsive} Rutherford scattering casting a proverbial shadow, shielding (under appropriately defined conditions) an entire portion of space from admitting any charged particle trajectory. The form of this shadow was first investigated in the fixed-target frame and the center-of-mass frame, and was shown to be paraboloidal in both frames. Though the Rutherford scattering itself is a regular subject of (under)graduate physics courses, as a very cornerstone of nuclear physics, and though its shadowing effect is at the center of a material surface investigation method known as Low-Energy Ion Scattering Spectroscopy \cite{leis1,leis2}, the shadowing feature seems to be little known throughout the educational literature. And all this despite it being fully within the mathematical capabilities of any (under)graduate student in physical sciences. The latest attempt at rekindling the interest in this worthy educational subject has already attracted attention and has lead to further illuminating expositions~\cite{reaction1,reaction2}. We have high hopes that all these efforts will lead to a widespread recognition of this topic's deserved place in physical studies.

There \textit{have} been earlier isolated attempts at drawing attention to the Rutherford scattering shadow \cite{shadow_prb1,shadow_prb2,shadow_hyp,shadow_geom}, mostly limited to the fixed-target frame---an accelerated frame of the charged target itself, where the target is at rest at all times. One of the earliest such references, by Adolph \textit{et al.} \cite{shadow_prb1}, comments upon the particle trajectories in the laboratory frame---an inertial frame where the charged target is at rest only at the initial moment---stating that `\textit{the construction of the orbits} [in the laboratory frame] \textit{is beyond the reach of simple geometry}.' We will obtain these trajectories by a Galilean transformation of the trajectories from the fixed-target frame. Even more generally, we will analyze it within an entire class of \mbox{\textit{comoving}} frames, consisting of any inertial frame moving in an appropriate direction, with the constant speed relative to the center-of-mass frame, i.e. to the particle-target system as a whole.


We adopt here a classical nonrelativistic approach. We hope to demonstrate that many, to our knowledge new results may yet be gained within this approach. We will show that, as far as the Rutherford scattering shadow is concerned, the transition between the frames in relative motion is not just a technical challenge from which no further insight could be gained. Quite the contrary: in opposition to the naive idea that the scattering shadow in the comoving frame might be obtained by some simple manipulation of the parabolic shadow from the fixed-target frame, we will find that: (1)~several technical challenges appear, consisting of a divergent integral and the $5^\mathrm{th}$ degree polynomial; (2)~the resulting shadow is no longer parabolic; (3)~the scattering shadow cannot form in just any inertial frame; (4)~there appears a qualitative alteration in the shadow behavior, akin to a certain type of phase transition, consisting in a formation of a sharp edge along the shadow caustic.

We face some instructive challenges in the derivation of the projectile trajectories (restricted to Appendix~\ref{appendix}). The first challenge is the appearance of a divergent integral, that we overcome by a careful and disciplined parametrization of the emerging divergence. The second challenge is the necessity for finding numerical solutions to the $5^\mathrm{th}$ degree polynomial, since there exists no solution in radicals for a \textit{general} polynomial of a degree greater than 4. There is a certain educational benefit in the ability to demonstrate the practical utilization of the modern computer resources in solving a particular, very well defined physical problem, appropriate even at lower levels of the (under)graduate studies where the Rutherford scattering is a regular subject.



Returning to the issue of the nonrelativistic approach, there is a rich discussion to be had, carried out in Section~\ref{relativistic}. In Section~\ref{trajectories} we illustrate the procedure for obtaining the projectile trajectories in the comoving frame. The technical derivation is presented in Appendix~\ref{appendix}. In Section~\ref{shadow} a procedure for obtaining the scattering shadow from these trajectories is presented. Section~\ref{existence} addresses and identifies the necessary conditions for the existence of the scattering shadow. Section~\ref{lab_frame} focuses on the laboratory frame, as one of the most prominent examples of comoving frames. Section~\ref{conclusion} summarizes the main conclusions of this work.

This paper is accompanied by the Supplementary note, expanding upon the main material presented herein. We stress that this paper is self-contained and that addressing the Supplementary note is by no means necessary for following the main content. Still, the Supplementary note offers deep, exciting and---to our knowledge---many novel expositions of various aspects of the repulsive Rutherford scattering in the comoving frame, to be appreciated by an interested reader.

\section{The (non)relativistic treatment}
\label{relativistic}

In the laboratory frame the charged target is put into motion by the recoil, which leads both to the transformation of its electric field and the additional induction of the magnetic field. The electric field transforms not only due to the target's non-zero speed, but also obtains a radiative component due to the target's acceleration. This is clearly seen from a well known, relativistically correct expression for the electric field of an arbitrarily moving point charge \cite{field1,field2}:
\pulleqa
\begin{equation}
\mathbf{E}(\mathbf{r},t)=\frac{q}{4\pi\epsilon_0}\left[\frac{\n-\boldsymbol{\beta}}{\gamma^2K^3R^2}+\frac{\n\times((\n-\boldsymbol{\beta})\times\mathbf{a})}{c^2K^3R}\right]_\tau ,
\label{E}
\pulleqb
\end{equation}
where $q$ is the value of the charge and $\epsilon_0$ is the vacuum permittivity. With $\mathbf{R}$ as a position-vector of a point at which the field is to be calculated ($\mathbf{r}$) relative to the position of the point charge ($\mathbf{r}'$): \mbox{$\mathbf{R}=\mathbf{r}-\mathbf{r}'=R\,\n$}, the terms $R$ and $\n$ appearing in~(\ref{E}) are its norm and unit direction, respectively: $R=|\mathbf{R}|$ and $\n=\mathbf{R}/R$. Alongside \mbox{$\mathbf{v}=\dd\mathbf{r}'/\dd t$} as the velocity of the point charge, \mbox{$\mathbf{a}=\dd\mathbf{v}/\dd t$} as its acceleration and $c$ as the speed of light in vacuum, \mbox{$\boldsymbol{\beta}=\mathbf{v}/c$} is the standard relativistic notation, together with the Lorentz factor \mbox{$\gamma=(1-\boldsymbol{\beta}\cdot\boldsymbol{\beta})^{-1/2}$} and \mbox{$K=1-\n\cdot\boldsymbol{\beta}$}. Finally, $[\cdot]_\tau$ denotes that all quantities within the square brackets are to be calculated at the retarded time $\tau$ such that \mbox{$\tau+R(\tau)/c=t$}, since it takes finite time for the information to propagate from $\mathbf{r}'$ to $\mathbf{r}$. It is to be noted that the first term in square brackets (\mbox{$\propto1/R^2$}) is the field transformation solely due the motion of the charge, while the second one (\mbox{$\propto1/R$}) is the radiative component due to its acceleration. This separation of contributions to the electric field due to the `levels' of motion is even more clearly seen from an equivalent Feynman's formula \cite{feynman}:
\pulleqa
\begin{equation}
\mathbf{E}(\mathbf{r},t)=\frac{q}{4\pi\epsilon_0}\left(\left[\frac{\n}{R^2}\right]_\tau +\frac{[R]_\tau}{c}\frac{\dd}{\dd t}\left[\frac{\n}{R^2}\right]_\tau+\frac{1}{c^2}\frac{\dd^2[\n]_\tau}{\dd t^2} \right),
\end{equation}
where the first term is evidently the pure Coulomb field (electrostatic in form), the second term takes into account the general motion of the charge, while only the third term may produce the dependence upon the charge acceleration. The associated magnetic field of the point charge may be calculated from its electric field as:
\pulleqa
\begin{equation}
\mathbf{B}(\mathbf{r},t)=\frac{\left[\n\right]_\tau \times \mathbf{E}(\mathbf{r},t)}{c}.
\label{B}
\pulleqb
\end{equation}
It is worth noting that in case of the charge moving with the constant velocity, (\ref{B}) may also be expressed as:
\pulleqa
\begin{equation}
\mathbf{B}_{\mathbf{a}=\mathbf{0}}=\frac{\boldsymbol{\beta} \times \mathbf{E}_{\mathbf{a}=\mathbf{0}}}{c}.
\label{Bconst}
\pulleqb
\end{equation}
This is easily seen since applying the vector products from either relation leaves only $\boldsymbol{\beta}\times\n$ in place of the first term from~(\ref{E}). In fact, the relation from~(\ref{Bconst}) holds not only for the point charge, but for any charge distribution moving with the constant velocity \cite{jefimenko}.

If we were to calculate the scattering trajectories by solving the relativistic equations of motion in the laboratory (or any other inertial) frame, we would simply use the correct field expressions (\ref{E}) and (\ref{B}), properly taking into account all aspects of the field transformations (the departure from the electrostatic form, the induction of the magnetic field and the appearance of the radiative component). However, would we have to account for these effects if we treated the problem nonrelativistically, staying within the confines of Galilean mechanics? This can be judged based on the relative magnitude between the electric and magnetic forces in a given frame. In the nonrelativistic case any effect from the charge acceleration upon the electric field from~(\ref{E}) is suppressed by $1/c^2$, thus being negligible. In the absence of this term the magnetic field may be expressed as in~(\ref{Bconst}), meaning that for the nonrelativistic charge it always holds \mbox{$\mathbf{B}\approx \mathbf{v}\times\mathbf{E}/c^2$}. Now it is simple enough to inspect the relative magnitude between the forces exerted upon the charged projectile~($\pro$) by the charged target~($\tar$):
\pulleqa
\begin{equation}
\frac{F_\mathbf{B}}{F_\mathbf{E}}=\frac{|q_\pro\mathbf{v}_\pro\times\mathbf{B}_\tar|}{|q_\pro\mathbf{E}_\tar|}\propto\frac{v_\pro v_\tar}{c^2}.
\pulleqb
\label{evsb}
\end{equation}
Thus, in the nonrelativistic limit (\mbox{$v_\pro,v_\tar\ll c$}) the Lorentz force is negligible, relative to an electric one, fully justifying the Galilean treatment that we adopt in this work.

Once the nonrelativistic treatment has been justified and adopted based on~(\ref{evsb}), the Lorentz force \textit{must not} be taken into account (assuming that the magnetic field appears due to the transformation of the electric field between the frames in relative motion). The reason is the Galilean invariance of force, combined with the fact that---within the Galilean framework---the electric field retains the electrostatic form \mbox{$\mathbf{E}(\mathbf{r},t)=(q/4\pi\epsilon_0)\n/R^2$} in all frames, which is easily seen from the nonrelativistic limit of~(\ref{E}). The short argument is this. Consider the electrostatic force exerted upon the point charge~$q$: \mbox{$\mathbf{F}=q\mathbf{E}$}. When the transition is made to a frame moving with the relative velocity $\mathbf{v}$, the force is furnished by an additional, Lorentz component: \mbox{$\mathbf{F}'=q\mathbf{E}'+q\mathbf{v}\times\mathbf{B}'$}. However, from the demonstrated nonrelativistic invariance of the electrostatic field (\mbox{$\mathbf{E}=\mathbf{E}'$}) and the Galilean invariance of force (\mbox{$\mathbf{F}=\mathbf{F}'$}) it follows that \mbox{$q\mathbf{v}\times\mathbf{B}'=\mathbf{0}$}, i.e. there is no room left for any kind of effect by the magnetic field.

The previous argument is closely related to one of the two independent Galilean limits to the classical electrodynamics \cite{galilean1,galilean2,galilean3}, the so-called electric limit wherein the electric effects are dominant (\mbox{$|\mathbf{E}|\gg c|\mathbf{B}|$}). It was formally shown in a famous paper by Le Bellac and L\'{e}vy-Leblond \cite{galilean1} that if the Galilean invariance is to be preserved in the electric limit, one must indeed contend with the magnetic field exerting no force upon the electric charge, rather than taking any kind of low-velocity limit of the electric field that would account for the necessity of an additional, Lorentz force (\mbox{$q\mathbf{v}\times\mathbf{B}'\neq\mathbf{0}$} due to \mbox{$\mathbf{E}\neq\mathbf{E}'$}). However, this argument is not strictly applicable to our case, since the Galilean limits to the classical electrodynamics apply to the \textit{inertial} frames in relative motion. We, on the other hand, will be concerned with the transition between the accelerated (fixed-target) frame and the inertial (comoving) frame.

Related to the approach that we adopt in this work---boosting the charged particle trajectories from the fixed-target into the comoving frame by means of a Galilean transformation---let us suppose for a moment that we attempted to perform this procedure relativistically. If we managed to obtain the relativistic particle trajectories in the fixed-target frame, we would have to perform the relativistic boost into the comoving frame by employing the generalized Lorentz transformations for the noninertial frames \cite{noninertial}. Although the correct field transformations between the frames would be implicitly accounted for by thus transformed trajectories, it would make little sense attempting to perform a relativistic boost of the classical hyperbolic trajectories---as they are not the relativistic solutions themselves---unless one were to use them as the reasonable approximations to the fully relativistic trajectories in the fixed-target frame.

\section{Coulomb trajectories in the comoving frame}
\label{trajectories}

We will obtain the charged particle trajectories in the comoving frame by a Galilean boost of well known hyperbolic trajectories from the fixed-target frame. We remind the reader of the basic steps leading to these hyperbolic solutions. With projectile and target charges $Z_\pro$ and $Z_\tar$, respectively, in units of the elementary charge $e$, and $\epsilon_0$ as the vacuum permittivity, one starts from a Coulomb force $\mathbf{F}_{\tar\rightarrow\pro}$ exerted upon the charged projectile:
\begin{equation}
\mathbf{F}_{\tar\rightarrow\pro}=\frac{Z_\pro Z_\tar e^2}{4\pi\epsilon_0}\frac{\mathbf{r}_\pro-\mathbf{r}_\tar}{|\mathbf{r}_\pro-\mathbf{r}_\tar|^3},
\end{equation}
and performs a standard separation of variables by introducing the target-relative projectile position\footnote{
Our term for the `fixed-target frame' comes from a definition of a relative position $\mathbf{r}$: in a frame where we can equate the projectile position with $\mathbf{r}$, the target is by construction at rest, fixed at the origin of the frame. Therefore, the `fixed-target' term should not be confused with target being infinitively massive or held in place by an external force. For a finitely massive target the fixed-target frame is accelerated, as the target is continuously being recoiled from the incoming projectile. An alternative, somewhat mouthful term to be found in literature for this frame is the `instantaneous rest frame (of the charge)'.
} \mbox{$\mathbf{r}\equiv\mathbf{r}_\pro-\mathbf{r}_\tar$} and the center-of-mass position \mbox{$\mathbf{R}\equiv(m_\pro\mathbf{r}_\pro+m_\tar\mathbf{r}_\tar)/(m_\pro+m_\tar)$}. In doing so the motion of the system as a whole ($\mathbf{R}$) decouples from he relative motion ($\mathbf{r}$). The equation of the relative motion then reads:
\begin{equation}
\ddot{\mathbf{r}}=\frac{Z_\pro Z_\tar e^2}{4\pi\epsilon_0\mu}\frac{\mathbf{r}}{r^3},
\end{equation}
where the reduced mass $\mu$ of a projectile-target system appears, determined by the projectile and target masses $m_\pro$ and $m_\tar$ as \mbox{$\mu^{-1}\equiv m_\pro^{-1}+m_\tar^{-1}$}. With the appropriate set of initial conditions expressed in cylindrical coordinates:
\begin{align}
&\mathbf{r}(t=0)=\R_0 \hat{\boldsymbol{\rho}}-\Big(\lim_{z_0\to\infty}z_0\Big)\hat{\mathbf{z}},
\label{r000}\\
&\dot{\mathbf{r}}(t=0)=v_0\hat{\mathbf{z}},
\label{v000}
\end{align}
the solution for the radial component of the target-relative projectile position $\mathbf{r}$ reduces to:
\begin{equation}
r(\theta)=\frac{\R_0^2}{2(\R_0\tan\tfrac{\theta}{2}-\X)\cos^2\tfrac{\theta}{2}},
\label{master}
\end{equation}
with this particular form being the most convenient to this work. The polar angle $\theta$ is a conventionally defined spherical coordinate relative to the $z$-axis oriented along the projectile's initial velocity. An impact parameter $\R_0$ determines a specific projectile trajectory and corresponds to the initial distance from the $z$-axis. The central parameter $\X$ is defined as:
\begin{equation}
\X\equiv\frac{Z_\pro Z_\tar e^2}{4\pi\epsilon_0 \mu v_0^2},
\label{x}
\end{equation}
with $v_0$ as the initial relative speed between the target and projectile, that remains invariant under Galilean transformations. The infinity from the $z$-component of the initial relative position~(\ref{r000}) will propagate into later calculations, therefore we need to carefully parameterize so as to formally keep it under control. In this work we choose to parameterize it by a \textit{positive} parameter $z_0$.


In the absence of external forces a center-of-mass position $\mathbf{R}$ satisfies the equation of motion \mbox{$\ddot{\mathbf{R}}=\mathbf{0}$}, meaning that the system as a whole cannot accelerate spontaneously. In other words, the total linear momentum of the isolated system is conserved. Immediately introducing the shorthands:
\begin{equation}
\eta_{\pro,\tar}\equiv\frac{m_{\pro,\tar}}{m_\pro+m_\tar},
\end{equation}
the definitions of $\mathbf{r}$ and $\mathbf{R}$ may be inverted in order to recover the absolute projectile and target coordinates in any reference frame:
\pulleqa
\begin{align}
&\mathbf{r}_\pro=\mathbf{R}+\eta_\tar\mathbf{r},
\label{rp}\\
&\mathbf{r}_\tar=\mathbf{R}-\eta_\pro\mathbf{r},
\label{rt}
\end{align}
where the motion of the frame itself---or, equivalently, of the entire physical system within the given frame---is reflected solely through $\mathbf{R}$. Given the initial center-of-mass position $\mathbf{R}_0$, the solution to \mbox{$\ddot{\mathbf{R}}=\mathbf{0}$} is a rectilinear motion with the constant velocity $\mathbf{\V}_\cm$:
\pulleqa
\begin{equation}
\mathbf{R}(t)=\mathbf{R}_0+\mathbf{\V}_\cm t.
\label{rcm}
\pulleqb
\end{equation} 
Among all possible inertial frames, we limit our attention only to those moving along the $z$-axis (\mbox{$\mathbf{\V}_\cm=\V_\cm\hat{\mathbf{z}}$}), corresponding to the projectile's initial direction of motion. In addition, the origin of the coordinate frame will coincide with the target's initial position. Specifically, the final solution in the laboratory frame---where the target is at rest at the initial moment, so that \mbox{$\dot{\mathbf{r}}_\tar^{(\lab)}(t=0)=\mathbf{0}$} and \mbox{$\dot{\mathbf{r}}_\pro^{(\lab)}(t=0)=v_0 \hat{\mathbf{z}}$}---may always be recovered from the general solution by taking \mbox{$\V_\cm^{(\lab)}=\eta_\pro v_0$}.

In order to specify that the target is initially at the origin of the comoving frame, we complement the initial relative position~(\ref{r000}) by a consistent set of initial absolute positions:
\pulleqa
\begin{align}
&\mathbf{r}_\pro(t=0)=\R_0 \hat{\boldsymbol{\rho}}-\Big(\lim_{z_0\to\infty}z_0\Big)\hat{\mathbf{z}},
\label{rp0}\\
&\mathbf{r}_\tar(t=0)=\mathbf{0}.
\label{rt0}
\end{align}
From a definition of the center-of-mass position it now trivially follows that: \mbox{$\mathbf{R}_0=\eta_\pro\mathbf{r}_\pro(t=0)$}.

Though compelling due to several interesting technical changeless, the derivation of the Coulomb trajectories in the comoving frame is rather tedious and results in somewhat lengthy expressions. However, it is of central importance to this work so we present it in Appendix~\ref{appendix} (instead of the Supplementary note). For conciseness we only sketch here the general procedure and present the final results.
\begin{itemize}\itemsep0em
\item Starting from the known projectile trajectories~(\ref{master}) in the fixed-target frame, use the nonrelativistic kinematics from~(\ref{rp}) and (\ref{rt}) in order to obtain a Galilean boost into the comoving frame.
\item In performing a Galilean transformation, a divergent integral appears; carefully parameterize this divergence by means of a well defined limit:
\begin{equation}
\Z_0= \left(\frac{\V_\cm}{v_0}-\eta_\pro\right)\lim_{z_0\to\infty}z_0+\frac{\V_\cm}{v_0}\X \lim_{z_0\to\infty} \ln \frac{2z_0}{\e\LL},
\label{Z0}
\end{equation}
in order to isolate it from the relevant part of expression. In that, an arbitrary length scale $\LL$ appears, formally required for an argument of the logarithm to be dimensionless. A natural logarithm base $\e$ also appears here, not to be confused with the unit charge $e$.
\item Separate the parameterized divergence from the $z$-component $z_\pro$ of the particle trajectory such that:
\begin{equation}
z_\pro=\Z_\pro+\Z_0,
\label{z_sep}
\end{equation}
which is equivalent to the shift in coordinate origin by $\Z_0$. Continue calculations with the remaining, finite part of the expression:
\begin{align}
\begin{split}
\Z_\pro(\rho_\pro)=&\frac{\V_\cm}{v_0}\X\left(\frac{\rho_\pro}{\eta_\tar\R_0}+\ln\frac{\LL(\rho_\pro-\R_0)}{\eta_\tar\X\R_0}-\frac{\eta_\pro}{\eta_\tar}\right)+\\
&\left(\frac{\V_\cm}{\eta_\tar v_0}+1\right)\left(\frac{\R_0(\rho_\pro-\R_0)}{2\X}-\frac{\X(\rho_\pro-\eta_\pro\R_0)^2}{2\R_0(\rho_\pro-\R_0)}\right),
\end{split}
\label{zeta}
\end{align}
corresponding to the axial component of a projectile trajectory in the comoving frame where the center of mass moves along the $z$-axis with the speed $\V_\cm$. The boosted projectile trajectory is now fully determined:
\begin{equation}
\mathbf{r}_\pro(\rho_\pro)=\rho_\pro \hat{\boldsymbol{\rho}}+[\Z_\pro(\rho_\pro)+\Z_0]\hat{\mathbf{z}}
\end{equation}
as a function of a radial distance $\rho_\pro$ from the $z$-axis.
\end{itemize}


\section{Scattering shadow}
\label{shadow}

We now ask: at the radial distance $\rho_\pro$ from the $\Z$-axis, which trajectory reaches an extremal distance along the same axis (i.e. the extremal distance from the $xy$-plane), thus defining the point along the shadow caustic?

\noindent The problem boils down to finding the extremum\footnote{
The only (and inconsequential) difference in respect to the extremization procedure from~\cite{zugec} is that the angular parameter $\theta$ was kept constant therein, as in both the fixed-target and the center-of-mass frame it corresponds to a true angular coordinate. In any particular frame the extremization procedure must be performed by keeping some appropriate geometric parameter \textit{from the same frame} constant, as we are interested in the point of the extremal approach (among all possible trajectories) to a given point, axis or plane \textit{within that frame}. Since we already have an explicit dependence $\Z_\pro(\rho_\pro)$ from~(\ref{zeta}), we can directly extremize the trajectories' distance from the $xy$-plane by keeping $\rho_\pro$ constant, instead of first having to find the angular coordinate \mbox{$\theta_\pro=\acot(\Z_\pro/\rho_\pro)$} and then having to extremize the distance from the coordinate origin by keeping $\theta_\pro$ constant. In that, it should be noted that the extremization procedure from~\cite{zugec} was a \textit{minimization} of the distance from the coordinate origin, while~(\ref{derivative}) leads to the \textit{maximization} of the trajectories' reach in the $\Z$-direction, and only because of the selected direction of the initial projectile velocity.
} of $\Z_\pro(\rho_\pro)$ in respect to the impact parameter $\R_0$, for a constant $\rho_\pro$. This is done by finding the zero of the associated derivative, i.e. by solving:
\begin{align}
\begin{split}
\frac{\dd\Z_\pro}{\dd\R_0}\bigg|_{\tilde{\R}_0}=&\Big\{(\eta_\tar v_0+\V_\cm)\tilde{\R}_0^2(\rho_\pro-2\tilde{\R}_0)(\rho_\pro-\tilde{\R}_0)^2+\\
&\big[\eta_\tar(\eta_\tar v_0+\V_\cm)\tilde{\R}_0-(\eta_\tar v_0-\V_\cm)(\rho_\pro-\tilde{\R}_0)\big]\times\\
&\X^2\rho_\pro(\eta_\pro\tilde{\R}_0-\rho_\pro) \Big\}\Big/ \left[2\eta_\tar v_0\X\tilde{\R}_0^2(\rho_\pro-\tilde{\R}_0)^2\right]=0
\end{split}
\label{derivative}
\end{align}
for $\tilde{\R}_0$. We see that, in general case, we need to find the zero(s) of the $5^\mathrm{th}$ degree polynomial in $\tilde{\R}_0$. The general solution cannot be expressed in radicals. Even if it could, already the general solutions to the $3^\mathrm{rd}$ and $4^\mathrm{th}$ degree polynomial (Cardano formula and Ferrari method, respectively) are excessively long and incomprehensible. Therefore, we need to proceed numerically from this point. 

In order to avoid the confusion between the trajectory equation $\Z_\pro(\rho_\pro;\R_0)$ and the shadow equation, we will use the notation $\ZZ_\pro(\rho_\pro)$ for shadow caustic, in a sense:
\begin{equation}
\ZZ_\pro(\rho_\pro)\equiv \Z_\pro[\rho_\pro;\tilde{\R}_0(\rho_\pro)].
\label{zz}
\end{equation}
Evidently, the shadow caustic is obtained by adopting the extremizing value $\tilde{\R}_0(\rho_\pro)$ from~(\ref{derivative}). There \textit{is} one constraint upon the sought solution:
\begin{equation}
0\le \tilde{\R}_0(\rho_\pro) \le \rho_\pro,
\label{confine}
\end{equation}
that one might hope to use in eliminating the spurious solutions. It follows from purely geometric considerations: in order to have reached the radial distance $\rho_\pro$, the projectile must have started from a lesser $\tilde{\R}_0$, due to the repulsive scattering \textit{away} from the $z$-axis. However, there may still be multiple branches $\tilde{\R}_0^{(i)}$ consistent with~(\ref{confine}), among which only one yields a sought solution at any given point. A detailed numerical investigation leads to a simple procedure for obtaining the shadow boundary under such conditions:
\begin{equation}
\ZZ_\pro(\rho_\pro)=\max_i\big\{\Z_\pro[\rho_\pro;\tilde{\R}_0^{(i)}(\rho_\pro)]\big\},
\label{max}
\end{equation}
where $i$ enumerates all the solutions consistent with~(\ref{confine}). The basic reasoning behind this procedure may be easily understood. All nonnegative solutions to the extremization problem~(\ref{derivative}) do indeed yield \textit{some} meaningful, local extremum. However, the shadow caustic is determined by the \textit{global} maximum, beyond which no additional trajectories are to be found (we recall that the \textit{maximum} is the relevant extremum only because of the selected direction of initial projectile velocity). Thus, beyond the maximum of all local extrema no other extremum, as a candidate for a point on a shadow caustic, can be found.



\vspace*{-4mm}

\section{Existence conditions}
\label{existence}

Let us try to anticipate any possible conditions for the existence of the scattering shadow in a given comoving frame. In attempting this, a \textit{shadow vertex}---a single shadow point lying on the \mbox{$\Z$-axis}, i.e. \mbox{$\ZZ_\pro(\rho_\pro=0)$}---will be of special importance. To this end let us consider what happens with the projectile trajectory impinging frontally upon the target (\mbox{$\R_0=0$}). It is to be noted that the shadow vertex is completely determined by precisely this one trajectory, which is entirely confined along the \mbox{$\Z$-axis}. If, in a given comoving frame, the projectile with the impact parameter \mbox{$\R_0=0$} can be backscattered (recoiled backwards off the target), then the shadow vertex stays at some finite position along the \mbox{$\Z$-axis}. On the other hand, if the backscattering is kinematically impossible (due to the projectile being too massive or the forward center-of-mass speed $\V_\cm$ being too high), the projectile keeps moving in a forward direction, implying that the shadow vertex escapes to infinity, even after the primary shift to infinity by $\Z_0$!

It is now a simple matter to argue that if the \mbox{$\R_0=0$} trajectory is entirely forward directed, than all other \mbox{$\R_0>0$} trajectories also retain the forward motion without ever bending backwards. To this end we first consider the motion in the center-of-mass frame. At the initial moment all projectiles are put into motion with the velocity \mbox{$\mathbf{v}_\pro^{(\cm)}(0)=\eta_\tar v_0\hat{\mathbf{z}}$}. This speed value also corresponds to a maximum speed component \mbox{$\mathcal{V}_z^{(\mathrm{max})}$} along the same axis, for any trajectory \textit{and} among all trajectories (for any $\R_0$):
\begin{equation}
\mathcal{V}_z^{(\mathrm{max})}\equiv \max_{t,\R_0} \big[\mathbf{v}_\pro^{(\cm)}(t;\R_0)\big]_z=\big[\mathbf{v}_\pro^{(\cm)}(0;\R_0)\big]_z=\eta_\tar v_0.
\end{equation}
The same speed is also reached asymptotically by all projectile trajectories: \mbox{$v_\pro^{(\cm)}(\infty)=\eta_\tar v_0$}. Yet, due to the particular scattering angle, only for the frontal trajectory (\mbox{$\R_0=0$}) is the final velocity directed entirely along the $z$-axis, meaning that this particular case yields the minimum speed component \mbox{$\mathcal{V}_z^{(\mathrm{min})}$}, for this particular trajectory \textit{and} among all possible trajectories:
\begin{equation}
\mathcal{V}_z^{(\mathrm{min})}\equiv \min_{t,\R_0} \big[\mathbf{v}_\pro^{(\cm)}(t;\R_0)\big]_z=\big[\mathbf{v}_\pro^{(\cm)}(\infty;0)\big]_z=-\eta_\tar v_0.
\end{equation}
Therefore, if the center-of-mass speed $\V_\cm$ is sufficient to boost forward the asymptotic state of the frontal trajectory (i.e. if \mbox{$\mathcal{V}_z^{(\mathrm{min})}+\V_\cm\ge0$}, being equivalent to \mbox{$\V_\cm\ge\eta_\tar v_0$}) then all other trajectories will also be forward directed. This has a remarkable consequence: there will be no back-bending of any trajectory, so that when the frontal trajectory escapes to infinity (beyond $\Z_0$), there is a continuum of entirely-forward-directed trajectories sweeping the entire geometric space. Thus, the scattering shadow cannot exist at all in comoving frames with \mbox{$\V_\cm\ge\eta_\tar v_0$}!

There is another extreme to this condition. As \mbox{$\mathcal{V}_z^{(\mathrm{max})}$} is the maximum speed along the $z$-axis for any and all trajectories, consider what happens when \mbox{$\mathcal{V}_z^{(\mathrm{max})}+\V_\cm\le0$}, i.e. when \mbox{$\V_\cm\le-\eta_\tar v_0$}. In such frames all the projectiles are immediately boosted backwards and the entire geometric space beyond their initial position is shielded from their trajectories. This means that the scattering shadow spans the entire geometric space. Its form may still be properly parameterized, being trivial\footnote{
It is not as trivial to obtain it formally as a limit \mbox{$\lim_{\V_\cm\to-\eta_\tar v_0}\ZZ_\pro(\rho_\pro;\V_\cm)$} of the procedure from the Section~\ref{shadow}. Taking another glance at~(\ref{derivative}), we may notice that precisely in the limiting case \mbox{$\V_\cm=-\eta_\tar v_0$} the extremization condition simplifies from the 5$^\mathrm{th}$ degree to the 2$^\mathrm{nd}$ degree polynomial, i.e. the quadratic equation in $\tilde{\R}_0$, yielding the two solutions: \mbox{$\tilde{\R}_0^{(1)}=\rho_\pro$} and \mbox{$\tilde{\R}_0^{(2)}=\rho_\pro/\eta_\pro$}. The second solution is clearly unacceptable, as seen from~(\ref{confine}), since \mbox{$0\le\eta_\pro\le 1$}. Though the first solution is also unacceptable---as \mbox{$\tilde{\R}_0=\rho_\pro$} can only be for the frontal trajectory (\mbox{$\R_0=0$})---it clearly represents the limit of the acceptable solutions, as $\V_\cm$ approaches \mbox{$-\eta_\tar v_0$}.  However, one can easily check that plugging this solution into~(\ref{zz}) does not help in identifying the trivial-shadow equation~(\ref{flat}), due to no apparent connection between \mbox{$\rho_\pro-\tilde{\R}_0$} and $z_0$ in the limit \mbox{$\V_\cm\to-\eta_\tar v_0$}. Finally, it is interesting to explicitly state the physical meaning behind the frame defined by \mbox{$\V_\cm=-\eta_\tar v_0$}. From the definition of the center-of-mass speed we can easily see that in this frame, instead of the target, the projectile is at rest at the initial moment. Thus, we may think of it as the \textit{inverse-laboratory frame}.
} in itself. One only needs to note that the shadow caustic is now a plane spanning the geometric place of all initial projectile positions. Ever since~(\ref{rp0}), we have had these positions parameterized as: \mbox{$z_\pro(t=0;\R_0)=-\lim_{z_0\to\infty}z_0$}. By the virtue of~(\ref{z_sep}) and (\ref{zz}) we may immediately write:
\vspace*{-1.5mm}
\begin{equation}
\ZZ_\pro(\rho_\pro;\V_\cm\le-\eta_\tar v_0)=-\lim_{z_0\to\infty}z_0-\Z_0,
\label{flat}
\vspace*{-2mm}
\end{equation}
which we call a \textit{trivial} scattering shadow.



In summary, the scattering shadow in the comoving frame can exist and take a nontrivial form if and only if:
\vspace*{-4mm}
\begin{equation}
-\eta_\tar v_0<\V_\cm<\eta_\tar v_0 \quad\Leftrightarrow\quad \eta_\tar>\frac{|\V_\cm|}{v_0}.
\label{condition}
\vspace*{-2mm}
\end{equation}
The left expression is to be interpreted as the condition upon the center-of-mass speed, given the projectile and target masses. The right expression is the condition upon their respective masses, given the center-of-mass speed. Since \mbox{$0\le \eta_\tar\le1$}, these conditions mean that there cannot possibly exist a (nontrivial) scattering shadow in the comoving frames such that \mbox{$|\V_\cm|\ge v_0$}, regardless of the selection of the projectile and target masses.



\section{Laboratory frame}
\label{lab_frame}

\begin{figure*}[t!]
\centering
\includegraphics[width=1\linewidth,keepaspectratio]{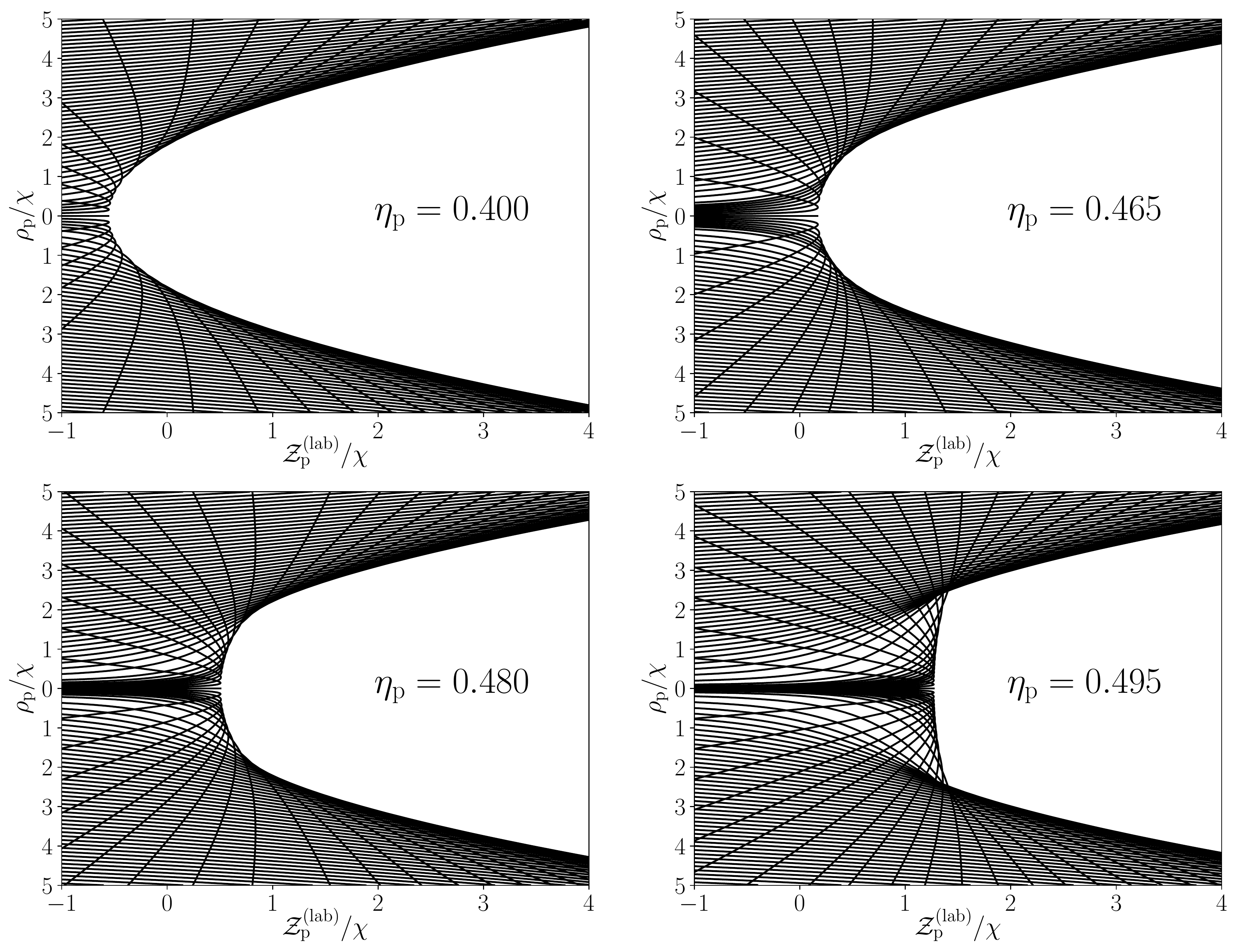}
\pull
\caption{Examples of the projectile trajectories in the laboratory frame for several selected values of $\eta_\pro$. The trajectories' envelope forms a scattering shadow caustic. For visual purposes the impact parameters are not all selected as equidistant, thus the plots do not display the true density of the trajectories. For each separate value of $\eta_\pro$, the origin shift $\Z_0$ from~(\ref{Z0}) is different, therefore the care should be taken in interpreting the transformed coordinate $\Z_\pro^{(\lab)}$. The scattering shadow in the laboratory frame exists only for $\eta_\pro<0.5$, otherwise the trajectories sweep the entire geometric space.}
\pullc
\label{fig1}
\end{figure*}

In this section we commit ourselves to the laboratory frame, where the target is at rest at the initial moment: \mbox{$v_\tar^{(\lab)}(0)=0$}. Since the projectile carries all of the initial relative speed: \mbox{$v_\pro^{(\lab)}(0)=v_0$}, the center-of-mass speed in the laboratory frame is: \mbox{$\V_\cm^{(\lab)}=\eta_\pro v_0$}. Plugging this into~(\ref{zeta}) and choosing \mbox{$\LL=\X$} for an arbitrary length scale, we obtain the particle trajectories in the laboratory frame:
\begin{align}
\begin{split}
\Z_\pro^{(\lab)}(\rho_\pro)=&\frac{\X(\rho_\pro-\eta_\pro\R_0)[(\eta_\pro-\eta_\tar)\rho_\pro-\eta_\pro\R_0]}{2\eta_\tar\R_0(\rho_\pro-\R_0)}\\
&+\frac{\R_0(\rho_\pro-\R_0)}{2\eta_\tar\X}+\eta_\pro\X\ln\frac{\rho_\pro-\R_0}{\eta_\tar\R_0}.
\end{split}
\label{zeta_lab}
\end{align}
It should be noted that after selecting the natural scale \mbox{$\LL=\X$}, the trajectory and and all the results following from it may be expressed in a scaled, dimensionless coordinates \mbox{$\bar{x}\equiv x/\X$}, where \mbox{$x\in\{\R_0,\rho_{\pro,\tar},z_{\pro,\tar},\Z_{\pro,\tar},\ZZ_{\pro,\tar},\dots\}$}. Though this universal form---independent of the underlying parameters from $\X$---may already be applied to a general case from~(\ref{zeta}), for illustrative purposes we only give the example of a scaled version of~(\ref{zeta_lab}):
\begin{align}
\begin{split}
\bar{\Z}_\pro^{(\lab)}(\bar{\rho}_\pro)=&\frac{(\bar{\rho}_\pro-\eta_\pro\bar{\R}_0)[(\eta_\pro-\eta_\tar)\bar{\rho}_\pro-\eta_\pro\bar{\R}_0]}{2\eta_\tar\bar{\R}_0(\bar{\rho}_\pro-\bar{\R}_0)}\\
&+\frac{\bar{\R}_0(\bar{\rho}_\pro-\bar{\R}_0)}{2\eta_\tar}+\eta_\pro\ln\frac{\bar{\rho}_\pro-\bar{\R}_0}{\eta_\tar\bar{\R}_0}.
\end{split}
\label{Z_scaled_lab}
\end{align}
We use such scaled coordinates for displaying all graphical results. In that, figure~\ref{fig1} shows examples of the projectile trajectories in the laboratory frame---according to~(\ref{Z_scaled_lab})---for several values of~$\eta_\pro$. According to the existence conditions from~(\ref{condition}) the scattering shadow in the laboratory frame exists only for $\eta_\pro<0.5$.



\begin{figure}[t!]
\centering
\includegraphics[width=1\linewidth,keepaspectratio]{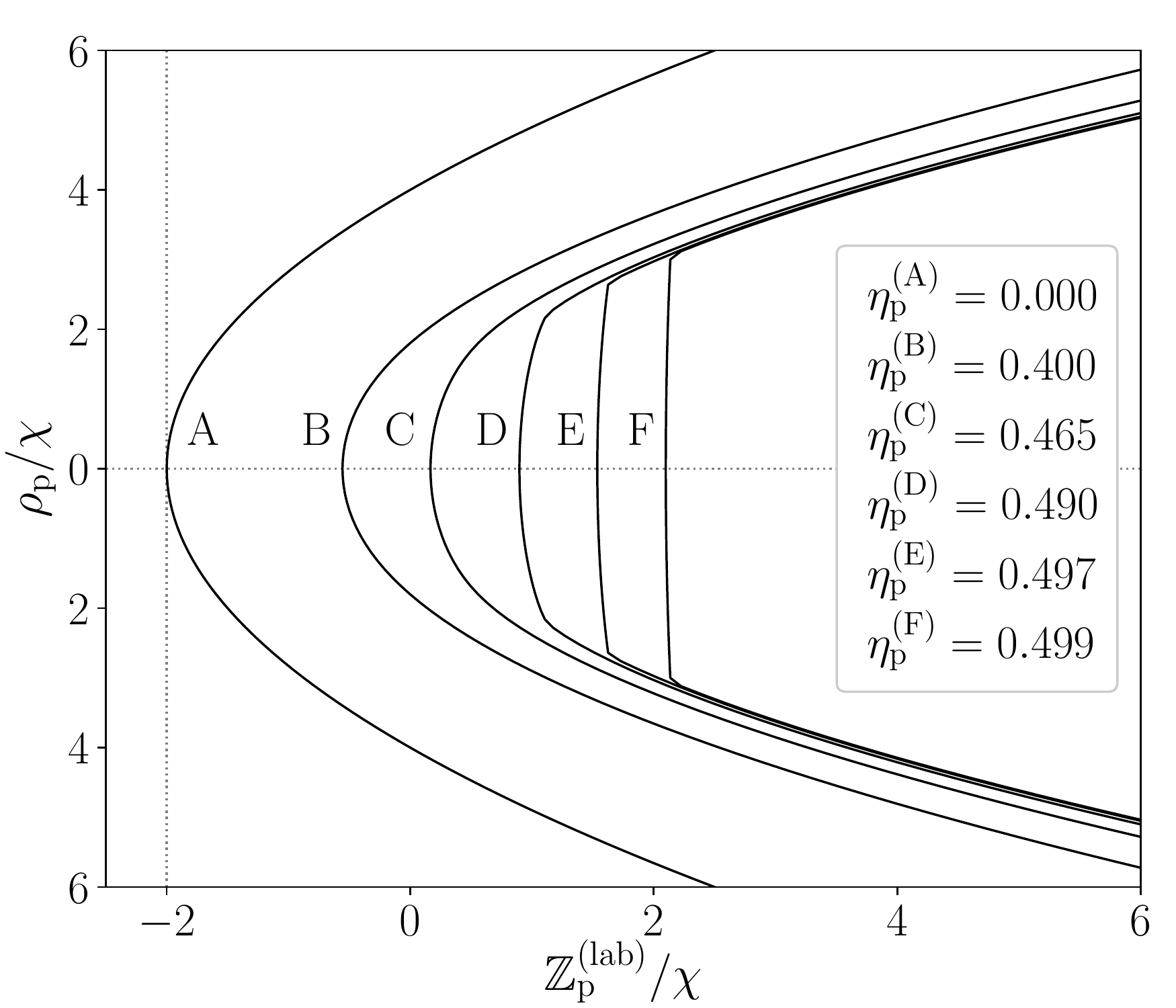}
\pull
\caption{Examples of the scattering shadow in the laboratory frame, obtained by solving~(\ref{lab_r0}) for $\tilde{\R}_0$ and employing~(\ref{max}). The shadow exists in the laboratory frame only for \mbox{$\eta_\pro<0.5$}. The relative positioning of shadows in a transformed coordinate $\ZZ_\pro^{(\lab)}$ does not reflect their true geometric positioning. The case \mbox{$\eta_\pro^{(\mathrm{A})}=0$} corresponds to an infinitely massive target, when the laboratory frame coincides both with the fixed-target and the center-of-mass frame, with the shadow taking a simple paraboloidal form \mbox{$\ZZ_\pro^{(\lab)}/\X=(\rho_\pro/\X)^2/8-2$} from~\cite{zugec}.}
\vspace*{-4mm}
\label{fig2}
\end{figure}

Figure~\ref{fig2} recovers several selected shadow forms directly, by following all the steps required by the shadow determination procedure from~(\ref{max}). The central part of this procedure is the numerical identification of the relevant roots to the $5^\mathrm{th}$ degree polynomial from the associated extremization condition for determining $\tilde{\R}_0$:
\begin{align}
\begin{split}
\frac{\dd \Z_\pro^{(\lab)}}{\dd \R_0}\bigg|_{\tilde{\R}_0}=&\Big\{\tilde{\R}_0^2(\rho_\pro-2\tilde{\R}_0)(\rho_\pro-\tilde{\R}_0)^2-\X^2\rho_\pro\times\\
& \big[\eta_\pro(\eta_\pro-2\eta_\tar)\R_0^2+2(\eta_\tar-\eta_\pro^2)\R_0\rho_\pro+\\
&(\eta_\pro-\eta_\tar)\rho_\pro^2\big] \Big\}\Big/ \left[2\eta_\tar \X\tilde{\R}_0^2(\rho_\pro-\tilde{\R}_0)^2\right]=0 .
\end{split}
\label{lab_r0}
\end{align}
It should be noted that both in figure~\ref{fig1} and \ref{fig2} each particular shadow has been shifted by a separate value of $\Z_0$, so that in a transformed coordinate $\Z_\pro$ the separate shadows do not reflect their true relative geometric positioning, as they do in a true spatial coordinate $z_\pro$. 





Something remarkable may be observed in figures~\ref{fig1} and \ref{fig2}. For the values of $\eta_\pro$ above approximately 0.49 there is a discontinuity among the projectile trajectories which contribute to the formation of a shadow caustic, causing the scattering shadow to exhibit a sharp edge. This is indeed the case and not just the visual artifact. The sudden breakdown of the shadow smoothness is directly related to a qualitative change in the behavior of solutions to the extremization problem~(\ref{derivative}) and the appearance of the multiple roots $\tilde{\R}_0^{(i)}$ consistent with~(\ref{confine}). Singling out the relevant root is the precise purpose of the method from~(\ref{max}). Based on these observations, one can rightly claim that this effect has a mathematical form of a \textit{phase transition}---a new and by no means obvious phenomenon that does not manifest itself either in the fixed-target or the center-of-mass frame.

\begin{figure}[t!]
\centering
\includegraphics[width=1\linewidth,keepaspectratio]{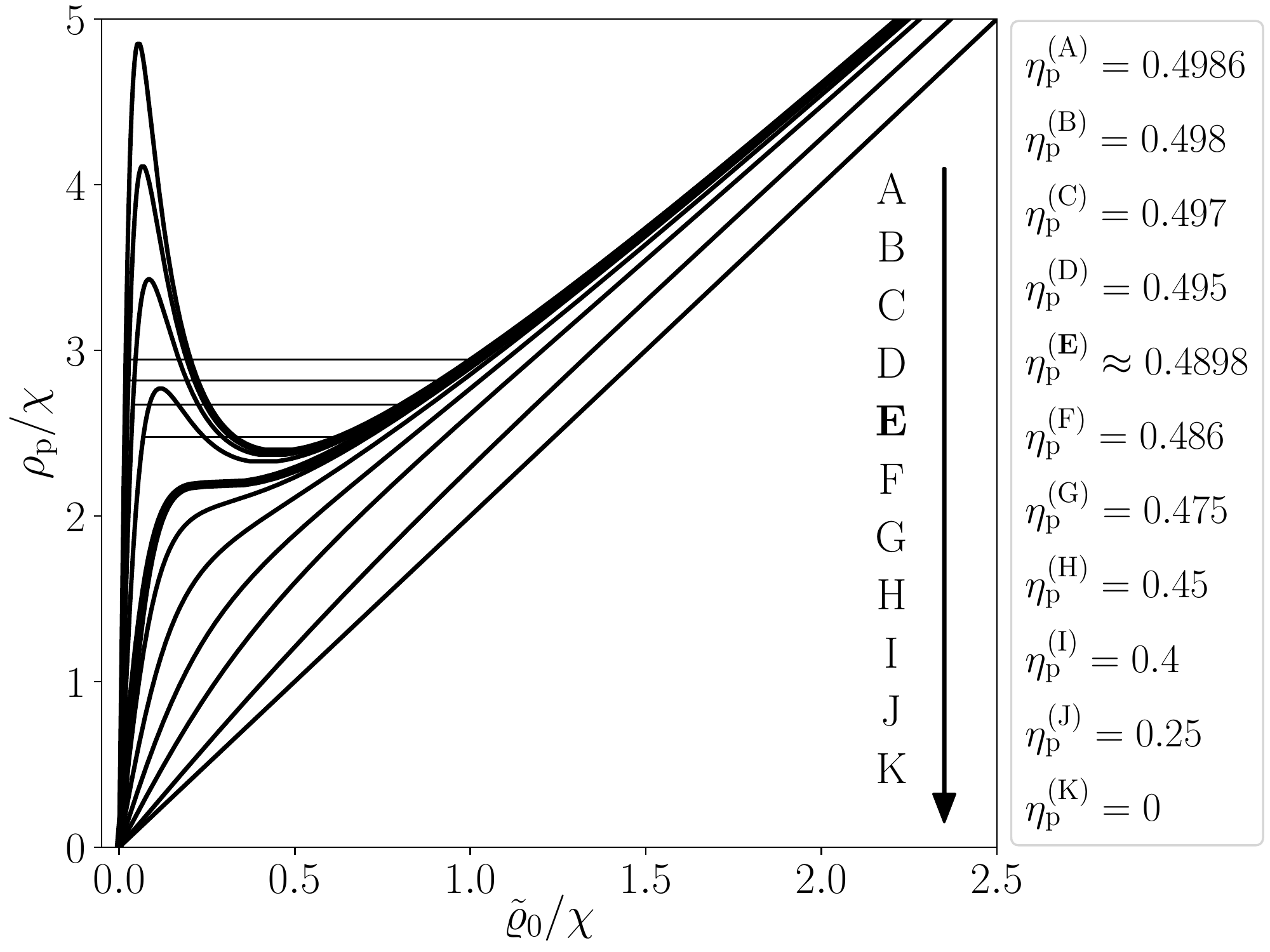}
\pull
\caption{Solutions $\tilde{\R}_0(\rho_\pro)$---consistent with~(\ref{confine})---to the extremization condition~(\ref{lab_r0}) from the laboratory frame, for different values of the projectile/target mass ratio, expressed through $\eta_\pro$. Inverse dependence $\rho_\pro(\tilde{\R}_0)$ is shown in order to illustrate the analogy with the thermodynamic van der Waals model. Cases $\mathrm{A}$--$\mathrm{D}$ show a split shadow phase; cases $\mathrm{F}$--$\mathrm{K}$ show a smooth shadow phase. Case~$\mathbf{E}$ (a thick line) corresponds to a critical value of $\eta_\pro$, when a phase transition between the shadow phase states occurs. Examples of the Maxwellian construction are shown by horizontal line segments.}
\pullc
\label{figx}
\end{figure}

For a few selected values of $\eta_\pro$, figure~\ref{figx} shows the behavior of the solutions $\tilde{\R}_0(\rho_\pro)$---consistent with~(\ref{confine})---to the extremization condition~(\ref{lab_r0}) from the laboratory frame. In order to reinforce the notion of a phase transition, at least by mathematical analogy, the inverse dependence $\rho_\pro(\tilde{\R}_0)$ is deliberately shown, so as to remind the reader of a phase transition in a well known van der Waals equation of state, describing the thermodynamic behavior of real gases. Within this analogy, each curve from figure~\ref{figx} is reminiscent of a particular isotherm from a pressure-volume diagram of the van der Waals model. Thus, the following analogies may be identified between its thermodynamic parameters---temperature $T$, (molar) volume $V$ and pressure $p$---and the Rutherford scattering parameters: \mbox{$\eta_{\pro/\tar}\leftrightarrow T$}, \mbox{$\tilde{\R}_0\leftrightarrow V$} and \mbox{$\rho_\pro\leftrightarrow p$}.


A phase transition in the van der Waals model clearly corresponds to a thermodynamic transition between the liquid and gaseous state of matter. A physical interpretation of a phase transition in the Rutherford scattering shadow is more difficult to identify. The reason is the nature of parameters governing the phase states. The phase transition in the scattering shadow can be achieved\footnote{
It is shown in Section~F of the Supplementary note that for a given mass ratio ($\eta_\pro$ or, equivalently, $\eta_\tar$) one can \textit{always} achieve a a phase transition by varying the ratio $V_\cm/v_0$. However, for a given $V_\cm/v_0$ one can achieve a phase transition by varying the mass ratio \textit{only if} \mbox{$\V_\cm/\eta_\tar v_0\notin[0,6\sqrt{15}/25]\approx[0,0.9295]$}. This demonstration is outside of the scope of calculations presented herein.
} (1)~by varying the projectile/target mass ratio from $\eta_\pro$, (2)~by varying the ratio \mbox{$V_\cm/v_0$} of relevant speeds, thus switching between the comoving frames, (3)~or by varying both ratios simultaneously, as in the case of all possible laboratory frames, defined by \mbox{$\V_\cm^{(\lab)}/v_0=\eta_\pro$}. Unlike the thermodynamic parameters from the van der Waals model, the projectile and target masses can hardly be varied---either continuously or at all---thus obscuring the physical interpretation of the phase transition by precluding its realization in practice. On the other hand, inducing a phase transition by switching between the comoving frames makes both the shadow itself and (the possibility of) its phase transition observer-dependent. This means that the shadow phase is determined by the observer's point of view, rather than being an intrinsic property of the scattering shadow itself, again obstructing the physical interpretation of its phase transition. For this reason it should be reemphasized that the phase transition in the scattering shadow is only a matter of a mathematical analogy, lacking in physically meaningful `parameters of state' that would uniquely determine the shadow state upon which all observers could agree. Rather, the phase transition occurs in the observer's own kinematic relation to (and in his/her own perception of) the portion of space shielded from the projectile trajectories. This is because---for simultaneously released projectiles---the shadow caustic is not formed simultaneously. It ensues from the intersection of the trajectories' \textit{geometric forms}, without them actually passing through the same point in space at the same point in time. For a moving observer different points alongside the projectile trajectory are shifted by a different amount (\mbox{$\mathbf{\V}_\cm t$}), thus affecting the shape of the entire trajectory, rather than just translating it between the frames in relative motion. Since the trajectories themselves deform, so does the locus of their geometric intersections. This justifies any and all shadow caustic distortions between the frames in relative motion, including the possibility of a sudden and significant qualitative change, mathematically manifested as a type of a phase transition.


We have already noted that after selecting the natural scale \mbox{$\LL=\X$} in general expressions~(\ref{Z0}) and (\ref{zeta}), just as we have done in (\ref{zeta_lab}), the Rutherford scattering problem becomes scale invariant. Van der Waals equation also exhibits the scale invariance when expressed in terms of the so-called reduced thermodynamic variables: \mbox{$(\bar{p}+3/\bar{V}^2)(3\bar{V}-1)=8\bar{T}$}, each scaled by an appropriate critical value: \mbox{$\bar{T}=T/T_\mathrm{c}$}, \mbox{$\bar{p}/p_\mathrm{c}$}, \mbox{$\bar{V}=V/V_\mathrm{c}$}. By now it should not be surprising at all that that the phase transition in the scattering shadow also features its own (analogies of) critical parameters. Axes labels in figure~\ref{figx} help in observing that for both $\rho_\pro$ and $\tilde{\R}_0$ the analogy of the critical, scaling parameter is the length scale $\X$. The existence and the value of the critical projectile/target mass ratio is dependent on a particular comoving frame, but when it exists it is a critical value in a true sense. The constraint \mbox{$\V_\cm^{(\lab)}=\eta_\pro v_0$} defining the laboratory frame for a particular value of $\eta_\pro$ is such that the critical value of $\eta_\pro$ does exist among all possible laboratory frames (for different $\eta_\pro$). In figure~\ref{figx} the critical value \mbox{$\eta_\pro\approx0.489756$}, i.e. \mbox{$m_\pro/m_\tar\approx0.959846$} is illustrated by a thick line corresponding to the case~$\mathbf{E}$.\linebreak

The parallels with the van der Waals model go so far that the phase transition in the scattering shadow even features its own analogy of the Maxwell construction. Within the van der Waals model the Maxwell construction consists of a manual correction of the smooth but unphysical $p(V)$ dependence below the critical temperature $T_\mathrm{c}$, when the phase transition just becomes possible to achieve. The analogy of the Maxwell construction for the scattering shadow---that we call a \textit{Maxwellian construction}---comes about as a consequence of the recipe~(\ref{max}) for selecting the appropriate, shadow related solution $\tilde{\R}_0$. In figure~\ref{figx} the examples of the Maxwellian construction are shown by horizontal segments, appearing beyond (and only beyond) the critical value of $\eta_\pro$. In case of the scattering shadow the solution $\tilde{\R}_0(\rho_\pro)$ discontinuously switches between the leftmost and rightmost branch, so that no solution $\tilde{\R}_0$, i.e. no projectile trajectory within the range of Maxwellian construction takes part the shadow formation. Precisely this discontinuity leads to the appearance of a sharp edge along the shadow caustic.

There are two related differences in respect to the van der Waals model that need to be clarified, so as to avoid any confusion. The first, inessential one is a matter of nomenclature: what we regard as a phase transition, i.e. what we consider as distinct phases in the scattering shadow. The second, essential one is related to the attainability of the thermodynamic states along the Maxwell construction and of the solutions $\tilde{\R}_0(\rho_\pro)$ along the Maxwellian construction. In the van der Waals model the $p(V)$ dependences from each side of the Maxwell construction correspond to the separate, liquid and gaseous phase states. The Maxwell construction corresponds to a phase transition itself, representing a realistic mixture of these phase states, through which the evolution of a thermodynamic system progresses in reality. In case of the scattering shadow we consider the entire $\tilde{\R}_0(\rho_\pro)$ dependences for particular values of $\eta_\pro$ as separate phases, based on the (non)existence of the sharp edge along the shadow caustic. Recalling the shadow existence condition~(\ref{condition}), we distinguish between four phase states: a \textit{smooth shadow} phase, a \textit{split shadow} phase, a \textit{trivial shadow} phase and \textit{no shadow} phase. In this sense, only the critical line (case~$\mathbf{E}$) from figure~\ref{figx} represents the phase transition, between a split (cases $\mathrm{A}$--$\mathrm{D}$) and a smooth (cases $\mathrm{F}$--$\mathrm{K}$) shadow. It also bears repeating that the solutions $\tilde{\R}_0(\rho_\pro)$ along the Maxwellian construction are excluded from the shadow formation, i.e. are not attainable in a sense of a liquid-gas mixture from the van der Waals model. In a direct analogy with the van der Walls model, the smooth shadow would represent a pure gaseous state, the only one present above the critical temperature. Within the split shadow phase the sharp edge along the shadow caustic---manifested in the Maxwellian construction---would correspond to the mixture of phase states, while the two split shadow branches would correspond to a pure liquid and a pure gaseous state. In that, the branch containing the shadow vertex (\mbox{$\rho_\pro=0$}) is reminiscent of a liquid state---as its portion increases by driving the relevant $\eta_\pro$ parameter away from the smooth shadow phase---leaving the other, asymptotically paraboloidal branch to represent a gaseous state. If one's imagination was let to run free, one could make further comparisons of a trivial and no shadow phase with the plasma and solid state of matter, respectively.




\vspace*{-2mm}

\section{Conclusion}
\label{conclusion}

We have established a method for obtaining the Rutherford scattering shadow in the comoving frame---an inertial frame moving along the initial projectile direction, with the charged target initially being at the origin of the frame. The laboratory frame fits this categorization perfectly, being defined by an additional requirement that the target be initially at rest. The method itself consists of the extermization procedure related to the projectile trajectories in the comoving frame and involves solving the $5^\mathrm{th}$ degree polynomial. We have identified the condition for the existence of the (nontrivial) scattering shadow in a given comoving frame, which puts a limitation on its speed of motion, relative to the center-of-mass frame. The trivial scattering shadow refers to the entire geometric space being shadowed, when all the projectile trajectories are immediately boosted backwards due to the excessive center-of-mass speed in a backward direction. In the laboratory frame, in particular, the scattering shadow exists only if the target is more massive than the projectile ($m_\tar>m_\pro$). Otherwise, the projectile trajectories sweep the entire geometric space, not forming any shadow at all. A new phenomenon was identified, related to a transition between the comoving frames or, alternatively, to a varying ratio of the projectile and target masses. It consists of a phase transition between an entirely smooth type of scattering shadow and the one characterized by a loss of smoothness at the point where the shadow caustic splits, forming a sharp edge. This finding shows that the transition between the frames is not just a technical challenge to be carried out for the sake of completeness. Rather, it is a rewarding venture, offering a novel insight into otherwise well known scattering process.



{
\appendix

\vspace*{-3mm}

\section{Derivation of the Coulomb trajectories in the comoving frame}
\label{appendix}

\renewcommand{\theequation}{\thesection.\arabic{equation}}

We need to determine the time dependence of the trajectories in order to be able to boost them via the $\mathbf{R}(t)$ term from~(\ref{rcm}). We start from (see, for example~\cite{zugec}):
\vspace*{-1mm}
\begin{equation}
\frac{\dd\theta}{\dd t}=-\frac{\R_0 v_0}{r^2(\theta)} \quad\Rightarrow\quad \int_0^{t(\theta)} \hspace*{-4mm} \dd t'=-\frac{1}{\R_0 v_0}\lim_{\theta_0\to\pi}\int_{\theta_0}^\theta r^2(\theta')\dd\theta',
\end{equation}
which is equivalent to the conservation of the angular momentum. It is to be noted that the angular coordinate $\theta$ is still the one from the fixed-target frame and will remain so throughout the entirety of our calculations. Therefore, it is not to be confused with the actual angular coordinate from the given comoving frame. Since the initial angular coordinate of the projectile ($\theta_0\to\pi$) leads to the divergence in the right-hand-side integral, based on~(\ref{rp0}) we parameterize it as: \mbox{$\lim_{\theta_0\to\pi}\theta_0=\pi-\lim_{z_0\to\infty}\atan(\R_0/z_0)$}, so that:
\begin{equation}
t(\theta)=-\frac{1}{\R_0 v_0}\lim_{z_0\to\infty}\int_{\pi-\atan(\R_0/z_0)}^\theta \hspace*{-17mm} r^2(\theta')\dd\theta'.
\label{app_t}
\end{equation}
It should be stated that for finite $z_0$ (i.e. $\theta_0 \not\to\pi $) the solution~(\ref{master}) for $r(\theta)$ does not hold any more. However, we will keep $z_0$ infinite at all times, while this parameterization only serves to do so in a strictly controlled and formally correct manner. The antiderivative of $r^2(\theta)$ is:
\begin{align}
\begin{split}
\int^\theta r^2(\theta')d\theta'=&-\big(\R_0\cos\theta+\X\sin\theta\big) r(\theta)\\
&+\X\R_0\ln \left[\big(\R_0\tan\tfrac{\theta}{2}-\X\big)/\LL\right],
\end{split}
\end{align}
with $\LL$ as an arbitrary length scale, formally required for an argument of the logarithm to be dimensionless. In entering the lower integration bound from~(\ref{app_t}) we make use of the following limits:
\begin{align}
&\lim_{\theta_0\to\pi} r(\theta_0)\sin\theta_0=\R_0,
\label{app_lim_sin}\\
&\lim_{\theta_0\to\pi} r(\theta_0)\cos\theta_0=-\lim_{z_0\to\infty} z_0,
\label{app_lim_cos}\\
&\lim_{\theta_0\to\pi}\ln\left[\big(\R_0\tan\tfrac{\theta_0}{2}-\X\big)/\LL\right]=\lim_{z_0\to\infty} \ln \big(2z_0/\LL\big).
\label{app_lim_ln}
\end{align}
Using these in the context of~(\ref{app_t}), we have:
\begin{align}
\begin{split}
t(\theta)=\frac{1}{v_0}& \Big[\big(\cos\theta+\tfrac{\X}{\R_0}\sin\theta\big) r(\theta)\\
&-\X\ln \left[\big(\R_0\tan\tfrac{\theta}{2}-\X\big)/\LL\right]\\
&+\lim_{z_0\to\infty} z_0+\X\lim_{z_0\to\infty} \ln \big(2z_0/\e\LL\big)\Big].
\end{split}
\label{app_t_th}
\end{align}
Here we have absorbed an additional $-\X$ term---related to the limit~(\ref{app_lim_sin})---within the last logarithm by introducing the natural logarithm base $\e$ (not to be confused with the unit charge $e$).

Combining~(\ref{rp}) and (\ref{rp0}), and recalling that \mbox{$\mathbf{R}_0=\eta_\pro\mathbf{r}_\pro(t=0)$}, a particle trajectory in an arbitrary comoving frame where the center of mass moves along the $z$-axis with the velocity \mbox{$\mathbf{\V}_\cm=\V_\cm\hat{\mathbf{z}}$}:
\begin{equation}
\mathbf{r}_\pro(\theta)=\mathbf{R}_0+\mathbf{\V}_\cm t(\theta)+\eta_\tar\mathbf{r}(\theta)
\label{app_r_trans}
\end{equation}
is easily decomposed into the cylindrical components as:
\begin{equation}
\rho_\pro(\theta)=\mathbf{r}_\pro(\theta)\cdot\hat{\boldsymbol{\rho}}=\eta_\pro\R_0+\eta_\tar r(\theta)\sin\theta,
\label{app_rhop}
\end{equation}
together with:
\begin{align}
\begin{split}
z_\pro(\theta)=&\mathbf{r}_\pro(\theta)\cdot\hat{\mathbf{z}}=-\eta_\pro\lim_{z_0\to\infty}z_0 +\V_\cm t(\theta)+\eta_\tar r(\theta)\cos\theta\\
=&\Z_\pro(\theta)+\Z_0,
\end{split}
\label{app_zp}
\end{align}
where we have absorbed all the infinities into:
\begin{equation}
\Z_0\equiv \left(\frac{\V_\cm}{v_0}-\eta_\pro\right)\lim_{z_0\to\infty}z_0+\frac{\V_\cm}{v_0}\X \lim_{z_0\to\infty} \ln \frac{2z_0}{\e\LL},
\label{app_Z0}
\end{equation}
and isolated all the relevant dependence by:
\begin{align}
\begin{split}
\Z_\pro(\theta)\equiv&\left(\frac{\V_\cm}{v_0}+\eta_\tar\right) r(\theta)\cos\theta +\frac{\V_\cm}{v_0} \frac{\X}{\R_0} r(\theta)\sin\theta\\
&-\frac{\V_\cm}{v_0}\X\ln \left[\big(\R_0\tan\tfrac{\theta}{2}-\X\big)/\LL\right].
\end{split}
\label{app_zetap}
\end{align}
It bears repeating that the angle $\theta$ is still the one from the fixed-target frame, parameterizing the trajectory in a selected comoving frame. This is precisely why the projections of a target-relative projectile position $\mathbf{r}(\theta)$ may be and have been performed simply by taking \mbox{$r(\theta)\sin\theta$} and \mbox{$r(\theta)\cos\theta$}. From $\Z_0$ and \mbox{$\mathbf{r}_\tar=\mathbf{r}_\pro-\mathbf{r}$} we see that that the Coulomb interaction is strong enough that the target is pushed infinitely far by a recoil before the projectile manages to approach it at some finite distance (\mbox{$\theta<\pi$}). This may have been suspected, but not \textit{a priori} expected from the fixed-target and the center-of-mass frame. In all frames the infinite distance must be negotiated between the projectile and the target, but that does not mean \textit{in advance} that the target itself is shifted by an infinite distance from its initial position. Seeing now that it is in any of the comoving frames, we introduced the new coordinate origin $\Z_0$, thus defining the new, transformed coordinate \mbox{$\Z\equiv z-\Z_0$}. In other words, we are now observing the scattering around the point $\Z_0$ at an infinite distance from the target's initial position, so that the transformed coordinate $\Z$ is entirely under control. Since $\theta$ is an angular coordinate from a fixed-target frame, rather than from the comoving frame, no geometric redefinition of it needs to take place due to this origin shift.

We now have the projectile trajectory parameterized by the cylindrical coordinates from~(\ref{app_rhop}) and (\ref{app_zetap}), dependent on the angle $\theta$ from the fixed-target frame. In order to determine the scattering shadow in the comoving fame, we need to find the extremum of some specific trajectory distance by keeping fixed some geometric parameter \textit{from the same frame}. For example, we might extremize the distance from the origin by keeping fixed an angle relative to the $\Z$-axis, or the distance along the $\Z$-axis by keeping fixed the distance $\rho_\pro$ from the same axis. To this end, we aim to translate the dependence $\Z_\pro(\theta)$ from~(\ref{app_zetap}) into $\Z_\pro(\rho_\pro)$. For brevity of expressions we temporarily define the projectile distance from the $z$-axis in a fixed-target frame as:
\begin{equation}
\ro_\pro\equiv \rho_\pro^{(\fix)}= r(\theta)\sin\theta.
\label{app_xi_def}
\end{equation}
Using~(\ref{master}), one easily obtains:
\begin{equation}
\ro_\pro=\frac{\R_0^2\tan\tfrac{\theta}{2}}{\R_0\tan\tfrac{\theta}{2}-\X} \quad\Rightarrow\quad \tan\tfrac{\theta}{2}=\frac{\X\ro_\pro}{\R_0(\ro_\pro-\R_0)},
\end{equation}
so that:
\begin{equation}
r(\theta)\cos\theta=\frac{\R_0^2(1-\tan^2\tfrac{\theta}{2})}{2(\R_0\tan\tfrac{\theta}{2}-\X)}=\frac{\R_0^2(\ro_\pro-\R_0)^2-\X^2\ro_\pro^2}{2\X\R_0(\ro_\pro-\R_0)}.
\end{equation}
Combining these results within~(\ref{app_zetap}) and using the relation for $\ro_\pro(\rho_\pro)$ from~(\ref{app_rhop}):
\begin{equation}
\ro_\pro(\rho_\pro)=\frac{\rho_\pro-\eta_\pro\R_0}{\eta_\tar},
\label{app_xi}
\end{equation}
we finally arrive at:
\begin{align}
\begin{split}
\Z_\pro(\rho_\pro)=&\frac{\V_\cm}{v_0}\X\left(\frac{\rho_\pro}{\eta_\tar\R_0}+\ln\frac{\LL(\rho_\pro-\R_0)}{\eta_\tar\X\R_0}-\frac{\eta_\pro}{\eta_\tar}\right)+\\
&\left(\frac{\V_\cm}{\eta_\tar v_0}+1\right)\left(\frac{\R_0(\rho_\pro-\R_0)}{2\X}-\frac{\X(\rho_\pro-\eta_\pro\R_0)^2}{2\R_0(\rho_\pro-\R_0)}\right),
\end{split}
\label{app_zeta}
\end{align}
which is a projectile trajectory in the comoving frame where the center of mass moves along the $\Z$-axis with the speed $\V_\cm$.

}



\clearpage

\thispagestyle{empty}

\setcounter{page}{1}
\setcounter{section}{0}
\setcounter{footnote}{0} 

\renewcommand{\thefootnote}{S\arabic{footnote}}

\renewcommand{\thesection}{\Alph{section}}
\numberwithin{equation}{section}
\numberwithin{figure}{section}
\numberwithin{table}{section}

\renewcommand{\theequation}{\thesection\arabic{equation}}
\renewcommand{\thefigure}{\thesection\arabic{figure}}

\renewcommand{\thepage}{S\arabic{page}}

\onecolumngrid
\begin{center}

\textbf{\LARGE Supplementary note}\\[.5cm]

\textbf{\large A shadow of the repulsive Rutherford scattering in the laboratory frame}\\[.5cm]

Petar \v{Z}ugec$^{1*}$ and Dario Rudec$^1$\\[.1cm]
{\small
{\itshape
$^1$Department of Physics, Faculty of Science, University of Zagreb, Zagreb, Croatia\\}
$^*$Electronic address: pzugec@phy.hr\\[.5cm]
}

\begin{minipage}{400pt}
\small
This note presents the supplementary material to the main paper. The references to figures and equations not starting with the alphabetical character or starting with `A.'---such as~(1) or (A.1)---refer to those from the main paper and its Appendix, while those starting with the appropriate letter intermediately followed by a number---e.g.~(A1)---refer to those from this note.\\
\end{minipage}

\end{center}

\twocolumngrid

\section{The infinitely-massive-target and the center-of-mass frame}
\label{imt_com_frame}


Let us first confirm that the procedure from~(\ref{derivative}) successfully recreates the familiar solutions from~\cite{zugec_supp}. In that, we cannot reconstruct the general solution from a fixed-target frame, as for the finite mass target the fixed-target frame is accelerated due to the Coulomb recoil. However, in a special case of an infinitely massive target (\mbox{$m_\tar\to\infty$}) the fixed-target frame, the center-of-mass frame and the laboratory frame all coincide and are thus entirely within the domain of the earlier procedure. This case is realized by introducing \mbox{$\eta_\pro=0$} and \mbox{$\eta_\tar=1$} into previous equations, yielding as the first consequence: \mbox{$\mathbf{R}_0^{(\infty)}=\textbf{0}$} and \mbox{$\mathbf{\V}_\cm^{(\infty)}=\textbf{0}$}. Though we can immediately proceed to~(\ref{derivative}), let us first note that by virtue of plugging all these values into~(\ref{Z0}), it immediately follows: $\Z_0^{(\infty)}=0$, i.e. the geometry is immediately `renormalized' and we can switch back to the original coordinate $z_\pro^{(\infty)}(\rho_\pro)$, in place of $\Z_\pro^{(\infty)}(\rho_\pro)$ from~(\ref{zeta}):
\begin{equation}
z_\pro^{(\infty)}(\rho_\pro)=\frac{\R_0(\rho_\pro-\R_0)}{2\X}-\frac{\X\rho_\pro^2}{2\R_0(\rho_\pro-\R_0)}.
\end{equation}
This is a trajectory equation in the cylindrical coordinates from the infinitely-massive-target frame. The associated extremization condition is easily obtained from~(\ref{derivative}) by inserting \mbox{$\eta_\pro=0$}, \mbox{$\eta_\tar=1$} and $\V_\cm^{(\infty)}=0$:
\begin{equation}
\frac{\dd z_\pro^{(\infty)}}{\dd \R_0}\bigg|_{\tilde{\R}_0}=\frac{(\rho_\pro-2\tilde{\R}_0)(\tilde{\R}_0^4-2\tilde{\R}_0^3\rho_\pro+\tilde{\R}_0^2\rho_\pro^2+\X^2\rho_\pro^2)}{2\X\tilde{\R}_0^2(\rho_\pro-\tilde{\R}_0)^2}=0.
\label{inf_der}
\end{equation}
There is a single real solution to this equation\footnote{
Four remaining complex solutions to~(\ref{inf_der}) are:
\begin{equation*}
[\tilde{\R}_0^{(\infty)}]_{1,2,3,4}=\frac{\rho_\pro\pm\sqrt{\rho_\pro(\rho_\pro\pm 4\mathrm{i}\X)}}{2}.
\end{equation*}
}:
\begin{equation}
\tilde{\R}_0^{(\infty)}=\frac{\rho_\pro}{2},
\label{inf_r0}
\end{equation}
yielding a familiar shadow equation:
\begin{equation}
\ZZ_\pro^{(\infty)}(\rho_\pro)=z_\pro^{(\infty)}[\rho_\pro; \tilde{\R}_0^{(\infty)}(\rho_\pro)]=\frac{\rho_\pro^2}{8\X}-2\X,
\label{inf_shad}
\end{equation}
that corresponds to equation~(20) from~\cite{zugec_supp}.\pagebreak

As opposed to the fixed-target frame, the center-of-mass frame is partially within the domain of~(\ref{zeta}) and (\ref{derivative}), as it is certainly comoving with itself. However, there are some qualitative differences to be taken into account. By the very definition, the center of mass is at rest in and at the origin of the same frame: \mbox{$\mathbf{R}_0^{(\cm)}=\textbf{0}$} and \mbox{$\mathbf{\V}_\cm^{(\cm)}=\textbf{0}$}, leading again to \mbox{$\Z_0^{(\cm)}=0$} and the `renormalization' of geometry. It should be noted that---unlike the infinitely-massive-target frame, where \mbox{$\mathbf{R}_0^{(\infty)}=\textbf{0}$} was consistent with the initial conditions from~(\ref{rp0}) and (\ref{rt0}) due to \mbox{$m_\tar\to\infty$}---for finite $m_\tar$ the center-of-mass frame necessarily invalidates these conditions, since \mbox{$\eta_\pro\mathbf{r}_\pro^{(\cm)}=-\eta_\tar\mathbf{r}_\tar^{(\cm)}$} holds at any moment, including the initial one. Furthermore, we also need to take into account that the $z$-axis in the center-of-mass frame lies in between the projectile and the target, as opposed to passing through the target, as we have explicitly assumed in earlier calculations. This is reflected through the redefinition of the cylindrical radial coordinate from~(\ref{app_rhop}), so that the same radial coordinate in the center-of-mass frame (\mbox{$\Rho_\pro\equiv \rho_\pro^{(\cm)}$}, for brevity) reads: \mbox{$\Rho_\pro=\eta_\tar r(\theta)\sin\theta$}. Thus, alongside \mbox{$\V_\cm^{(\cm)}=0$}, one needs to use \mbox{$\ro_\pro(\Rho_\pro)=\Rho_\pro/\eta_\tar$} in place of~(\ref{app_xi}), in order to arrive at the trajectory in the center-of-mass coordinates:
\begin{equation}
z_\pro^{(\cm)}(\Rho_\pro)=\frac{\R_0(\Rho_\pro-\eta_\tar\R_0)}{2\X}-\frac{\X\Rho_\pro^2}{2\R_0(\Rho_\pro-\eta_\tar\R_0)}.
\end{equation}
The corresponding extremization condition is:
\begin{align}
\begin{split}
\frac{\dd z_\pro^{(\cm)}}{\dd \R_0}&\bigg|_{\tilde{\R}_0}=0\\
&=\frac{(\Rho_\pro-2\eta_\tar\tilde{\R}_0)(\eta_\tar^2\tilde{\R}_0^4-2\eta_\tar\tilde{\R}_0^3\Rho_\pro+\tilde{\R}_0^2\Rho_\pro^2+\X^2\Rho_\pro^2)}{2\X\tilde{\R}_0^2(\Rho_\pro-\eta_\tar\tilde{\R}_0)^2},
\end{split}
\label{cm_der}
\end{align}
yielding a single real solution\footnote{
Four remaining complex solutions to~(\ref{cm_der}) are:
\begin{equation*}
[\tilde{\R}_0^{(\cm)}]_{1,2,3,4}=\frac{\Rho_\pro\pm\sqrt{\Rho_\pro(\Rho_\pro\pm 4\mathrm{i}\eta_\tar\X)}}{2\eta_\tar}.
\end{equation*}
}:
\begin{equation}
\tilde{\R}_0^{(\cm)}=\frac{\Rho_\pro}{2\eta_\tar},
\label{cm_r0}
\end{equation}
ultimately leading to the familiar shadow equation in the center-of-mass frame:
\begin{equation}
\ZZ_\pro^{(\cm)}(\Rho_\pro)=z_\pro^{(\cm)}[\Rho_\pro; \tilde{\R}_0^{(\cm)}(\Rho_\pro)]=\frac{\Rho_\pro^2}{8\eta_\tar\X}-2\eta_\tar\X,
\label{cm_shad}
\end{equation}
in a perfect agreement with equation~(29) from~\cite{zugec_supp}.

\section{Scattering angle}

\begin{figure}[t!]
\centering
\includegraphics[width=1\linewidth,keepaspectratio]{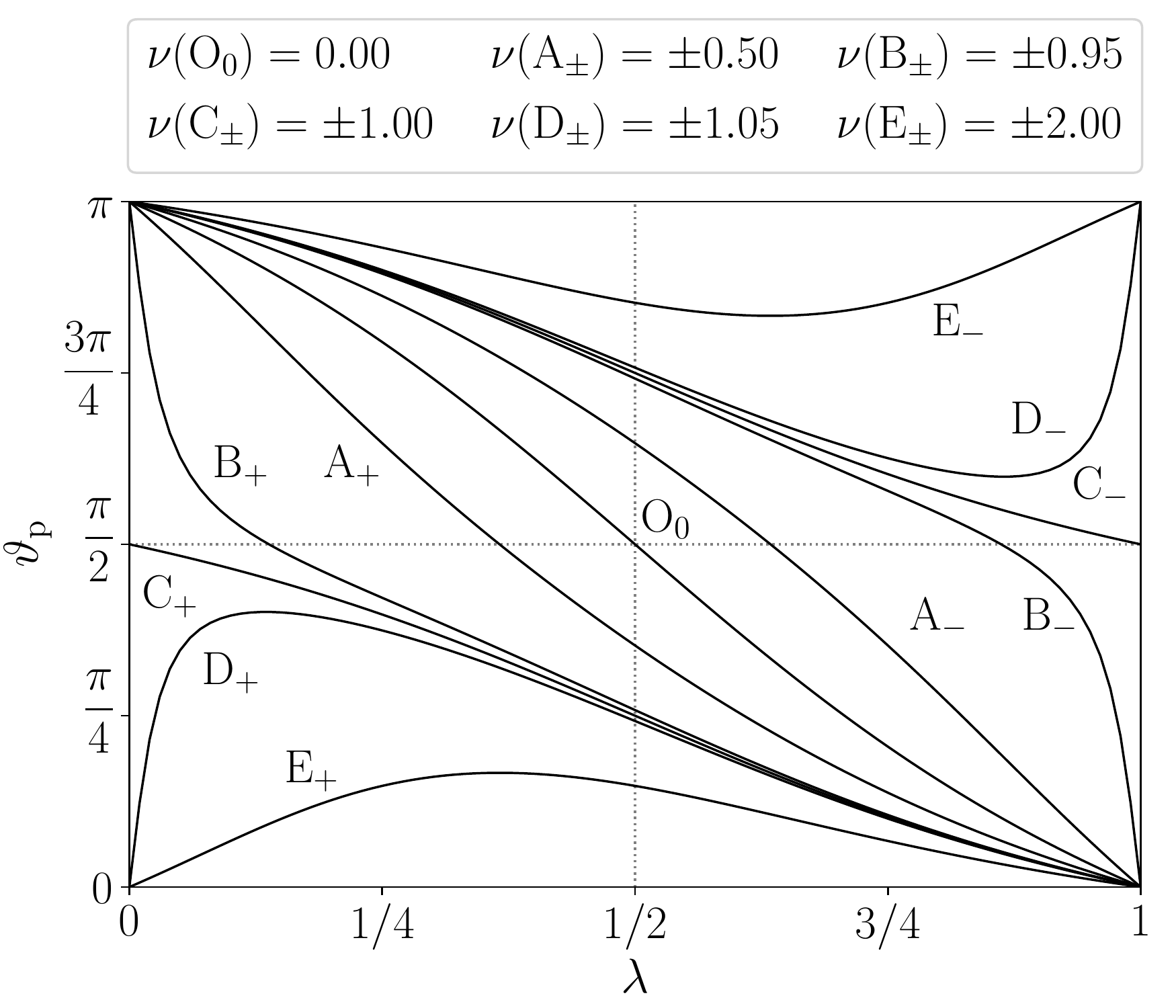}
\pull
\caption{Projectile scattering angle dependence from~(\ref{cot}) in a transformed variable $\lambda$ from~(\ref{lambda}), for a few selected comoving frames defined by \mbox{$\nu=\V_\cm/\eta_\tar v_0$}. A remarkable symmetry may be observed for the values of $\nu$ of opposing signs (i.e. for the same center-of-mass-relative speed, but in opposite directions), which is captured by~(\ref{symmetry}).}
\pullc
\label{fig3}
\end{figure}

Let us investigate the projectile scattering angle in the comoving frame, and its connection to the scattering shadow. From a trajectory derivative over the relevant geometric variable $\rho_\pro$:
\begin{align}
\begin{split}
\frac{\dd \Z_\pro}{\dd \rho_\pro}=&\Big\{(\eta_\tar v_0+\V_\cm)\R_0^2(\rho_\pro-\R_0)^2+\X^2(\rho_\pro-\eta_\pro\R_0)\times\\
&\big[\eta_\tar(\eta_\tar v_0+\V_\cm)\R_0-(\eta_\tar v_0-\V_\cm)(\rho_\pro-\R_0)\big]\Big\}\Big/\\
&  \left[2\eta_\tar v_0\X\R_0(\rho_\pro-\R_0)^2\right],
\end{split}
\label{der_rhop}
\end{align}
the asymptotic scattering angle $\vartheta_\pro$ relative to the $\Z$-axis is quickly obtained as\footnote{
The scattering angle from~(\ref{scat_angle}) may also be obtained by a kinematic boost from the center-of-mass frame. It is a commonly recognized fact (see, for example, \cite{goldstein}) that the projectile scattering angle in the center-of-mass frame:
\begin{equation*}
\vartheta_\pro^{(\cm)}=2\:\atan(\X/\R_0)
\end{equation*}
is equal to a well known scattering angle from a fixed-target frame, corresponding to equation~(13) from~\cite{zugec_supp}. This is due to the transformation \mbox{$\mathbf{r}_\pro^{(\cm)}=\eta_\tar\mathbf{r}$} from~(\ref{rp}) not affecting the definition of the geometric angle $\theta$ between those frames, by the virtue of \mbox{$\mathbf{R}^{(\cm)}=\mathbf{0}$}. Knowing the initial projectile velocity \mbox{$\mathbf{v}_\pro^{(\cm)}(t=0)=\eta_\tar v_0\hat{\mathbf{z}}$}, one can easily write out the final, asymptotic velocity:
\begin{equation*}
\mathbf{v}_\pro^{(\cm)}(t\to\infty)=\eta_\tar v_0\left(\sin\vartheta_\pro^{(\cm)}\hat{\boldsymbol{\rho}}+\cos\vartheta_\pro^{(\cm)}\hat{\mathbf{z}}\right)
\end{equation*}
from a symmetry between the initial and final states of motion in the center-of-mass frame. Galilean boost into the desired comoving frame at any point in time is now trivially performed as: \mbox{$\mathbf{v}_\pro=\mathbf{v}_\pro^{(\cm)}+\mathbf{\V}_\cm$}, with \mbox{$\mathbf{\V}_\cm=\V_\cm\hat{\mathbf{z}}$}, leading to the asymptotic velocity:
\begin{equation*}
\mathbf{v}_\pro(t\to\infty)=\eta_\tar v_0\sin\vartheta_\pro^{(\cm)}\hat{\boldsymbol{\rho}}+\left(\eta_\tar v_0\cos\vartheta_\pro^{(\cm)}+\V_\cm\right)\hat{\mathbf{z}}.
\end{equation*}
The cotangent of the scattering angle $\vartheta_\pro$ in a comoving frame is now read out as the ratio of the velocity components:
\begin{equation*}
\cot \vartheta_\pro=\frac{\eta_\tar v_0\cos\vartheta_\pro^{(\cm)}+\V_\cm}{\eta_\tar v_0\sin\vartheta_\pro^{(\cm)}}.
\end{equation*}
Plugging in $\vartheta_\pro^{(\cm)}$ yields the result identical to~(\ref{scat_angle}).
}:
\begin{equation}
\cot \vartheta_\pro=\!\lim_{\rho_\pro\to\infty}\!\frac{\dd\Z_\pro}{\dd\rho_\pro}=\frac{\big(\eta_\tar v_0+\V_\cm\big)\R_0^2-\big(\eta_\tar v_0-\V_\cm\big)\X^2}{2\eta_\tar v_0\X\R_0}.
\label{scat_angle}
\end{equation}
In an attempt to display this dependence for all possible trajectories, we introduce the transformation:
\begin{equation}
\frac{\R_0}{\X}=\frac{\lambda}{1-\lambda} \quad\Leftrightarrow\quad \lambda=\frac{\R_0/\X}{1+\R_0/\X},
\label{lambda}
\end{equation}
which maps an infinite domain \mbox{$\varrho_0\in[0,\infty\rangle$} onto a finite range \mbox{$\lambda\in[0,1\rangle$}. Together with the following shorthand:
\begin{equation}
\nu\equiv\frac{\V_\cm}{\eta_\tar v_0},
\label{nu}
\end{equation}
equation~(\ref{scat_angle}) takes on the more universal form:
\begin{equation}
\cot \vartheta_\pro=\frac{\nu[1-2\lambda(1-\lambda)]+\lambda-(1-\lambda)}{2\lambda(1-\lambda)}.
\label{cot}
\end{equation}
We have purposefully elected to write it in this particular manner, for reasons that will soon become evident. Using a conventional definition of the arcus cotangent function---equivalent to \mbox{$\acot\,x=\pi/2-\atan\,x$}, thus yielding the values \mbox{$\vartheta_\pro\in\langle0,\pi\rangle$}---figure~\ref{fig3} shows thus transformed dependence of the scattering angle $\vartheta_\pro$ itself, for a few values of $\nu$ selected symmetrically around 0. We now see that the transition from $\R_0$ to $\lambda$ was most fortuitous, as even a purely visual inspection of figure~\ref{fig3} reveals a particular symmetry between the scattering angles for values of $\nu$ of opposing signs, allowing us to draw the following conjecture:
\begin{equation}
\vartheta_\pro(\lambda;\nu)+\vartheta_\pro(1-\lambda;-\nu)=\pi.
\label{symmetry}
\end{equation}
This  conjecture is readily proven to be true, as it implies:
\begin{equation}
\cot[\vartheta_\pro(1-\lambda;-\nu)]=-\cot[\vartheta_\pro(\lambda;\nu)],
\end{equation}
which is easily confirmed even by a simple glance at~(\ref{cot}).

\begin{figure}[t!]
\centering
\includegraphics[width=1\linewidth,keepaspectratio]{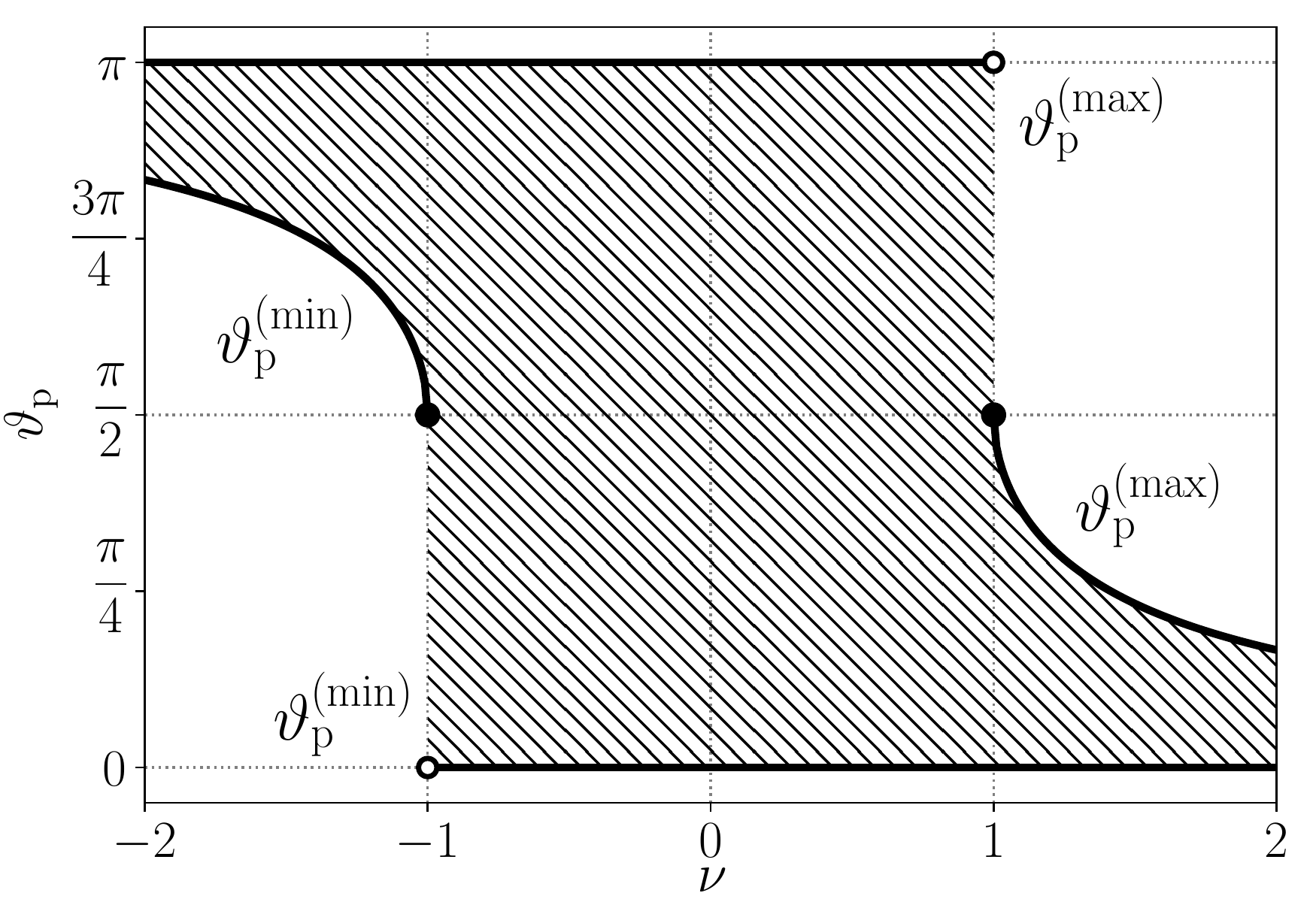}
\pull
\caption{Projectile scattering angles accessible in a given comoving frame. The (nontrivial) scattering shadow exists only within the range \mbox{$-1<\nu<1$}, i.e. only when the entire solid angle is accessible to the scattered projectiles.}
\pullc
\label{fig4}
\end{figure}

Figure~\ref{fig3} also impels us to identify the minimum and the maximum projectile scattering angle in a given comoving frame. This is easily done by a simple extremization procedure:
\begin{equation}
\frac{\dd\vartheta_\pro}{\dd\R_0}\Big|_{\R_0^\star}=0 \quad\Rightarrow\quad \R_0^\star=\X\sqrt{\frac{\nu-1}{\nu+1}},
\end{equation}
yielding, upon little deliberation:
\begin{align}
\vartheta_\pro^{(\mathrm{min})}&=\left\{\begin{array}{cl}
\acot\left(-\sqrt{\nu^2-1}\,\right)&\ifif \; \nu\le-1\\
0&\ifif \; \nu>-1\\
\end{array}\right. ,\\
\vartheta_\pro^{(\mathrm{max})}&=\left\{\begin{array}{cl}
\pi&\ifif \; \nu<1\\
\acot\left(\sqrt{\nu^2-1}\,\right)&\ifif \; \nu\ge1\\
\end{array}\right. .
\end{align}
Figure~\ref{fig4} shows the range of the scattering angles accessible in a given comoving frame. We see that the entire angular range \mbox{$0\le\vartheta_\pro\le\pi$} is accessible to the projectile only for \mbox{$-1<\nu<1$}, which is equivalent to the (nontrivial) shadow existence condition from~(\ref{condition}). Therefore, the (nontrivial) scattering shadow in a comoving frame exists if and only if the scattered projectiles sweep the entire solid angle!

\vspace{-\baselineskip}

\section{Shadow vertex}
\label{vertex}

We now analyze in some detail the behavior of the shadow vertex, i.e. the shadow characteristics in an immediate vicinity of the \mbox{$z$-axis}. We aim to investigate its position, sharpness and curvature. To this end we consider the extremization condition from~(\ref{derivative}) in the limit \mbox{$\rho_\pro\to0$}, in order to obtain the limiting behavior of the solution $\tilde{\R}_0(\rho_\pro)$. The fact that this extremization condition is a polynomial in  $\tilde{\R}_0$, whose own coefficients are polynomials in $\rho_\pro$, in conjunction with the obvious fact that \mbox{$\tilde{\R}_0(0)=0$}, necessitates the limiting dependence to follow some specific power-law: \mbox{$\tilde{\R}_0(\rho_\pro\to0)\propto\rho_\pro^n$}. This can be thought of as the smallest power from a proverbial series expansion (not necessarily in integer powers) of $\tilde{\R}_0(\rho_\pro)$. Since there are two separate length scales present: $\rho_\pro$ and $\X$ (alongside the sought $\tilde{\R}_0$), we must allow for the general dependence: \mbox{$\tilde{\R}_0(\rho_\pro\to0)=\K_0\X^{1-n}\rho_\pro^n$}, following from purely dimensional considerations. However, we may immediately impose a physical constraint \mbox{$n\ge1$}, as the sought impact parameter must be an increasing function of $\rho_\pro$ (eliminating \mbox{$n\le0$}), but never greater than $\rho_\pro$, thus not increasing faster than $\rho_\pro$ (eliminating \mbox{$n<1$} for \mbox{$\rho_\pro\to0$}).  Using the abbreviations \mbox{$\alpha_\pm\equiv(\eta_\tar v_0\pm\V_\cm)/2\eta_\tar v_0$} and \mbox{$\bar{\rho}_\pro\equiv\rho_\pro/\X$}, and carefully examining the limit of~(\ref{derivative}) within \textit{a priori} permissible range of exponents:
\begin{equation}
\lim_{\rho_\pro\to0}\frac{\dd\Z_\pro}{\dd\R_0}\!=\!\left\{\begin{array}{lcc}
\!\!\dfrac{(\eta_\pro\K_0-1)[(\eta_\tar\alpha_++\alpha_-)\K_0-\alpha_-]}{\K_0^2(1-\K_0)^2\bar{\rho}_\pro} &\ifif \; n=1\\
\!\!\dfrac{\alpha_--\eta_\tar\alpha_+\K_0\bar{\rho}_\pro^{n-1}}{\K_0^2\bar{\rho}_\pro^{2n-1}} &\ifif \; n>1\\
\end{array}\right. 
\label{der_0}
\end{equation}
we see that $n=1$ remains the only admissible solution, as for \mbox{$n>1$} the extremization condition (the vanishing of the derivative) cannot be satisfied for any finite $\K_0$ \textit{independent} of $\rho_\pro$. Among the two solutions following from \mbox{$n=1$} case, \mbox{$\K_0=1/\eta_\pro$} is unacceptable one, as due to \mbox{$\eta_\pro\le1$} it would lead to $\tilde{\R}_0(\rho_\pro)$ increasing faster than $\rho_\pro$. Therefore, a unique solution remains:
\begin{equation}
\K_0=\frac{\eta_\tar v_0-\V_\cm}{(1+\eta_\tar)\eta_\tar v_0-\eta_\pro\V_\cm},
\end{equation}
meaning that:
\begin{equation}
\lim_{\rho_\pro\to0}\tilde{\R}_0(\rho_\pro)=\K_0\rho_\pro.
\label{asym_0}
\end{equation}

Now it is simple enough to obtain any shadow property at its vertex. We start with the vertex position $\vtx_\pro$, defined in accordance with~(\ref{zz}) as:
\begin{equation}
\vtx_\pro\equiv \ZZ(0)=\lim_{\rho_\pro\to0}\Z_\pro(\rho_\pro;\K_0\rho_\pro).
\label{vtx_def}
\end{equation}
With little calculation it is easily shown that\footnote{
There are many alternative methods for obtaining the vertex position~(\ref{vtx_eq}). One of them is taking the limit of the back-bending curves, as prescribed by~(\ref{vtx_lim}) from Section~\ref{back_bend}. The other is a direct integration of the equation of motion for a frontally impinging projectile (\mbox{$\R_0=0$}), which is rather straightforward, save for one sensitive step. In order to obtain a result consistent with~(\ref{vtx_eq}), the following limit---appearing during the evaluation of integrals---must be carefully taken:
\begin{equation*}
\lim_{z_0\to\infty}\sqrt{z_0(z_0-2\X)}=\lim_{z_0\to\infty}z_0-\X,
\end{equation*}
where the delicate step consists in not discarding the finite term~$\X$.
}:
\begin{equation}
\vtx_\pro=-2\eta_\tar\X+\frac{\V_\cm}{v_0}\X\ln\frac{\LL(\eta_\tar v_0+\V_\cm)}{\X(\eta_\tar v_0-\V_\cm)}.
\label{vtx_eq}
\end{equation}
As for the vertex sharpness $\vtx_\pro'$:
\begin{equation}
\vtx_\pro'\equiv \frac{\dd\ZZ}{\dd\rho_\pro}\bigg|_0=\lim_{\rho_\pro\to0}\frac{\dd\Z_\pro(\rho_\pro;\K_0\rho_\pro)}{\dd\rho_\pro},
\end{equation}
it simply follows as:
\begin{equation}
\vtx_\pro'=\lim_{\rho_\pro\to0}\frac{(\eta_\tar v_0-\V_\cm)(\eta_\tar v_0+\V_\cm)^2}{v_0\X[(1+\eta_\tar)\eta_\tar v_0-\eta_\pro\V_\cm]^2}\rho_\pro=0.
\end{equation}
Therefore, the scattering shadow is smooth at its vertex\footnote{
There is no \textit{a priori} reason to expect that the shadow should not end sharply at $\rho_\pro=0$, as the vertex is at the very edge of radial coordinate domain \mbox{$\rho_\pro\in[0,\infty\rangle$}. In other words, the function $\ZZ_\pro(\rho_\pro)$ does not extend into negative arguments, so it cannot be expected \textit{in advance} that it should reach the vertex smoothly.
}, as already suggested by the shadow examples from figure~\ref{fig2}. Finally, the vertex curvature $\vtx_\pro''$:
\begin{equation}
\vtx_\pro''\equiv \frac{\dd^2\ZZ}{\dd\rho_\pro^2}\bigg|_0=\lim_{\rho_\pro\to0}\frac{\dd^2\Z_\pro(\rho_\pro;\K_0\rho_\pro)}{\dd\rho_\pro^2}
\end{equation}
is directly obtained as:
\begin{equation}
\vtx_\pro''=\frac{(\eta_\tar v_0-\V_\cm)(\eta_\tar v_0+\V_\cm)^2}{v_0\X[(1+\eta_\tar)\eta_\tar v_0-\eta_\pro\V_\cm]^2}.
\label{vtx_cur}
\end{equation}

\begin{figure}[t!]
\centering
\includegraphics[width=1\linewidth,keepaspectratio]{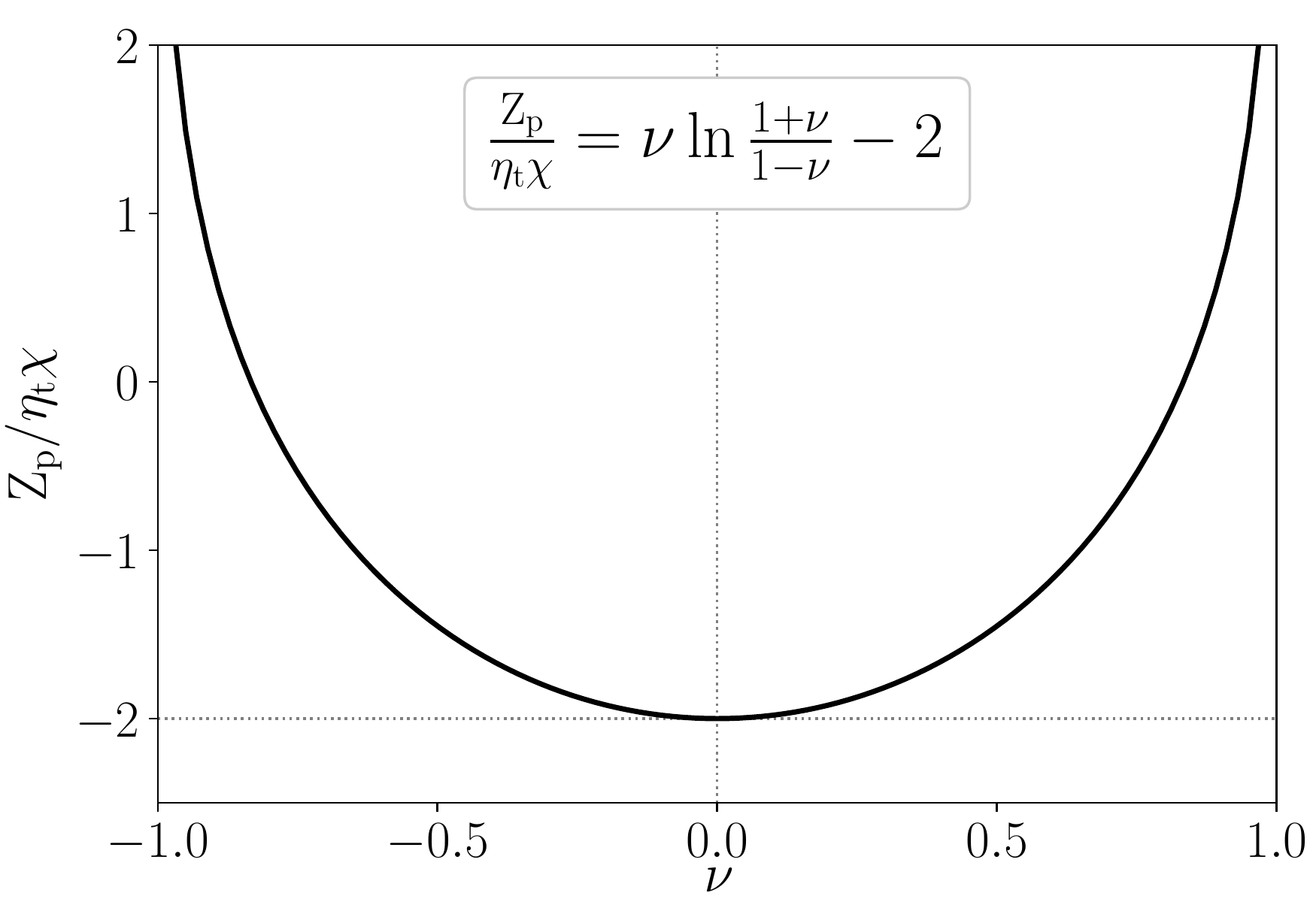}
\pull
\caption{Vertex position in a particular comoving frame, as determined by a frame speed $\V_\cm$ from \mbox{$\nu=\V_\cm/\eta_\tar v_0$}.}
\pullc
\label{fig5}
\end{figure}

Figure~\ref{fig5} shows the dependence of the vertex position~(\ref{vtx_eq}), assuming the specific but natural selection \mbox{$\LL=\X$}. It may be observed that the function is even in $\nu$ (though only for \mbox{$\LL=\X$}), meaning that the vertex is at the same position for the frames moving with the same speed $|\V_\cm|$, but in opposite directions. It should be noted that this dependence is by no means of some fundamental physical significance or meaning, but just an additional, finite remnant after the subtraction of the initial, arbitrarily extracted divergence $\Z_0$ from~(\ref{Z0}). The connection is also to be made with the long-since-recognized fact that the (nontrivial) shadow form exists only for \mbox{$-1<\nu<1$}, clearly reflected through the vertex' escape to infinity for \mbox{$\nu=\pm1$}, even beyond the initially subtracted $\Z_0$.

\begin{figure}[t!]
\centering
\includegraphics[width=1\linewidth,keepaspectratio]{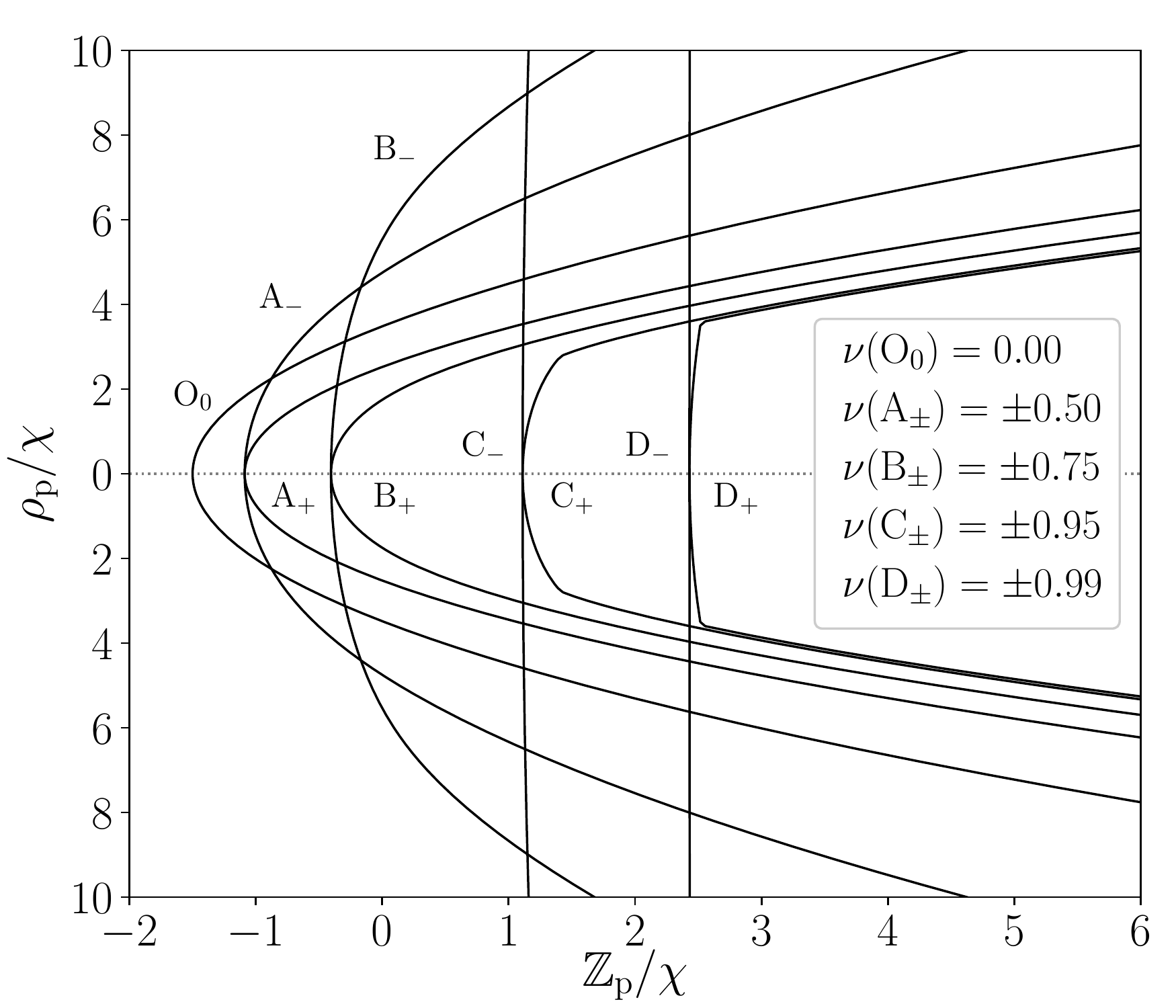}
\pull
\caption{Scattering shadow for \mbox{$\eta_\pro=0.25$}, as seen from different comoving frames. As \mbox{$\nu\to-1$}, the shadow flattens towards the (trivial) asymptotic form from~(\ref{flat}), i.e.~(\ref{flat_nu}).}
\pullc
\label{fig6}
\end{figure}

In figure~\ref{fig2} we have already shown several shadow examples for varying $\eta_\pro$, while the definition of the laboratory frame required the parameter $\nu$ from~(\ref{nu}) to vary accordingly: \mbox{$\nu^{(\lab)}=\eta_\pro/\eta_\tar$}. Only now, after understanding the behavior of the vertex position, do we wish to inspect the shadow both in forward-moving (\mbox{$\nu>0$}) and backward-moving  (\mbox{$\nu<0$}) frames. Figure~\ref{fig6} shows several such examples for freely varied $\nu$, the fixed value of \mbox{$\eta_\pro=0.25$} and \mbox{$\LL=\X$}. From the fact that the vertex position is even in $\nu$ (as observed in figure~\ref{fig5}), we now understand why the vertices for the same absolute value of $\nu$ in figure~\ref{fig6} coincide. In addition, for systematically decreasing values of $\nu$---most prominently for \mbox{$\nu(\mathrm{C}_-)=-0.95$} and \mbox{$\nu(\mathrm{D}_-)=-0.99$}---one may observe the significant shadow flattening, consistent with the expected (trivial) asymptotic state from~(\ref{flat}). While in a true geometric coordinate $z_\pro$ this asymptotic form stays at the initial, infinitely distant projectile position \mbox{$z_\pro(0)=-\lim_{z_0\to\infty}z_0$} at the \textit{negative} side of the \mbox{$z$-axis}, both figures~\ref{fig5} and \ref{fig6} suggest that in the transformed coordinate $\Z_\pro$ the same asymptotic form escapes to the \textit{positive} infinity. This may be easily corroborated by plugging $\Z_0$ from~(\ref{Z0}) into~(\ref{flat}):
\begin{equation}
\ZZ_\pro(\rho_\pro;\nu\le-1)=-\eta_\tar\lim_{z_0\to\infty}\left((\nu+1)z_0+\nu\X  \ln \frac{2z_0}{\e\LL}\right)
\label{flat_nu}
\end{equation}
and noting that the entire expression under the limit is negative due to $\nu\le1$, thus making the final result positive\footnote{
From~(\ref{flat_nu}) one may also observe a subtle difference between the general \mbox{$\nu<-1$} and the specific \mbox{$\nu=-1$} case:
\begin{equation*}
\ZZ_\pro(\rho_\pro;\nu=-1)=\eta_\tar\X\lim_{z_0\to\infty}\ln \frac{2z_0}{\e\LL},
\end{equation*}
yielding the asymptotic parametrization of the vertex position from~(\ref{vtx_eq}). Since the canceling of the term \mbox{$\lim_{z_0\to\infty}z_0$} occurs only for $\nu=-1$, as soon as this value is exceeded, the shadow in a transformed $\Z_\pro$ coordinate is immediately pushed to `farther infinity'. However, there is no such discontinuous behavior in a true geometric coordinate $z_\pro$, where the trivial shadow caustic stays at he initial projectile position \mbox{$z_\pro(0)=-\lim_{z_0\to\infty}z_0$} for every \mbox{$\nu\le-1$}.
}.

\begin{figure}[t!]
\centering
\includegraphics[width=1\linewidth,keepaspectratio]{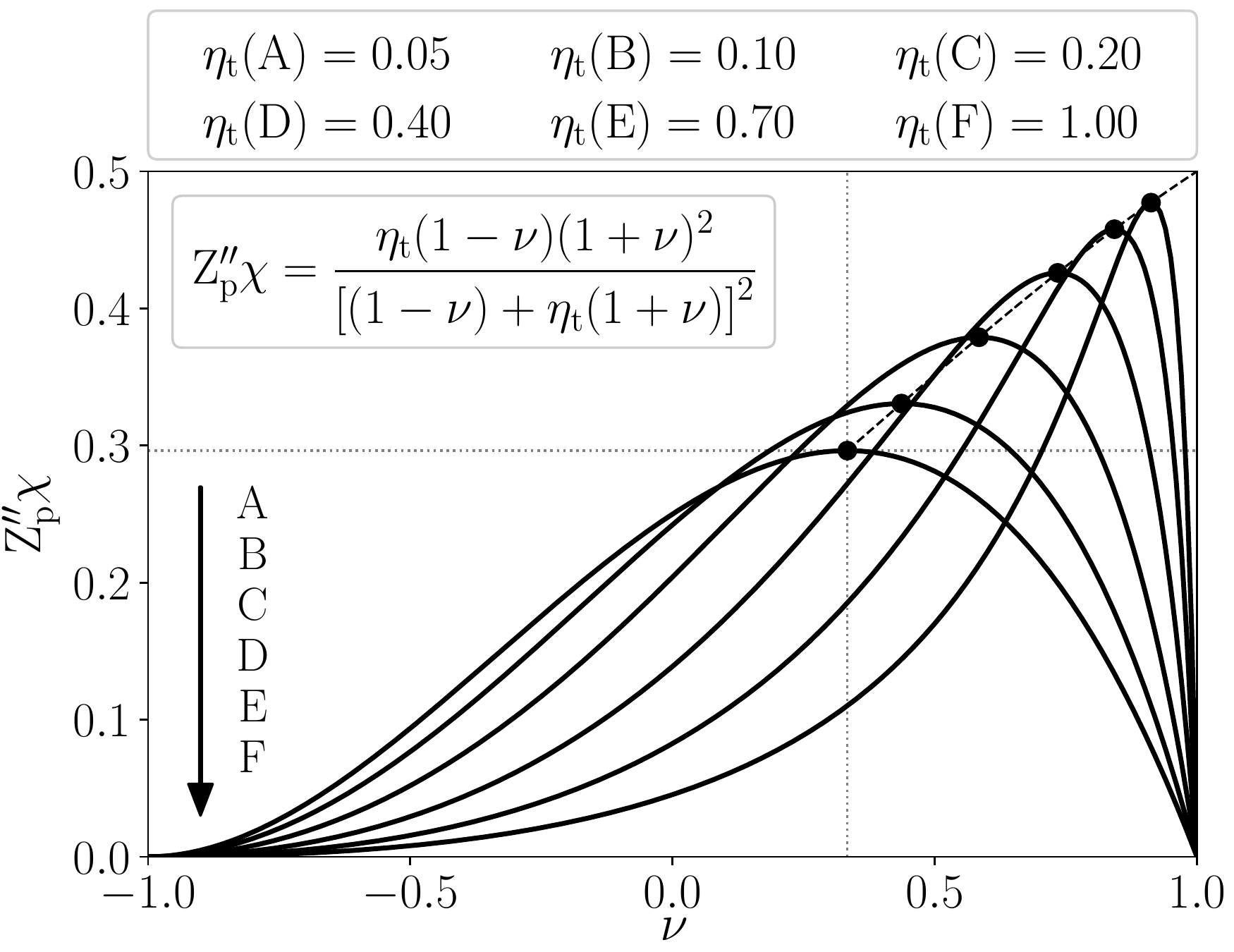}
\pull
\caption{Vertex curvature dependence for several selected values of $\eta_\tar$. The scaled-curvature maxima $\gamma(\tilde{\nu})$---each at the position $\tilde{\nu}$ from~(\ref{crv_nu})---are confined along the dashed curve \mbox{$\gamma(\tilde{\nu})=(3-\tilde{\nu})(1+\tilde{\nu})^2/16$}. The specially designated coordinates for the $\eta_\tar=1$ maximum are: \mbox{$(\tilde{\nu},\gamma)=(1/3,8/27)$}.}
\pullc
\label{fig7}
\end{figure}

Finally, figure~\ref{fig7} shows the vertex curvature from~(\ref{vtx_cur}). Unlike the vertex position, the curvature dependence on $\eta_\tar$ cannot be factored out and its dependence on $\nu$ remains contingent on a particular value of $\eta_\tar$. We may observe that the \mbox{$\nu=0$} case bears no specific relevance for the vertex curvature. Though, it may be used in conjunction with an infinitely massive target (\mbox{$\eta_\tar=1$}) in order to confirm that the result from~(\ref{vtx_cur}) recovers the known result from~(\ref{inf_shad}): \mbox{$\vtx_\pro''(\nu=0;\eta_\tar=1)=1/4\X$}. It is to be noted that the general result for \mbox{$\nu=0$}: \mbox{$\vtx_\pro''(\nu=0)=\eta_\tar/\X(1+\eta_\tar)^2$} cannot be directly compared with~(\ref{cm_shad}) from the center-of-mass frame, due to the shift in the coordinate origin from the center-of-mass frame (see Section~\ref{imt_com_frame}). For \mbox{$\nu=\pm1$} the shadow flattens at its vertex, as seen from the fact that in both cases its curvature drops to 0. This is easily understood from figure~\ref{fig6}---most prominently from examples $\mathrm{C}_\pm$ and $\mathrm{D}_\pm$---corroborating that the flattening at the vertex is the consequence of the \textit{entire} shadow gradually flattening. Figure~\ref{fig7} clearly shows that, for a given $\eta_\tar$, the vertex curvature reaches a maximum which is easily determined by a simple extremization procedure:
\vspace*{-1mm}
\begin{equation}
\frac{\dd\vtx_\pro''}{\dd\nu}\Big|_{\tilde{\nu}}=0 \quad\Rightarrow\quad \tilde{\nu}=\frac{2+\eta_\tar-\sqrt{1+8\eta_\tar}}{1-\eta_\tar},
\label{crv_nu}
\vspace*{-1mm}
\end{equation}
yielding a maximum dimensionless curvature $\gamma$:
\vspace*{-1mm}
\begin{equation}
\gamma\equiv\max_\nu\left(\vtx_\pro''\X\right)=\frac{1-20\eta_\tar-8\eta_\tar^2+\sqrt{(1+8\eta_\tar)^3}}{4(1-\eta_\tar)^3}.
\label{crv_max}
\end{equation}
Equation~(\ref{crv_nu}) reveals that the maximizing value $\tilde{\nu}$ is confined between \mbox{$\tilde{\nu}(\eta_\tar\to1)=1/3$} and \mbox{$\tilde{\nu}(\eta_\tar\to0)=1$}, yielding the maximum curvature between \mbox{$\gamma(\eta_\tar\to1)=8/27$} and \mbox{$\gamma(\eta_\tar\to0)=1/2$}. By solving~(\ref{crv_nu}) for $\eta_\tar$ and plugging the result \mbox{$\eta_\tar=(\tilde{\nu}^2-4\tilde{\nu}+3)/(1+\tilde{\nu})^2$} into~(\ref{crv_max}), an explicit dependence $\gamma(\tilde{\nu})$ may be obtained: \mbox{$\gamma(\tilde{\nu})=(3-\tilde{\nu})(1+\tilde{\nu})^2/16$}. This curve, along which the curvature maxima slide, is shown in figure~\ref{fig7} by a dashed line. Already the simple visual inspection of figure~\ref{fig7} reveals that the absolute maximum vertex curvature equals \mbox{$\gamma_\mathrm{max}=1/2$}, that is:
\vspace*{-1mm}
\begin{equation}
\max_{\eta_\tar,\nu}\left(\vtx_\pro''\right)=\lim_{\eta_\tar\to0}\lim_{\nu\to1}\vtx_\pro''=\frac{\X}{2}.
\vspace*{-5mm}
\end{equation}

\section{Asymptotic form}

Let us investigate the asymptotic shadow form, as \mbox{$\rho_\pro\to\infty$}. For the same reason as the limiting dependence at the vertex---due to the extremization condition from~(\ref{derivative}) being polynomial in $\tilde{\R}_0$ with coefficients as polynomials in $\rho_\pro$---the asymptotic dependence $\tilde{\R}_0(\rho_\pro)$ must follow some specific power-law: \mbox{$\tilde{\R}_0(\rho_\pro\to\infty)\propto\rho_\pro^n$}. Again, we must allow for a general dependence of the form: \mbox{$\tilde{\R}_0(\rho_\pro\to\infty)=\K_\infty\X^{1-n}\rho_\pro^n$}, due to two separate length scales $\rho_\pro$ and $\X$ being present. This time we may immediately impose a physical constraint \mbox{$0<n\le1$}, as the sought impact parameter must still be an increasing function of $\rho_\pro$ (eliminating \mbox{$n\le0$}), while never greater than $\rho_\pro$, thus not increasing asymptotically faster than $\rho_\pro$ (eliminating \mbox{$n>1$}). Using the abbreviations \mbox{$\alpha_\pm\equiv(\eta_\tar v_0\pm\V_\cm)/2\eta_\tar v_0$} and \mbox{$\bar{\rho}_\pro\equiv\rho_\pro/\X$}, as in~(\ref{der_0}), and examining~(\ref{derivative}) in the limit~\mbox{$\rho_\pro\to\infty$}:
\vspace*{-0.5mm}
\begin{equation}
\lim_{\rho_\pro\to\infty}\frac{\dd\Z_\pro}{\dd\R_0}=\left\{\begin{array}{lcc}
\alpha_+\bar{\rho}_\pro+\alpha_-\bar{\rho}_\pro^{1-2n}/\K_\infty^2  &\ifif & 0<n\le\tfrac{1}{2}\\
\alpha_+\bar{\rho}_\pro&\ifif & \tfrac{1}{2}<n<1\\
(1-2\K_\infty)\alpha_+\bar{\rho}_\pro&\ifif & n=1
\end{array}\right. 
\label{der_inf}
\end{equation}
we see yet again that $n=1$ remains the only admissible solution, as neither of the alternative forms supports the extremization condition for any $\K_\infty$ independent of $\rho_\pro$. For \mbox{$n=1$} this requirement also provides the value \mbox{$\K_\infty=1/2$}, finally yielding\footnote{
It may be noted that though the extremization condition from~(\ref{derivative}) is the $5^\mathrm{th}$ degree polynomial in $\tilde{\R}_0$, it is only of the $3^\mathrm{rd}$ degree in $\rho_\pro$.  Thus, one could find an explicit inverse dependence $\rho_\pro(\tilde{\R}_0)$ by means of Cardano formula and obtain~(\ref{asym_0}) and (\ref{asym_inf}) by taking it to limits \mbox{$\rho_\pro\to0$} and \mbox{$\rho_\pro\to\infty$}. However, the general form of Cardano formula is long and tiresome. Furthermore, the approach from~(\ref{der_0}) and (\ref{der_inf}) is more general, as it may be carried out for a polynomial dependence of any degree, instead of relying on~(\ref{derivative}) to be at most the $4^\mathrm{th}$ degree polynomial either in $\rho_\pro$ or $\tilde{\R}_0$ (still solvable in radicals).
}:
\vspace*{-2mm}
\begin{equation}
\lim_{\rho_\pro\to\infty}\tilde{\R}_0(\rho_\pro)=\frac{\rho_\pro}{2}.
\label{asym_inf}
\vspace*{-2mm}
\end{equation}
It may be noted that the same dependence has already been encountered in~(\ref{inf_r0}), as the exact dependence from the infinitely-massive-target frame. Passing this solution into~(\ref{zz}), the asymptotic shadow form is immediately obtained as \mbox{$\Z_\pro(\rho_\pro;\rho_\pro/2)$}:
\vspace*{-1mm}
\begin{align}
\begin{split}
\lim_{\rho_\pro\to\infty} \!\ZZ_\pro(\rho_\pro)=&\frac{\eta_\tar v_0+\V_\cm}{8\eta_\tar v_0\X}\rho_\pro^2+\frac{\V_\cm}{v_0}\X\ln\frac{\LL}{\eta_\tar\X}-\\
&\frac{\X(1+\eta_\tar)[\eta_\tar(\eta_\tar v_0+\V_\cm)+(\eta_\tar v_0-\V_\cm)]}{2\eta_\tar v_0},
\end{split}
\label{asym}
\end{align}
evidently being is purely parabolic in $\rho_\pro$! Since the scattering shadow is also parabolic in the infinitely-massive-target frame, the asymptotic form must agree with it in its entirety. Indeed, it is easy to check that plugging \mbox{$\eta_\tar=1$} and \mbox{$\nu=0$} into~(\ref{asym}) immediately recovers a known result from~(\ref{inf_shad}).

\begin{figure}[t!]
\centering
\includegraphics[width=1\linewidth,keepaspectratio]{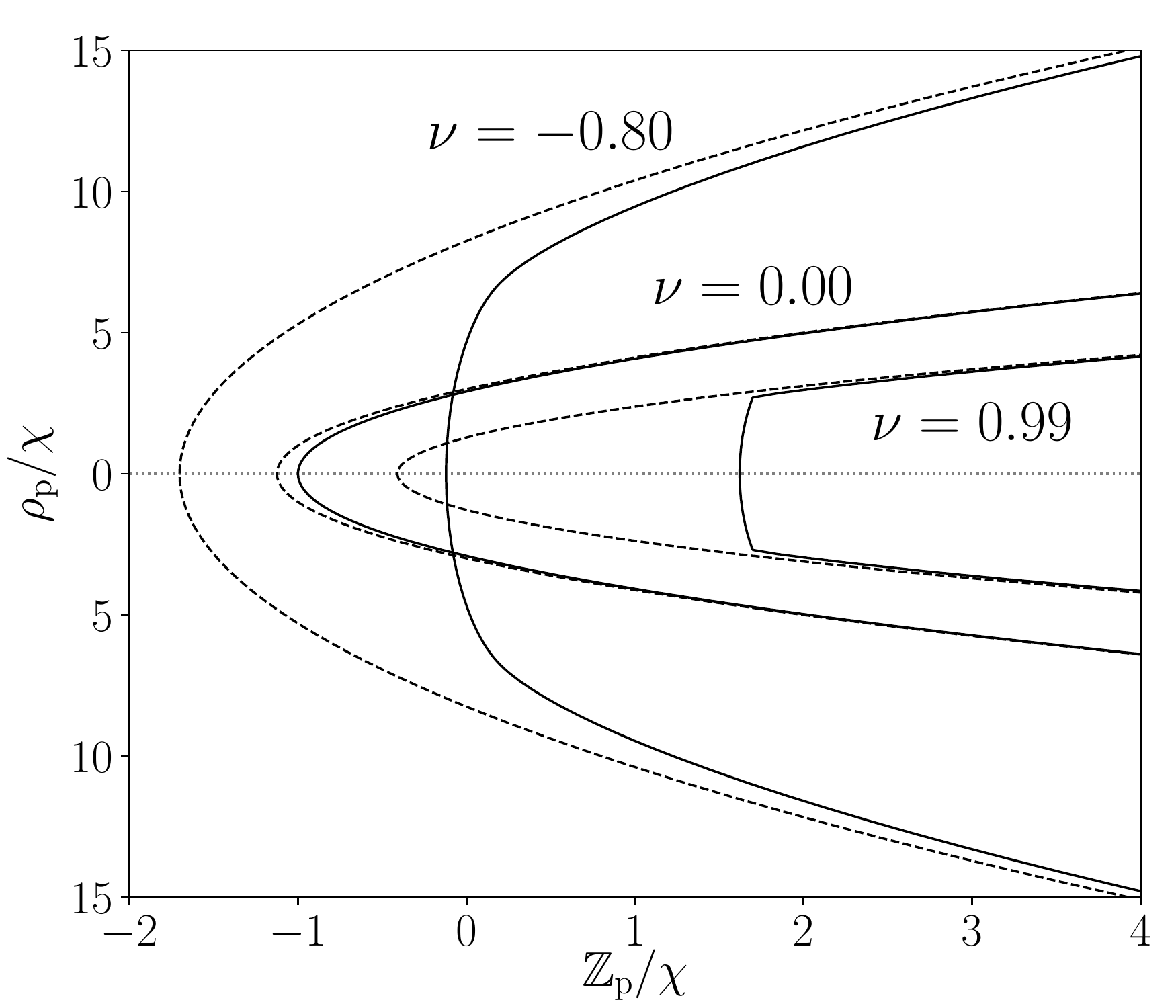}
\pull
\caption{Comparison between several scattering shadows (full line) and their parabolic asymptotic forms (dashed line) for \mbox{$\eta_\tar=0.5$}.}
\pullc
\label{fig8}
\end{figure}

Figure~\ref{fig8} shows a comparison between several true-shadow forms and the associated asymptotic ones, for \mbox{$\eta_\tar=0.5$} and \mbox{$\LL=\X$}. Expectedly, the asymptotic form does not reproduce accurately the true shadow caustic around the vertex, at least not for the extreme parameter values. It is also interesting to note from~(\ref{asym}) that as $\V_\cm\to\pm\eta_\tar v_0$ (i.e. \mbox{$\nu\to\pm1$}) the vertex of the asymptotic form does not even feature a true vertex' escape to infinity. Rather, in both cases it stays at some finite position along the \mbox{$\Z$-axis}. Finally, one can confirm from \mbox{$\nu=0$} case that not even in the \mbox{$\V_\cm=0$} frame is the shadow caustic parabolic; if it were, it would agree with the asymptotic form in its entirety.


\section{Erroneous procedure}
\label{error}

Instead of employing the correct extremization procedure from~(\ref{derivative}), we might be tempted to obtain the scattering shadow in the comoving frame by straightforward geometric manipulations of the known paraboloidal caustic from the fixed-target frame. However, this procedure is incorrect, as we will soon demonstrate. In order to understand why, consider some particular point on the shadow boundary as an intersection\footnote{
Formally speaking, the concept of the intersection of trajectories used here is more of the convergence of separate trajectories \textit{around} (infinitesimally close to), rather than \textit{through} the same point. This is easily understood from a fact that there is a unique solution to the extremizing impact parameter $\tilde{\R}_0$, corresponding to only one of these (infinitesimally close) trajectories.
} of different---if need be, infinitesimally separate---projectile trajectories. The problem lies in these trajectories not passing through their intersection point \textit{simultaneously}. In other words, the intersection of trajectories refers only to the crossing of their \textit{geometric shapes}, rather than their passing through the same point in space \textit{and} time. Thus, boosting the same intersection point, when treated as lying on separate trajectories, would have to be performed at different points in time, leading to its separation into distinct points. We may conclude that between the frames in relative motion, the shadow caustic is determined by an intersection of entirely different sets of trajectories.

\begin{figure}[b!]
\centering
\vspace*{-3mm}
\includegraphics[width=1\linewidth,keepaspectratio]{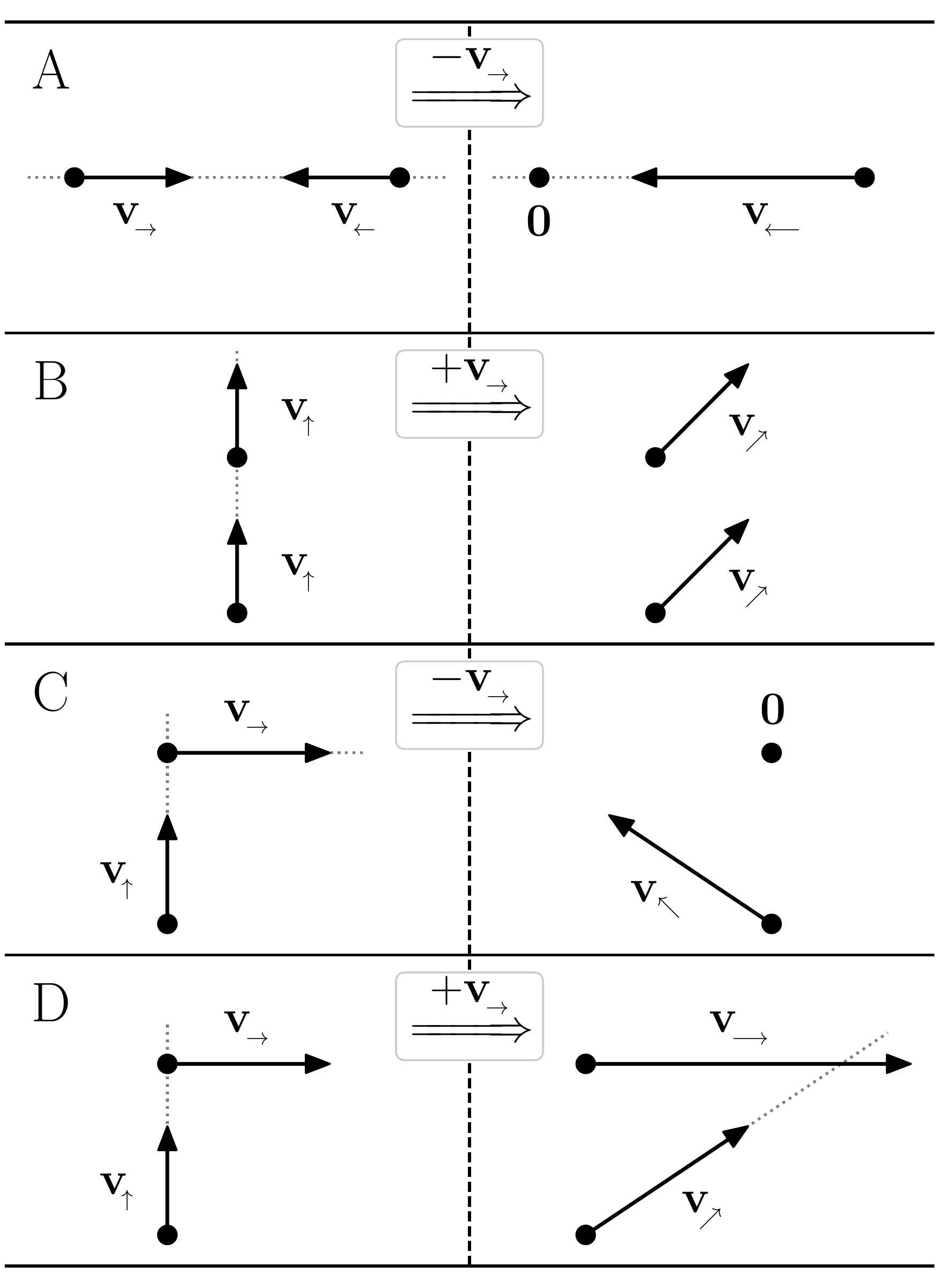}
\pull
\caption{Simple examples illustrating how the intersection of the trajectories' geometric shapes may change between the frames in relative motion when their crossing is not simultaneous. The cases show: (A) an infinite number of intersection points replaced by a single point; (B) an infinite number of intersection points replaced by no intersection at all; (C) a single intersection point replaced by no intersection at all; (D) a change in the position of the intersection point.}
\pullc
\label{figC}
\end{figure}

There are very simple examples illustrating not only (1)~that the two trajectories whose geometric shapes intersect at different points in time, may intersect at different points in space within the separate comoving frames, but also (2)~that the number of intersection points may vary, and drastically so between the frames, as well as (3)~that if they do intersect in one frame, they do not necessarily intersect in any other frame. Some of those examples are illustrated in figure~\ref{figC}. In all cases the boost velocity between the frames is clearly designated as $\pm\mathbf{v}_{\!_\rightarrow}$. The case (A) shows an example where the entire intersection line from one frame is replaced by a single intersection point in another. This one point endures in all other frames, as it is the only one where the two trajectories intersect simultaneously. The case (B) shows that the entire intersection line from one frame does not necessarily imply any intersection at all in any other frame. The case (C) shows how a single intersection point in one frame may lead to no intersection in another. The case (D), which is the most pertinent to our considerations, shows how the position of the intersection point may change between the frames in relative motion. This is most evident from the fact that the $\mathbf{v}_{\!_\uparrow}$-trajectory intersects the \textit{initial} position of the $\mathbf{v}_{\!_\rightarrow}$-trajectory in one frame, while some \textit{later} position on the boosted $\mathbf{v}_{\!_{\longrightarrow}}$-trajectory serves as the intersection point in another frame.

Returning to our attempt at demonstrating the erroneous shadow derivation, we first note that boosting the shadow from any starting frame does not alter the fact that all points on its boundary come from the specific particle trajectories. Thus, we may readily use the boosted trajectories from~(\ref{zeta}), that we already have at our disposal. The only item that may change in respect to a correct derivation procedure is the form of the extremizing value $\tilde{\R}_0(\rho_\pro)$, to be inserted into~(\ref{zz}). From~\cite{zugec_supp}---in particular, from its equations~(16) and (18), together with figure~B1 from its Supplementary note---we already know that the trajectories with the impact parameter $\tilde{\R}_0$ give rise to the shadow points at the radial distance:
\begin{equation}
\tilde{\ro}_\pro=2\tilde{\R}_0
\label{err_ass}
\end{equation}
from the $z$-axis in the fixed-target frame. As the parameter $\ro_\pro$ has been an integral part of a correct derivation ever since~(\ref{app_xi_def}), we introduce the previous value (supposing that it still holds in a comoving frame) into~(\ref{app_xi}), thus obtaining:
\begin{equation}
\rho_\pro(\tilde{\ro}_\pro)=\eta_\tar\tilde{\ro}_\pro+\eta_\pro\tilde{\R}_0 \quad \Rightarrow\quad \tilde{\R}_0(\rho_\pro)=\frac{\rho_\pro}{1+\eta_\tar},
\label{supp}
\end{equation}
which is a sought relation $\tilde{\R}_0(\rho_\pro)$ following from an erroneous assumption that the shadow in the comoving frame could be obtained by a straightforward geometric transformation between the frames in relative motion. Plugging this value into~(\ref{zz}) immediately yields a supposed shadow caustic as \mbox{$\ZZ(\rho_\pro)=\Z_\pro[\rho_\pro;\rho_\pro/(1+\eta_\tar)]$}:
\begin{equation}
\ZZ_\pro(\rho_\pro)=\frac{\eta_\tar v_0+\V_\cm}{2(1+\eta_\tar)^2 v_0\X}\rho_\pro^2+\frac{\V_\cm}{v_0}\X\ln\frac{\LL}{\X}-2\eta_\tar\X,
\label{wrong}
\end{equation}
which does not even agree with the correct asymptotic form~(\ref{asym})! It should be clearly stated that this procedure does not even correctly recover the shadow caustic from the \mbox{$\V_\cm=0$} frame, as~(\ref{wrong}) again disagrees with~(\ref{asym}). Moreover, recall from \mbox{$\nu=0$} case in figure~\ref{fig8} that in the \mbox{$\V_\cm=0$} frame the shadow caustic is not parabolic, to start with.



The only instance when this erroneous procedure agrees at least with the correct asymptotic form~(\ref{asym}) is the case of the infinitively massive target (\mbox{$\eta_\tar=1$}), since the supposed relation from~(\ref{supp}) reduces to \mbox{$\tilde{\R}_0(\rho_\pro)=\rho_\pro/2$}, as in~(\ref{asym_inf}). However, not even in that case does~(\ref{wrong}) agree with the \textit{general} shadow form, which is easily confirmed by observing that the supposed relation \mbox{$\tilde{\R}_0(\rho_\pro)=\rho_\pro/2$} is not the extremizing value consistent with~(\ref{derivative}):
\begin{equation}
\frac{\dd\Z_\pro}{\dd\R_0}\bigg|_{\rho_\pro/2}^{(\eta_\tar=1)}=-\frac{8\V_\cm\X}{v_0\rho_\pro}\neq0.
\end{equation}
The only case when the previous relation does yield zero is \mbox{$\V_\cm=0$}, when~(\ref{wrong}) reduces to a known~(\ref{inf_shad}) from the infinitely-massive-target frame. However, this is only because the infinitely-massive-target frame also corresponds to the fixed-target frame (as well as to the center-of-mass and the laboratory frame), so that the correct result is a simple consequence of no shadow transformation having been performed at all!

Finally, we may ask is there, for any other $\eta_\tar$, at least \textit{some} particular frame speed $\V_\cm$ for which~(\ref{derivative}) identically vanishes under the assumption of $\tilde{\R_0}(\rho_\pro)$ from~(\ref{supp})? In other words, could that relation be correct at least under some very stringent circumstances? Attempting to solve~(\ref{derivative}) for $\V_\cm$ in such case:
\begin{equation}
\frac{\dd\Z_\pro}{\dd\R_0}\bigg|_{\frac{\rho_\pro}{1+\eta_\tar}}=0 \quad\Rightarrow\quad \V_\cm=-\frac{\eta_\pro\eta_\tar v_0\rho_\pro^2}{4(1+\eta_\tar)^3\X^2+\eta_\pro\rho_\pro^2}
\end{equation}
reveals that the answer is negative, as the obtained value is not independent of $\rho_\pro$! Therefore, the erroneously obtained shadow from~(\ref{wrong}) is so utterly incorrect that it cannot even incidentally reproduce the correct solution!

\section{Identifying the correct root}
\label{root}

\begin{figure*}[t!]
\centering
\includegraphics[width=1\linewidth,keepaspectratio]{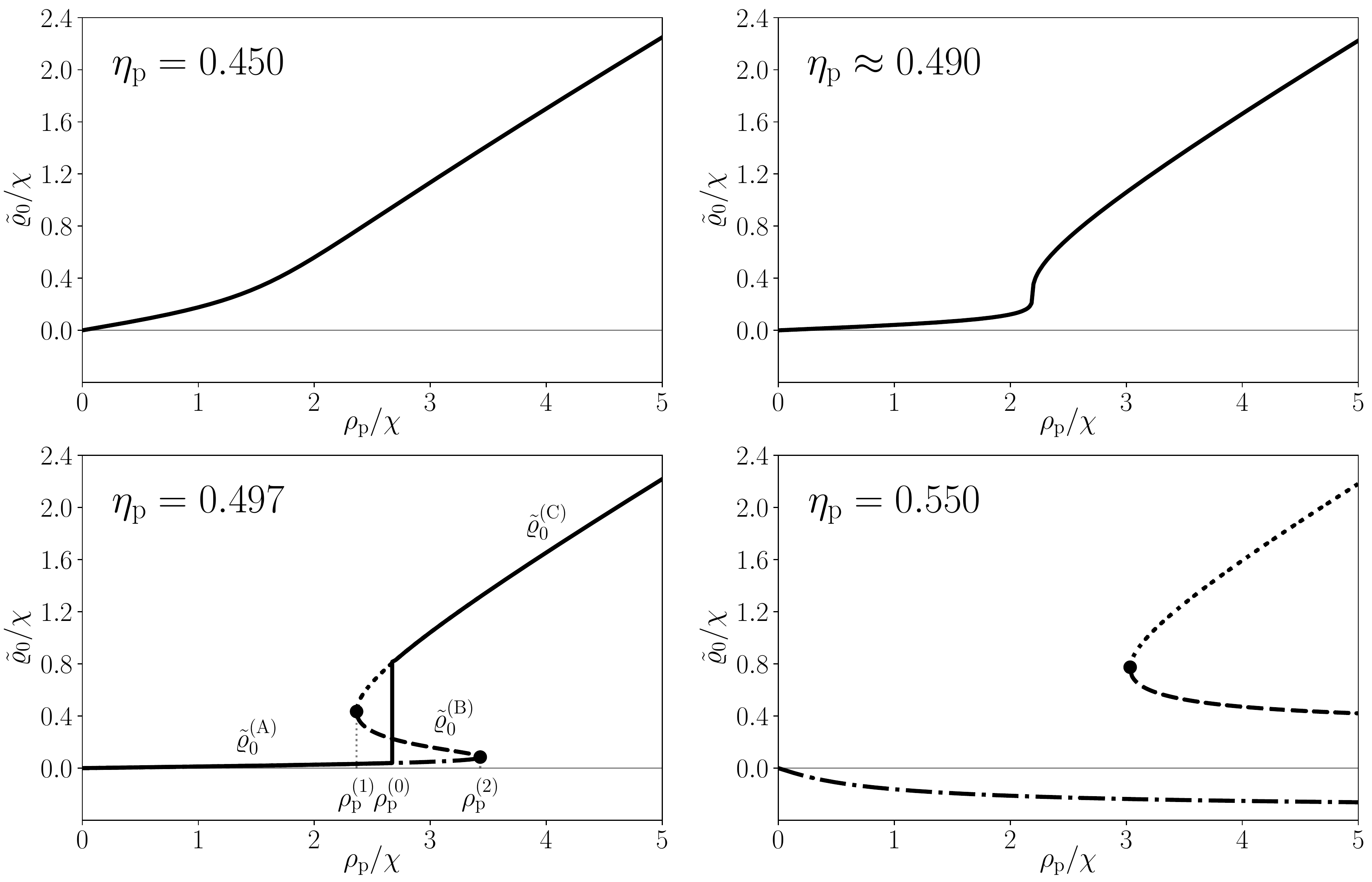}
\pull
\caption{Numerical investigation of the relevant roots to the extremization condition from~(\ref{lab_r0}) (the laboratory frame). Under the regular conditions (here: \mbox{$\eta_\pro=0.450$} and \mbox{$\eta_\pro\approx0.490$}) there is a single solution to the 5$^\mathrm{th}$ degree polynomial, that satisfies the geometric confine \mbox{$0\le \tilde{\R}_0 \le \rho_\pro$}. As $\eta_\pro$ starts approaching the shadow disappearance limit (the \mbox{$\eta_\pro=0.497$} example), the additional roots start appearing. As the shadow disappearance limit \mbox{$\eta_\pro=0.5$} is exceeded (the \mbox{$\eta_\pro=0.550$} example), the only branch providing the real solution at low $\rho_\pro$ drops to the negative values.}
\pullc
\label{figA1}
\end{figure*}

The central extremization condition from~(\ref{derivative}) for finding the scattering shadow in a comoving frame requires solving the 5$^\mathrm{th}$ degree polynomial in $\tilde{\R}_0$:
\begin{align}
\begin{split}
\mathcal{P}(\tilde{\R}_0)=&(1+\nu)(\rho_\pro-2\tilde{\R}_0)(\rho_\pro-\tilde{\R}_0)^2\tilde{\R}_0^2+\\
&\X^2\rho_\pro(\eta_\pro\tilde{\R}_0-\rho_\pro)\big[\eta_\tar(1+\nu)\tilde{\R}_0-(1-\nu)(\rho_\pro-\tilde{\R}_0)\big].
\end{split}
\label{pol}
\end{align}
Searching for a unique solution, one would certainly hope that four out of five solutions $\tilde{\R}_0^{(i)}$  to this equation could systematically be eliminated by turning out to be complex, negative or out of range set by~(\ref{confine}). Regrettably, this is not so. Thus, a special care should be taken if one were interested in identifying the correct root $\tilde{\R}_0^{(i)}$ itself, which is, thanks to a procedure from~(\ref{max}), not even necessary for obtaining the correct shadow form.

Figure~\ref{figA1} shows the behavior of the real roots $\tilde{\R}_0^{(i)}$ to the extremization condition from~(\ref{lab_r0}), related to the laboratory frame examples from Section~\ref{lab_frame} of the main paper. With the exception of the last plot (for $\mbox{$\eta_\pro=0.550$}$), only the solutions satisfying the confine \mbox{$0\le \tilde{\R}_0^{(i)} \le \rho_\pro$} are shown. One can observe that under certain conditions there is indeed only a single acceptable root (\mbox{$\eta_\pro=0.450$} and \mbox{$\eta_\pro\approx0.490$} examples). In that, the \mbox{$\eta_\pro\approx0.490$} example (more precisely \mbox{$\eta_\pro\approx0.489756$}, a value that will be discussed later) is the borderline case, just before the additional roots start appearing, as demonstrated by the \mbox{$\eta_\pro=0.497$} example. The full line shows the `flow' of the correct shadow-related root, that switches discontinuously between the two branches designated as $\tilde{\R}_0^{(\mathrm{A})}$ and $\tilde{\R}_0^{(\mathrm{C})}$, and does so at the point $\rho_\pro^{(0)}$ that is in no easily discernible manner related to the branches' ending points $\rho_\pro^{(1)}$ and $\rho_\pro^{(2)}$. Finally, the \mbox{$\eta_\pro=0.550$} case shows the example of the roots' behavior above the shadow existence limit (\mbox{$\eta_\pro=0.5$} in the laboratory frame). We see that the shadow disappearance is signaled when the only branch providing the real solution at low $\rho_\pro$ drops to the negative values, leaving no root consistent with \mbox{$0\le \tilde{\R}_0 \le \rho_\pro$} within that range.

\begin{figure}[t!]
\centering
\includegraphics[width=1\linewidth,keepaspectratio]{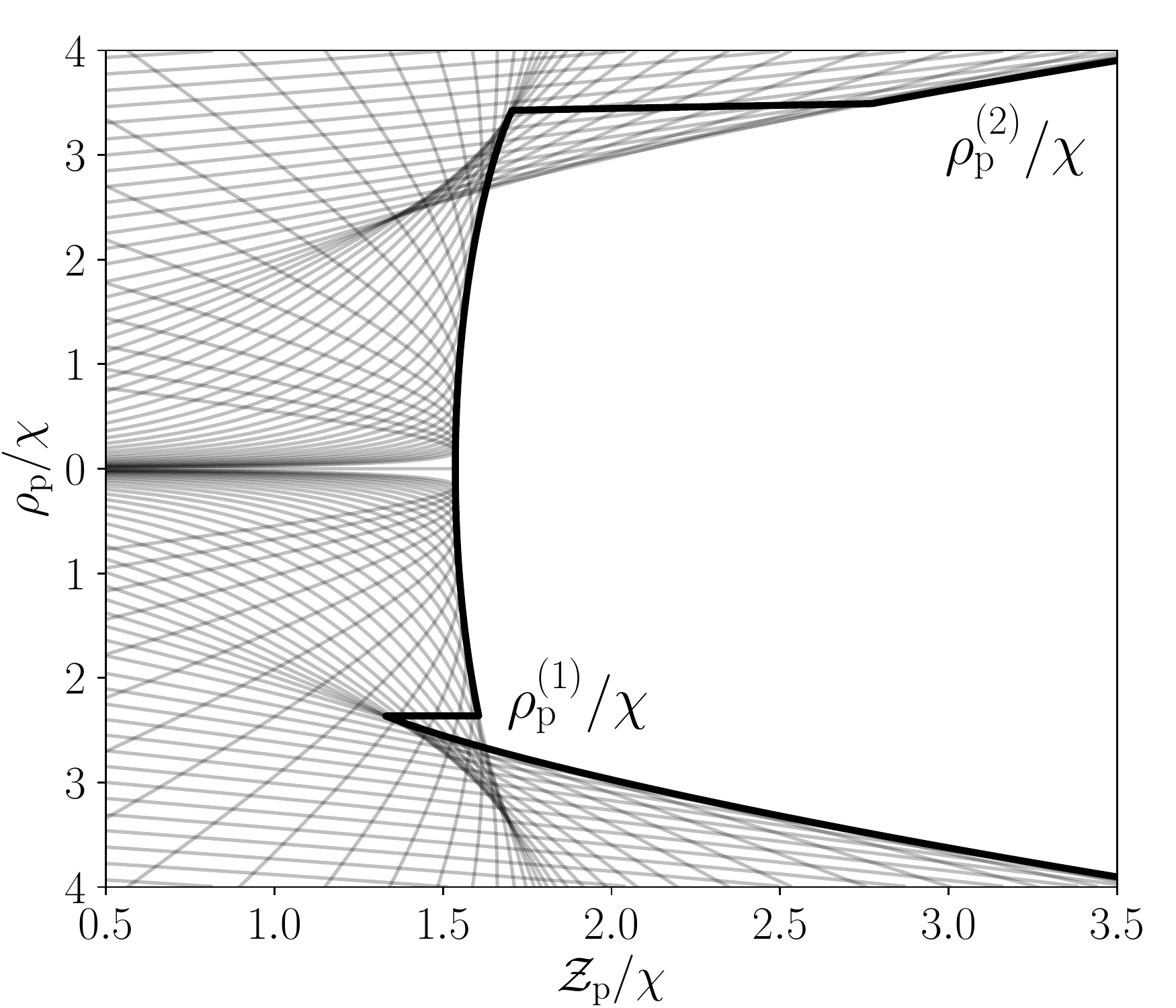}
\pull
\caption{Projectile trajectories in the laboratory frame for \mbox{$\eta_\pro=0.497$}, in comparison with the shadow forms predicted by the root branches $\tilde{\R}_0^{(\mathrm{A})}$ and $\tilde{\R}_0^{(\mathrm{C})}$ from figure~\ref{figA1}, if these branches were switched between either at $\rho_\pro^{(1)}$ (the lower shadow form) or at at $\rho_\pro^{(2)}$ (the upper shadow form). The switch needs to be made somewhere in between, at the point $\rho_\pro^{(0)}$ where the two shadow forms would intersect if overlapped. Precisely these observations are the basis behind~(\ref{max}).}
\pullc
\label{figA2}
\end{figure}

Let us pay a closer attention to the \mbox{$\eta_\pro=0.497$} example from figure~\ref{figA1}, representative of other such cases appearing for the extreme values of $\eta_\pro$ and $\nu$. Three separate branches consistent with  \mbox{$0\le \tilde{\R}_0^{(i)} \le \rho_\pro$} are clearly designated as $\tilde{\R}_0^{(\mathrm{A})}$, $\tilde{\R}_0^{(\mathrm{B})}$ and $\tilde{\R}_0^{(\mathrm{C})}$. The first branch `flows' from 0 to $\rho_\pro^{(2)}$, the second one from $\rho_\pro^{(1)}$ to $\rho_\pro^{(2)}$ and the third one from $\rho_\pro^{(1)}$ to infinity. Figure~\ref{figA2}---which is a backbone of the simple method from~(\ref{max})---shows how the reconstructed shadow would look if, in search for the correct root $\tilde{\R}_0$, we were to switch between the branches $\tilde{\R}_0^{(\mathrm{A})}$ and $\tilde{\R}_0^{(\mathrm{C})}$ either at the point $\rho_\pro^{(1)}$ (the lower shadow form) or at $\rho_\pro^{(2)}$ (the upper shadow form). Each branch correctly describes the shadow within the range where they are the only acceptable solution: $\tilde{\R}_0^{(\mathrm{A})}$ up to $\rho_\pro^{(1)}$ and  $\tilde{\R}_0^{(\mathrm{C})}$ from $\rho_\pro^{(2)}$. However, within the range \mbox{$\rho_\pro^{(1)}<\rho_\pro<\rho_\pro^{(2)}$} one needs to switch between the branches at the point where the shadow features a \textit{sharp edge}, this point being precisely $\rho_\pro^{(0)}$ from figure~\ref{figA1}. We note that the behavior of solutions from figure~\ref{figA1} is characteristic of the \textit{phase transition}, wherein we have identified four possible phase states: a~\textit{smooth shadow} phase, a~\textit{split shadow} phase, a \textit{trivial shadow} phase, and \textit{no shadow} phase.


Figure~\ref{figA2} clearly shows that the shadow-unrelated portions of the solutions $\tilde{\R}_0^{(\mathrm{A})}$ and $\tilde{\R}_0^{(\mathrm{C})}$ are, in fact, meaningful. They are, indeed, \textit{local} extremes of $\Z_\pro(\rho_\pro;\R_0)$ for a given $\rho_\pro$, which can be seen from the trajectories piling up along the superfluous lines (inside the portion of space swept by the trajectories). However, the shadow caustic is determined by the \textit{global} extremes, beyond which no additional trajectories are admitted. This is the reasoning behind the simple procedure from~(\ref{max}): of all the points satisfying \textit{any} extremization condition, those yielding the \textit{highest} extremum are those beyond which no more trajectories can be found.

Since the shadow forms $\ZZ_\pro^{(\mathrm{A})}(\rho_\pro)$ and $\ZZ_\pro^{(\mathrm{C})}(\rho_\pro)$ predicted by branches  $\tilde{\R}_0^{(\mathrm{A})}$ and  $\tilde{\R}_0^{(\mathrm{C})}$ intersect at the switching point $\rho_\pro^{(0)}$, this point is to be found by solving the equation:
\mbox{$\ZZ_\pro^{(\mathrm{A})}\big(\rho_\pro^{(0)}\big)=\ZZ_\pro^{(\mathrm{C})}\big(\rho_\pro^{(0)}\big)$}, i.e. by finding the numerical solution to:
\begin{equation}
\Z_\pro\big[\rho_\pro^{(0)};\tilde{\R}_0^{(\mathrm{A})}\big(\rho_\pro^{(0)}\big)\big]=\Z_\pro\big[\rho_\pro^{(0)};\tilde{\R}_0^{(\mathrm{C})}\big(\rho_\pro^{(0)}\big)\big],
\label{rhoP0}
\end{equation}
as per the shadow definition from~(\ref{zz}). Since the sought solution for $\tilde{\R}_0$ features a discontinuous jump at $\rho_\pro^{(0)}$, the intermediate values of $\R_0$ do not contribute to the formation of the shadow caustic. This means that when the branches separate, i.e. when the shadow features a sharp edge (a split shadow phase), there exists a range of trajectories:
\vspace*{-2mm}
\begin{equation}
\R_0\in\left\langle\tilde{\R}_0^{(\mathrm{A})}\big(\rho_\pro^{(0)}\big) \,,\, \tilde{\R}_0^{(\mathrm{C})}\big(\rho_\pro^{(0)}\big)\right\rangle
\vspace*{-2mm}
\end{equation}
which do not make contact with the shadow caustic at any point, i.e. do not take part in its formation!


Let us now identify the conditions for the existence of multiple branches $\tilde{\R}_0^{(i)}$, together with the procedure for obtaining $\rho_\pro^{(1)}$ and $\rho_\pro^{(2)}$, and thus $\rho_\pro^{(0)}$. As is evident from figure~\ref{figA1}, in particular from the \mbox{$\eta_\pro=0.497$} case, when the multiple branches exist, there are points where the two branches meet (precisely $\rho_\pro^{(1)}$ and $\rho_\pro^{(2)}$), making these points \textit{multiple roots} to the polynomial $\mathcal{P}(\tilde{\R}_0)$ from~(\ref{pol}).

\begin{figure}[t!]
\centering
\includegraphics[width=1\linewidth,keepaspectratio]{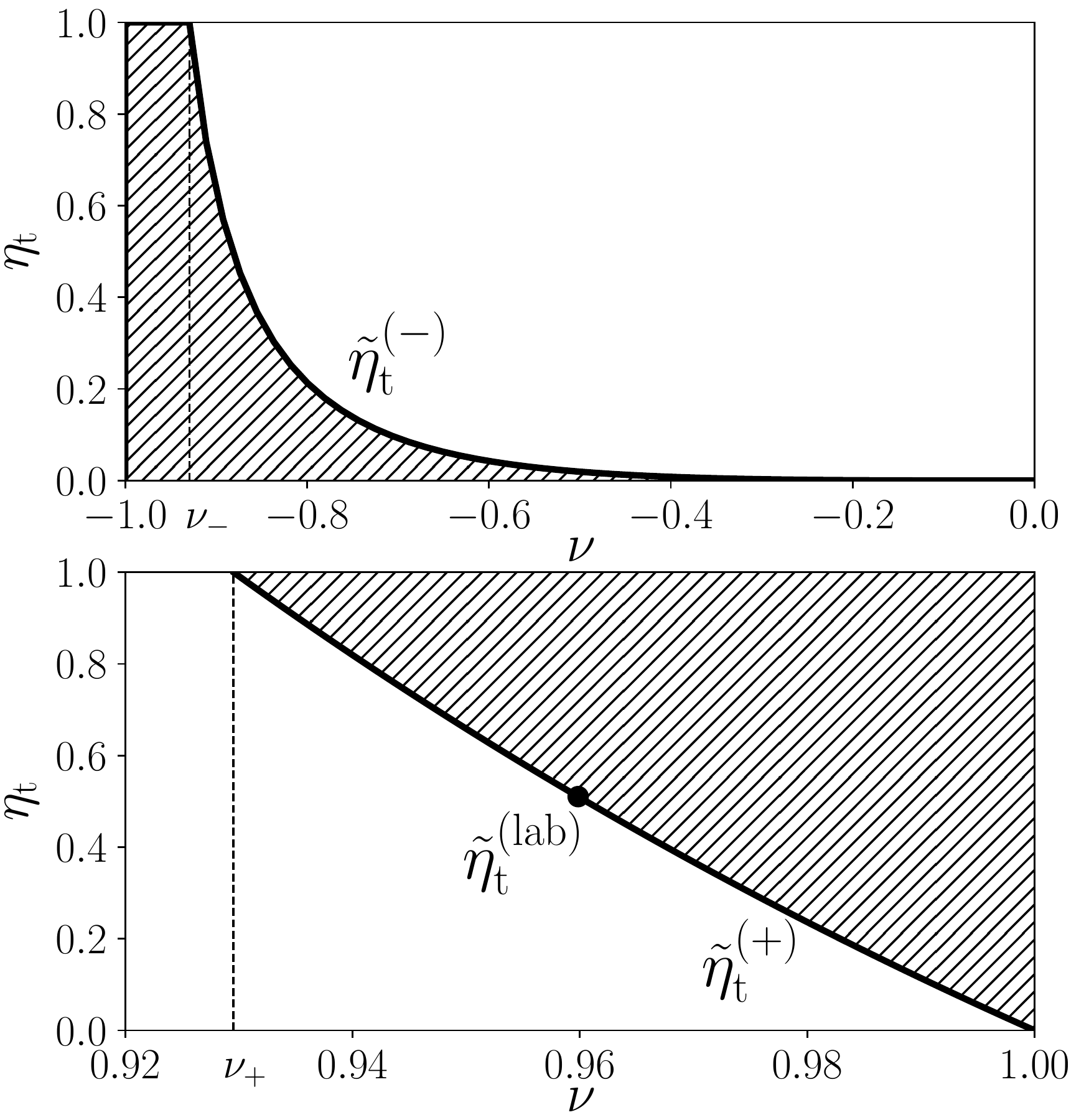}
\pull
\caption{Range of parameters $\nu$ and $\eta_\tar$ (the shaded area) for which there are multiple branches $\tilde{\R}_0^{(i)}$ satisfying the confine \mbox{$0\le \tilde{\R}_0^{(i)} \le \rho_\pro$}. The laboratory frame is represented by a point at \mbox{$\tilde{\eta}_\tar^{(\lab)}\approx0.510244$}, i.e. \mbox{$\nu_0^{(\lab)}\approx0.959846$}.}
\pullc
\label{figA3}
\end{figure}

\noindent Since the discriminant of a polynomial vanishes whenever the polynomial exhibits multiple roots, we observe the associated discriminant $\mathbb{D}_{\mathcal{P}(\tilde{\R}_0)}$:
\vspace*{-1.5mm}
\begin{equation}
\mathbb{D}_{\mathcal{P}(\tilde{\R}_0)}(\rho_\pro)=\rho_\pro^8\mathcal{Q}(\rho_\pro^2)
\label{disP}
\vspace*{-1mm}
\end{equation}
and obtain $\rho_\pro^{(1)}$ and $\rho_\pro^{(2)}$ by finding its zeros, i.e. by solving the equation:
\vspace*{-2mm}
\begin{equation}
\mathbb{D}_{\mathcal{P}(\tilde{\R}_0)}\big(\rho_\pro^{(1,2)}\big)=0.
\label{disP0}
\vspace*{-1mm}
\end{equation}
In~(\ref{disP}) we have indicated the structure of the discriminant without writing it out in its entirety, due to the expression being quite long (the full expression may easily be obtained and manipulated with the help of any available symbolic programming code, such as \textit{sympy} from \textit{Python}, to name just one). As the polynomial $\mathcal{P}$ is a function of both $\tilde{\R}_0$ and $\rho_\pro$, in treating it as a polynomial in $\tilde{\R}_0$ its discriminant remains a function of $\rho_\pro$, of indicated form. In that, $\mathcal{Q}(\rho_\pro^2)$ is the $4^\mathrm{th}$ degree polynomial in $\rho_\pro^2$ (a `biquartic' function). Evidently, any nontrivial solution to~(\ref{disP0}) is a root of a polynomial $\mathcal{Q}$, i.e. the sought points $\rho_\pro^{(1)}$ and $\rho_\pro^{(2)}$ follow from: \mbox{$\mathcal{Q}\big[\big(\rho_\pro^{(1,2)}\big)^2\big]=0$}. We further pose the question: when does this equation have any acceptable (positive) solutions, i.e. when do the multiple branches $\tilde{\R}_0^{(i)}$ even appear? Consider what happens at the limit, just before the multiple branches separate (the \mbox{$\eta_\pro\approx0.490$} case from figure~\ref{figA1}): $\rho_\pro^{(1)}$ and $\rho_\pro^{(2)}$ start separating from the same point. In other words, they start as the multiple roots to~(\ref{disP0})! Therefore, we might identify the limiting case by observing \textit{the discriminant of the discriminant} from~(\ref{disP}). However, as $\rho_\pro=0$ is always its multiple (octuple) root, the discriminant of~$\mathbb{D}_{\mathcal{P}(\tilde{\R}_0)}$ is always zero. Hence, we must isolate the discriminant of that part from which the solutions $\rho_\pro^{(1)}$ and $\rho_\pro^{(2)}$ originate\footnote{
Ideally, instead of the discriminant of the entire polynomial $\mathcal{Q}(\rho_\pro^2)$, we would only need a discriminant of a reduced polynomial \mbox{$\big(\rho_\pro-\rho_\pro^{(1)}\big)\big(\rho_\pro-\rho_\pro^{(2)}\big)$} yielding the relevant solutions $\rho_\pro^{(1)}$ and $\rho_\pro^{(2)}$. However, the sporadic solutions to $\mathcal{Q}(\rho_\pro^2)$ are not easily factored out, so we identify the relevant zeros of $\mathbb{D}_{\mathcal{Q}(\rho_\pro^2)}$ by a detailed numerical analysis.
}: of the polynomial $\mathcal{Q}(\rho_\pro^2)$. Without writing it out in its fullness, we again indicate the form of the nontrivial part of this discriminant:
\vspace*{-1.5mm}
\begin{equation}
\mathbb{D}_{\mathcal{Q}(\rho_\pro^2)}\left(\eta_\tar,\nu\right)\propto [\eta_\tar(1+\nu)-(1-\nu)]\mathcal{R}^2(\eta_\tar,\nu)\mathcal{S}^3(\eta_\tar,\nu),
\label{disQ}
\end{equation}
where $\mathcal{R}(\eta_\tar,\nu)$ is a polynomial of the $4^\mathrm{th}$ degree in both $\eta_\tar$ and $\nu$, while $\mathcal{S}(\eta_\tar,\nu)$ is a polynomial of the $4^\mathrm{th}$ degree in $\eta_\tar$ and the $5^\mathrm{th}$ degree in $\nu$. Finding the relevant zero of this discriminant---i.e. solving the equation \mbox{$\mathbb{D}_{\mathcal{Q}(\rho_\pro^2)}\left(\tilde{\eta}_\tar,\nu\right)=0$} as a condition for $\rho_\pro^{(1)}$ and $\rho_\pro^{(2)}$ starting to appear as multiple roots---one obtains the value $\tilde{\eta}_\tar(\nu)$ at which, for a given $\nu$, multiple branches $\tilde{\R}_0^{(i)}$ start separating. The detailed numerical investigation shows that the sought $\tilde{\eta}_\tar(\nu)$ is a solution to the polynomial $\mathcal{S}$ from~(\ref{disQ}). In that, one of the four solutions to $\mathcal{S}$ yields a correct $\tilde{\eta}_\tar(\nu)$ for \mbox{$\nu>0$}, while the separate solution applies for \mbox{$\nu<0$}. We denote these two relevant branches as $\tilde{\eta}_\tar^{(+)}$ and $\tilde{\eta}_\tar^{(-)}$ (for positive and negative $\nu$, respectively) and display them in figure~\ref{figA3}, so that the readers following our methodology may immediately recognize them among their own solutions. 

\pagebreak

The full description of figure~\ref{figA3} requires some further explanations. In their part, the branches $\tilde{\eta}_\tar^{(\pm)}$ yield the solutions greater than~1, while $\eta_\tar$ by its definition cannot exceed this value. Therefore, we wish to identify the values of $\nu$, at which the solutions for $\tilde{\eta}_\tar$ need to be truncated from $\tilde{\eta}_\tar^{(\pm)}$ to unity. This is easily done by observing the relevant condition $\mathcal{S}(\tilde{\eta}_\tar,\nu)=0$ at \mbox{$\tilde{\eta}_\tar=1$}, where the polynomial $\mathcal{S}$ takes the simple form: \mbox{$\mathcal{S}(\tilde{\eta}_\tar=1,\nu)=16\nu(125\nu^2-108)$}, yielding two relevant solutions:
\vspace*{-0.5mm}
\begin{equation}
\nu_\pm=\pm\frac{6\sqrt{15}}{25}\approx \pm0.929516.
\end{equation}
Since \mbox{$\tilde{\eta}_\tar^{(-)}>1$} for \mbox{$\nu<\nu_-$} and \mbox{$\tilde{\eta}_\tar^{(+)}>1$} for \mbox{$\nu<\nu_+$}, the physical constraint \mbox{$\tilde{\eta}_\tar=1$} must be forced for \mbox{$-1<\nu<\nu_-$} and \mbox{$0<\nu<\nu_+$}. Thus, we may write the final form for $\tilde{\eta}_\tar$ as:
\begin{equation}
\tilde{\eta}_\tar(\nu)=\left\{\begin{array}{cl}
1&\ifif \; -1<\nu\le\nu_-\\
\tilde{\eta}_\tar^{(-)}&\ifif \; \nu_-<\nu<0\\
1&\ifif \; 0<\nu\le\nu_+\\
\tilde{\eta}_\tar^{(+)}&\ifif \; \nu_+<\nu<1
\end{array}\right. .
\label{eta_tilde}
\end{equation}
 A numerical investigation further reveals that for \mbox{$\nu>0$} the multiple branches $\tilde{\R}_0^{(i)}$ appear when \mbox{$\eta_\tar>\tilde{\eta}_\tar$}, while for \mbox{$\nu<0$} when \mbox{$\eta_\tar<\tilde{\eta}_\tar$}, which is a range of values shown in figure~\ref{figA3} by the shaded area. It is easily seen that for \mbox{$-1<\nu\le\nu_-$} there are multiple branches for every $\eta_\tar$, while for \mbox{$0<\nu\le\nu_+$}, and in particular for \mbox{$\nu=0$}, there is always a single branch such that \mbox{$0\le\tilde{\R}_0\le\rho_\pro$}. All these claims may be elegantly compacted by saying that the separation of multiple branches appears if and only if \mbox{$\eta_\tar\,\sgn(\nu)>\tilde{\eta}_\tar\,\sgn(\nu)$}, with the \textit{signum} function $\sgn(\cdot)$ returning the sign of the argument ($-1$, $0$, $1$).

In a sense of a phase transition, figure~\ref{figA3} represents a phase diagram---clear and shaded area corresponding to a smooth and split shadow, respectively---with $\tilde{\eta}_\tar(\nu)$ as a critical line separating the phase states. In that, figure~\ref{figA3} additionally shows the position of the point $\tilde{\eta}_\tar^{(\lab)}$ corresponding to a phase transition in the laboratory frame. Since \mbox{$\nu^{(\lab)}=\eta_\pro/\eta_\tar=(1-\eta_\tar)/\eta_\tar$} by definition of the laboratory frame, one needs to solve the equation \mbox{$\mathcal{S}\big[\tilde{\eta}_\tar^{(\lab)},\big(1-\tilde{\eta}_\tar^{(\lab)}\big)/\tilde{\eta}_\tar^{(\lab)}\big]=0$} in a single variable. Numerical procedure yields the value \mbox{$\tilde{\eta}_\tar^{(\lab)}\approx0.510244$}---as the solution to the sextic equation $31x^6+623x^5+742x^4-322x^3-373x^2+195x-32=0$---thus also providing \mbox{$\nu_0^{(\lab)}\approx0.959846$}, which defines a laboratory frame in which the shadow starts featuring a sharp edge, i.e. the phase transition occurs. This value of $\tilde{\eta}_\tar^{(\lab)}$ is in a direct connection with the \mbox{$\eta_\pro\approx0.490$} case from figure~\ref{figA1}, which more precisely corresponds to \mbox{$\eta_\pro=1-\tilde{\eta}_\tar^{(\lab)}\approx0.489756$}.

\begin{figure}[b!]
\centering
\includegraphics[width=1\linewidth,keepaspectratio]{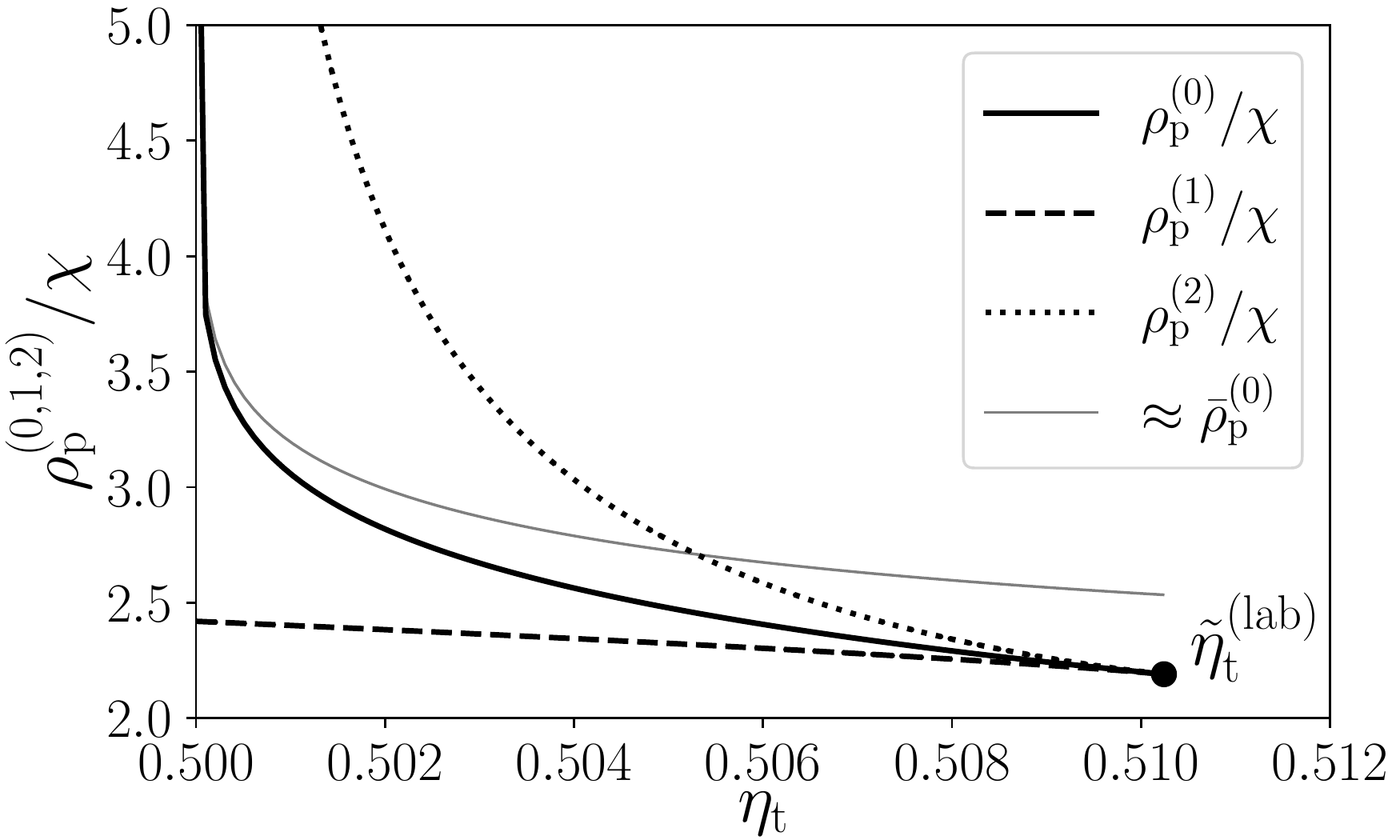}
\pull
\caption{Boundary points $\rho_\pro^{(1)}$ and $\rho_\pro^{(2)}$ of the multiple root-branches $\tilde{\R}_0^{(i)}$ in the laboratory frame (\mbox{$\tilde{\eta}_\tar^{(\lab)}\approx0.510244$}), together with the shadow `breaking' point $\rho_\pro^{(0)}$, where the branches discontinuously interchange. As \mbox{$\eta_\tar\to0.5$} (the shadow disappearance limit in the laboratory frame), $\rho_\pro^{(0)}$ and $\rho_\pro^{(2)}$ diverge, while: \mbox{$\lim_{\eta_\tar\to0.5}\rho_\pro^{(1)}/\X\approx2.42$}. The approximation to $\rho_\pro^{(0)}$ from~(\ref{r0_approx}) is also shown, revealing the divergence of the form: \mbox{$\lim_{\eta_\tar\to0.5}\rho_\pro^{(0)}\propto\sqrt{-\ln(2\eta_\tar-1)}$}.}
\pullc
\label{figA4}
\end{figure}

Once it has been determined whether the multiple $\tilde{\R}_0^{(i)}$ branches appear, one can find their bounds $\rho_\pro^{(1)}$ and $\rho_\pro^{(2)}$ by solving~(\ref{disP0})---more precisely, the equation \mbox{$\mathcal{Q}\big[\big(\rho_\pro^{(1,2)}\big)^2\big]=0$}---and identify the branches' interchange point $\rho_\pro^{(0)}$ by searching for a solution to~(\ref{rhoP0}) within the range \mbox{$\rho_\pro^{(1)}<\rho_\pro^{(0)}<\rho_\pro^{(2)}$}. Figure~\ref{figA4} shows an example of thus found solutions in a laboratory frame. The meaning of \mbox{$\tilde{\eta}_\tar^{(\lab)}$}---previously determined as \mbox{$\tilde{\eta}_\tar^{(\lab)}\approx0.510244$}---is clearly seen as a limiting value for which these solutions exist. One could have already intuited from the \mbox{$\eta_\pro=0.550$} case in figure~\ref{figA1}, when the scattering shadow does not exist in the laboratory frame, that $\rho_\pro^{(0)}$ and $\rho_\pro^{(2)}$ must diverge as the underlying parameters approach the shadow disappearance limit \mbox{$|\nu|=1$} (i.e. \mbox{$\eta_\tar=0.5$} in the laboratory frame). It can also be confirmed from the same case in figure~\ref{figA1} that $\rho_\pro^{(1)}$ may indeed stay finite under the same conditions (being represented by the two branches' meeting point around \mbox{$\rho_\pro/\X\approx3$}, even though the shadow has long since disappeared), thus supporting the finite limit suggested by figure~\ref{figA4}. Returning to the divergence of the shadow `breaking' point $\rho_\pro^{(0)}$, where it features a sharp edge as two of the branches $\tilde{\R}_0^{(i)}$ discontinuously interchange, figure~\ref{figA4} suggests that its divergence may be quite sharp. In fact, the nature of this divergence may be clearly identified from the approximate expression for $\rho_\pro^{(0)}$. Figure~\ref{fig8} suggests that as \mbox{$\nu\to\pm1$}, the estimate for $\rho_\pro^{(0)}$ might be obtained by intersecting the asymptotic shadow form~(\ref{asym}) with the limiting form around the shadow vertex:
\begin{equation}
\Big[\lim_{\rho_\pro\to0} \ZZ_\pro(\rho_\pro)\Big]_{\rho_\pro^{(0)}}=\Big[\lim_{\rho_\pro\to\infty} \ZZ_\pro(\rho_\pro)\Big]_{\rho_\pro^{(0)}},
\end{equation}
where \mbox{$\lim_{\rho_\pro\to0} \ZZ_\pro(\rho_\pro)=\frac{1}{2}\vtx''_\pro\rho_\pro^2+\vtx_\pro$}, with the vertex position $\vtx_\pro$ and curvature $\vtx''_\pro$ given by~(\ref{vtx_eq}) and (\ref{vtx_cur}). Solving this simple equation yields an approximation:
\begin{align}
\begin{split}
\rho_\pro^{(0)}\approx\,&2\X\frac{1+\eta_\tar-\eta_\pro\nu}{|(1+\eta_\tar)\nu-\eta_\pro|}\times\\
&\sqrt{\frac{\eta_\pro[\eta_\pro-(1+\eta_\tar)\nu]+2\eta_\tar\nu\ln[\eta_\tar(1+\nu)/(1-\nu)]}{ 1+\nu }}.
\end{split}
\label{r0_approx}
\end{align}
Notice: $\lim_{\nu\to\eta_\pro/(1+\eta_\tar)}\rho_\pro^{(0)}=2\sqrt{2\eta_\tar(1+\eta_\tar)}$. Figure~\ref{figA4} reveals this approximation as rather poor, in general case. However, it does become increasingly appropriate as the shadow flattens (\mbox{$\nu\to\pm1$}), allowing us to identify the form of the divergence for \mbox{$\nu\to1$}:
\pulleqa
\begin{equation}
\lim_{\nu\to1}\rho_\pro^{(0)}\propto\sqrt{-\ln(1-\nu)},
\pulleqb
\end{equation}
as well as for \mbox{$\nu\to-1$}:
\pulleqa
\begin{equation}
\lim_{\nu\to-1}\rho_\pro^{(0)}\propto\sqrt{-\frac{\ln(1+\nu)}{1+\nu}}.
\end{equation}
In the laboratory frame, where \mbox{$\nu^{(\lab)}=\eta_\pro/\eta_\tar$}, this can be translated into: \mbox{$\lim_{\eta_\tar\to0.5}\rho_\pro^{(0)}\propto\sqrt{-\ln(2\eta_\tar-1)}$}.

In summary, even in the presence of multiple branches $\tilde{\R}_0^{(i)}$ satisfying \mbox{$0\le \tilde{\R}_0^{(i)} \le \rho_\pro$}, the entire procedure leading to the identification of the `breaking' point $\rho_\pro^{(0)}$ may be safely circumvented by the virtue of the very simple and efficient method from~(\ref{max}), based on the observations from figure~\ref{figA2}. However, if the procedure was followed to the letter---reflecting all the underlying computational intricacies---the calculation of the scattering shadow caustic in the comoving frame could be decomposed into the following steps:
\begin{enumerate}\itemsep0em
\item[(1)] determine if the (nontrivial) shadow exists by checking if \mbox{$-1<\nu<1$};
\item[(2)] determine if multiple relevant branches $\tilde{\R}_0^{(i)}$ appear by checking if \mbox{$\eta_\tar\,\sgn(\nu)>\tilde{\eta}_\tar\,\sgn(\nu)$} [see~(\ref{eta_tilde})];
\item[(3a)] if (2) is satisfied, determine the branches' boundary points  $\rho_\pro^{(1)}$ and $\rho_\pro^{(2)}$ by solving~(\ref{disP0}), in order to find  their interchange point $\rho_\pro^{(0)}$ by solving~(\ref{rhoP0}) within the range \mbox{$\rho_\pro^{(1)}<\rho_\pro^{(0)}<\rho_\pro^{(2)}$};
\item[(3b)] otherwise, use a single available branch such that \mbox{$0\le \tilde{\R}_0 \le \rho_\pro$};
\item[(4)] calculate the scattering shadow from~(\ref{zz}) using the relevant branch(es).
\end{enumerate}

\section{The closest approach}
\label{closest}

We now extend our analysis to the points of the closest approach between the projectile and target. In~\cite{zugec_supp} we have already shown that in the fixed-target frame the projectile comes closest to the target when passing at an angle:
\begin{equation}
\tht(\R_0)=\pi/2+\atan(\X/\R_0).
\label{tht}
\end{equation}
From~(\ref{app_rhop}) and (\ref{app_zetap}) the closest approach curve (i.e. the geometric place of all the closest approach points) in the comoving frame is easily obtained in a parametric form. We only need to use the inverse relation:
\begin{equation}
\R_0\big(\tht\big)=-\X\tan\tht
\end{equation}
in order to establish the connection between the particular projectile trajectory and its angle of the closest approach. The parametric equations for the coordinates $\cpr_\pro$ and $\cpz_\pro$ of the projectiles' points of the closest approach may now be defined as:
\begin{align}
&\cpr_\pro\big(\tht\big)\equiv\rho_\pro\big[\tht;\R_0\big(\tht\big)\big],
\label{cp_rp}\\
&\cpz_\pro\big(\tht\big)\equiv\Z_\pro\big[\tht;\R_0\big(\tht\big)\big].
\label{cp_zp}
\end{align}
Since the charged target is  put into motion by the recoil in the comoving frame, we may also observe \textit{its} closest approach curve, which is obtained using the relative projectile-target distance from~(\ref{master}):
\begin{align}
&\cpr_\tar\big(\tht\big)\equiv r\big(\tht\big)\sin\tht-\cpr_\pro\big(\tht\big),
\label{cp_rt}\\
&\cpz_\tar\big(\tht\big)\equiv \cpz_\pro\big(\tht\big) - r\big(\tht\big)\cos\tht.
\label{cp_zt}
\end{align}
It is to be noted from~(\ref{tht}) that~(\ref{cp_rp})--(\ref{cp_zt}) may be meaningfully evaluated only for \mbox{$\pi/2<\tht<\pi$}. Using~(\ref{tht}) we might also translate the parametric dependences over $\tht$ into equivalent dependences over $\R_0$:
\begin{align}
&\cpr_\pro(\R_0)=\R_0+\frac{\eta_\tar\X\R_0}{\sqrt{\X^2+\R_0^2}},
\label{cp_rp_ro}\\
&\cpz_\pro(\R_0)=-\eta_\tar\X\left(1+\frac{\X}{\sqrt{\X^2+\R_0^2}}\right)-\eta_\tar\nu\X\ln\frac{\sqrt{\X^2+\R_0^2}}{\LL},
\label{cp_zp_ro}\\
&\cpr_\tar(\R_0)=\frac{\eta_\pro\X\R_0}{\sqrt{\X^2+\R_0^2}},
\label{cp_rt_ro}\\
&\cpz_\tar(\R_0)=\eta_\pro\X\left(1+\frac{\X}{\sqrt{\X^2+\R_0^2}}\right)-\eta_\tar\nu\X\ln\frac{\sqrt{\X^2+\R_0^2}}{\LL}.
\label{cp_zt_ro}
\end{align}
Though these relations might further be used to obtain the explicit dependences between the coordinates themselves,~(\ref{cp_rp_ro}) leads to a quartic equation for $\R_0(\cpr_\pro)$, making the final expression for $\cpz_\pro(\cpr_\pro)$ long and tiresome.

\begin{figure}[t!]
\centering
\includegraphics[width=1\linewidth,keepaspectratio]{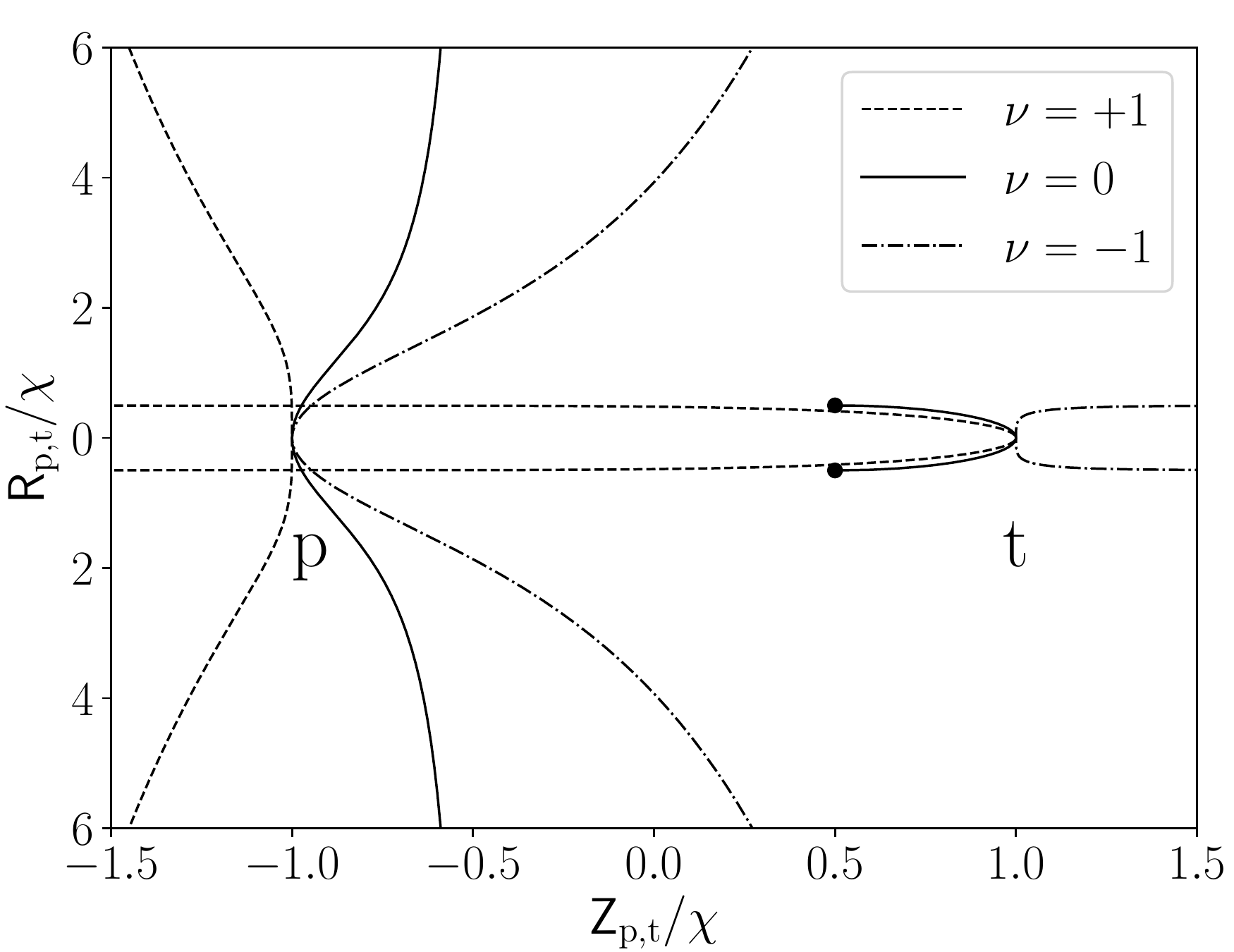}
\pull
\caption{Projectile and target closest approach curves for \mbox{$\eta_\pro=0.5$} and several values of $\nu$. The target curve for \mbox{$\nu=0$} is bounded by the dotted coordinates \mbox{$(\cpr_\tar/\X,\cpz_\tar/\X)=(\eta_\pro,\eta_\pro)$}.}
\pullc
\label{figD}
\end{figure}

Figure~\ref{figD} shows the closest approach curves for \mbox{$\eta_\tar=0.5$} and several selected values of $\nu$. It is to be noted that the closest approach curves are well defined for any $\nu$, not being limited by a scattering shadow existence condition \mbox{$-1<\nu<1$}. Equations~(\ref{cp_zp_ro}) and (\ref{cp_zt_ro}) reveal that only in the \mbox{$\nu=0$} frame (where the center of mass is at rest) the closest approach curves span a limited range along the $\Z$-axis. In that, the projectile curve has a vertical asymptote at:
\begin{equation}
\lim_{\R_0\to\infty}\cpz_\pro^{(\nu=0)}(\R_0)=-\eta_\tar\X,
\end{equation}
while the target curve is entirely bounded by:
\begin{equation}
\lim_{\R_0\to\infty}\cpr_\tar^{(\nu=0)}(\R_0)=\lim_{\R_0\to\infty}\cpz_\tar^{(\nu=0)}(\R_0)=\eta_\pro\X.
\label{endpts}
\end{equation}
For \mbox{$\nu\neq0$} both the projectile and target curves extend indefinitely, in a direction determined by $\nu$. The target curves still retain the horizontal asymptote from~(\ref{endpts}): \mbox{$\lim_{\R_0\to\infty}\cpr_\tar(\R_0)=\eta_\pro\X$}, regardless of $\nu$. Though figure~\ref{figD} might suggest that the projectile curves might have an asymptote under some particular skew angle, this is easily disproved by closely inspecting their asymptotic behavior. Since: \mbox{$\lim_{\R_0\to\infty}\cpr_\pro(\R_0)=\R_0$}, it follows that: \mbox{$\lim_{\cpr_\pro\to\infty}\cpz_\pro(\cpr_\pro)=-(\eta_\tar\nu\X/2)\ln\cpr_\pro$}, i.e. the projectile curves asymptotically feature a logarithmic behavior. Finally, it should be noted that the vertex of the projectiles' closest approach curve, \mbox{$\cpz_\pro(0)=-\eta_\tar\X[2+\ln(\X/\LL)]$}, does not correspond to a shadow vertex from~(\ref{vtx_eq}). The reason is that the shadow vertex is determined by a point where the frontal (\mbox{$\R_0=0$}) projectile comes to a rest in the \textit{comoving frame}, while the closest approach vertex by a point where the frontal projectile comes to a rest in the \textit{center-of-mass frame}.

\vspace*{-3mm}

\section{Back-bending}
\label{back_bend}

\vspace*{-2mm}

Here we investigate the back-bending of trajectories and the geometric place $\bnd_\pro(\R_0)$ of the back-bending points. The radial coordinate $\tilde{\rho}_\pro$ at which the back-bending occurs, i.e. the trajectory changes its direction, is to be found as the solution to the extremization problem:
\vspace*{-4mm}
\begin{equation}
\frac{\dd\Z_\pro}{\dd\rho_\pro}\bigg|_{\tilde{\rho}_\pro}=0.
\label{extrem_rp}
\vspace*{-1mm}
\end{equation}
The required derivative is already at hand from~(\ref{der_rhop}), yielding a quadratic equation for $\tilde{\rho}_\pro$. Out of the two solutions, only the one with the positive root-related sign:
\vspace*{-4mm}
\begin{equation}
\tilde{\rho}_\pro=\R_0+\eta_\tar\X\R_0\frac{\X\nu+\sqrt{\X^2-\R_0^2(1+\nu)^2}}{\X^2(1-\nu)-\R_0^2(1+\nu)}
\label{rho_plus}
\end{equation}
systematically satisfies \mbox{$\tilde{\rho}_\pro\ge\R_0$} around \mbox{$\R_0=0$}\footnote{
The negative sign counterpart $\tilde{\rho}_\pro^{(-)}$ to the positive root-related-sign solution from~(\ref{rho_plus})---that we denote here as \mbox{$\tilde{\rho}_\pro^{(+)}$}---is not systematically consistent with the requirement \mbox{$\tilde{\rho}_\pro\ge\R_0$}. This is easily confirmed by inspecting the behavior of the two roots around \mbox{$\R_0=0$}. To this end let us observe their derivatives, actively using the fact that the scattering shadow exists only for \mbox{$|\nu|<1$}:
\begin{align*}
\begin{split}
&\frac{\dd \tilde{\rho}_\pro^{(+)}}{\dd \R_0}\bigg|_0=1+\frac{1+\nu}{1-\nu}\eta_\tar\ge1,\\
&\frac{\dd \tilde{\rho}_\pro^{(-)}}{\dd \R_0}\bigg|_0=1-\eta_\tar\le1.
\end{split}
\end{align*}
As both roots start at \mbox{$\tilde{\rho}_\pro^{(\pm)}(0)=0$}, the negative sign solution behaves as: \mbox{$\lim_{\R_0\to0}\tilde{\rho}_\pro^{(-)}=(1-\eta_\tar)\R_0\le\R_0$}. It will be shown that under appropriate circumstances the solution $\tilde{\rho}_\pro^{(-)}$ may, in fact, provide a second bending point, which will be identified with the \textit{forward}-bending after the initial \textit{back}-bending, governed by~$\tilde{\rho}_\pro^{(+)}$.
}. The geometric place of the back-bending points is easily obtained by plugging $\tilde{\rho}_\pro$ into~(\ref{zeta}):
\vspace*{-1mm}
\begin{equation}
\bnd_\pro(\R_0)= \Z_\pro[\tilde{\rho}_\pro(\R_0);\R_0].
\label{sol_rp}
\end{equation}

\noindent As the expression is long an tedious, we do not write it here in its completeness. However, it is to be noted that by taking the limit \mbox{$\R_0\to0$}, we immediately recover the vertex position from~(\ref{vtx_eq}):
\begin{equation}
\vtx_\pro= \lim_{\R_0\to0}\bnd_\pro(\R_0)=\eta_\tar\X\left(\nu\ln\frac{\LL(1+\nu)}{\X(1-\nu)}-2\right).
\label{vtx_lim}
\end{equation}

Now we analyze the solution $\tilde{\rho}_\pro$ in order to determine which trajectories are subject to back-bending in the given comoving frame. At this point we purposefully rewrite it as:
\begin{align}
\begin{split}
&\tilde{\rho}_\pro=\frac{\N(\R_0)}{\D(\R_0)}=\\
&\frac{\R_0\big[\X^2(1-\eta_\pro\nu)-\R_0^2(1+\nu)+\eta_\tar\X\sqrt{\X^2-\R_0^2(1+\nu)^2}\big]}{\X^2(1-\nu)-\R_0^2(1+\nu)},
\end{split}
\end{align}
in order to be able to separate the numerator $\N(\R_0)$ from the denominator $\D(\R_0)$. As the discriminant vanishes for:
\begin{equation}
\R_0^{(\mathrm{dis})}\equiv \frac{\X}{1+\nu},
\label{r0_det}
\end{equation}
the positive discriminant condition for the existence of real solutions requires: \mbox{$\R_0<\R_0^{(\mathrm{dis})}$}. However, there may be a prior qualitative change in the solutions---signaled by the vanishing of the denominator---which occurs for:
\begin{equation}
\R_0^{(\mathrm{den})}\equiv \X\sqrt{\frac{1-\nu}{1+\nu}}
\label{r0_den}
\end{equation}
and commonly leads to the divergence and the abrupt change in the sign of the entire expression, \textit{unless} the numerator undergoes the same behavior at this point. We now ask: can such canceling of zeros actually occur, based on the value of the control parameter~$\nu$? At close inspection:
\begin{equation}
\N\big(\R_0^{(\mathrm{den})}\big)=\eta_\tar\X^3(\nu+|\nu|)\sqrt{\frac{1-\nu}{1+\nu}}
\end{equation}
we see that indeed it may: for every \mbox{$-1<\nu\le0$}! We have yet to confirm if under such circumstances the limit of the entire solution $\tilde{\rho}_\pro$ is finite, which is simply done by employing the l'H\^{o}pital rule:
\vspace*{-2mm}
\begin{align}
\begin{split}
\tilde{\rho}_\pro^{(\mathrm{den})}&\equiv\lim_{\R_0\to\R_0^{(\mathrm{den})}}\tilde{\rho}_\pro(\R_0;\nu\le0)=\frac{\dd \N/\dd\R_0}{\dd \D/\dd\R_0}\bigg|_{\R_0^{(\mathrm{den})}}\\
&=\X\frac{2\nu-\eta_\tar(1+\nu)}{2\nu}\sqrt{\frac{1-\nu}{1+\nu}}.
\end{split}
\label{rp_den}
\vspace*{-3mm}
\end{align}

\begin{figure}[t!]
\centering
\includegraphics[width=1\linewidth,keepaspectratio]{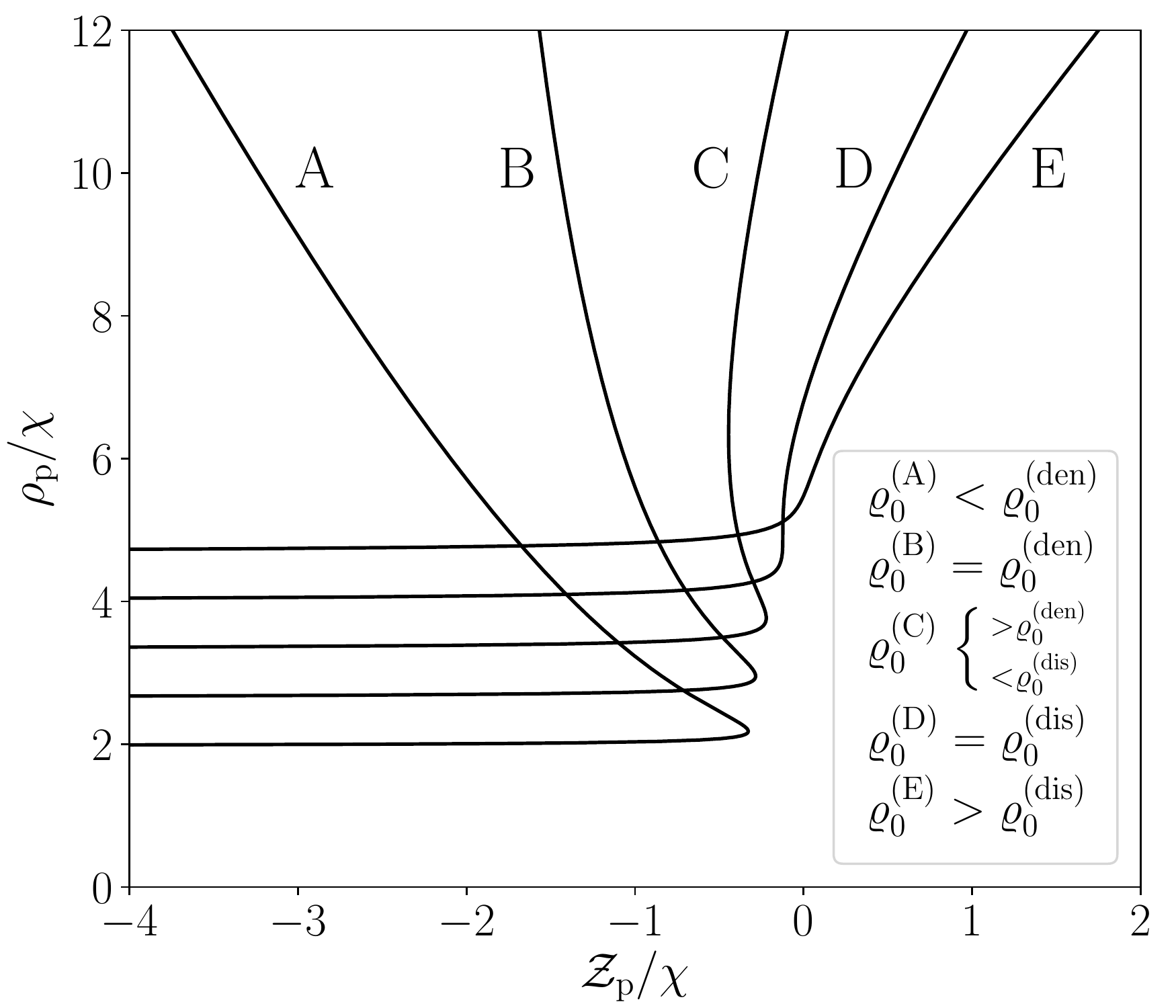}
\pull
\caption{Relevant trajectory examples in the backward-moving frame, for \mbox{$\eta_\tar=0.7$} and \mbox{$\nu=-0.75$}. The trajectory~$\mathrm{B}$ is a borderline backscattering case, its asymptotic scattering angle being \mbox{$\vartheta_\pro=\pi/2$}. The trajectory~$\mathrm{D}$ is a borderline back-bending case, its bending-point being the point of inflection. The trajectory~$\mathrm{C}$ is a prime example of a forward-scattered trajectory back-bending. The second bending point along the trajectory~$\mathrm{C}$ is the \textit{forward}-bending point, which we do not consider in this analysis.}
\pullc
\label{figB1}
\end{figure}

\noindent It is! Thus, for $-1<\nu<0$ no significant change takes place at $\R_0^{(\mathrm{den})}$, and $\tilde{\rho}_\pro$ continues unhindered until $\R_0^{(\mathrm{dis})}$. This suggests that there are two separate domains of validity, depending on value of $\nu$: one determined by the zero in denominator, the other by the zero in discriminant. On the other hand, let us consider the scattering angle $\vartheta_\pro$ from~(\ref{scat_angle}). If one were to expect only the backscattered trajectories (\mbox{$\vartheta_\pro>\pi/2$}) to undergo back-bending, then the implied requirement \mbox{$\cot\vartheta_\pro<0$} immediately leads to a unique condition: \mbox{$\R_0<\R_0^{(\mathrm{den})}$}, thus giving dominance to the zero in the denominator, whether or not its appearance has any effect upon the solution. In order to resolve this conundrum, we need to inspect the behavior of the projectile trajectories for some \mbox{$\nu<1$}. Figure~\ref{figB1} shows several representative trajectories for \mbox{$\eta_\tar=0.7$} and \mbox{$\nu=-0.75$}. The trajectory~$\mathrm{A}$ shows the expected case of the backscattered trajectory back-bending. The trajectory~$\mathrm{B}$ is the one asymptotically scattered under \mbox{$\vartheta_\pro=\pi/2$}, at the limit of the backscattering regime. The trajectory~$\mathrm{C}$ shows a clear case of a forward-scattered trajectory back-bending, thus invalidating the earlier assumption about the backscattered trajectories being a sole subject to back-bending\footnote{
It may be observed from figure~\ref{figB1} that within the domain \mbox{$\R_0^{(\mathrm{den})}<\R_0<\R_0^{(\mathrm{dis})}$} corresponding to the trajectory~$\mathrm{C}$, there are in fact two bending points. As the second one is also a solution to the extremization problem from~(\ref{extrem_rp}), it is described by the second solution to the associated quadratic equation, i.e. by the negative root-related-sign counterpart to~(\ref{rho_plus}). We do not consider them in the analysis, as they are \textit{forward}-bending, rather than the \textit{back}-bending points. Since the trajectories from this range are ultimately scattered forwards, only for the backscattered trajectories (i.e. for \mbox{$\R_0<\R_0^{(\mathrm{den})}$}) could the back-bending points be identified not only with the local, but also with the global maximum of the trajectory equation, in a sense: \mbox{$\bnd_\pro(\R_0)= \max_{\rho_\pro}[\Z_\pro(\rho_\pro;\R_0)]$}.
}. The trajectory~$\mathrm{D}$ shows the borderline case of back-bending, when the back-bending point becomes the point of inflection. The trajectory~$\mathrm{E}$ shows the typical case of a forward-scattered trajectory without back-bending. Based on this observations, we may safely conclude that there are indeed two separate conditions for the validity of $\tilde{\rho}_\pro$, determined by a maximum admissible argument:
\begin{equation}
\R_0^{(\mathrm{max})}=\left\{\begin{array}{cl}
\R_0^{(\mathrm{dis})}&\ifif \; \nu\le 0\\
\R_0^{(\mathrm{den})}&\ifif \; \nu\ge0
\end{array}\right. .
\label{r0_max}
\end{equation}

\begin{figure}[t!]
\centering
\includegraphics[width=1\linewidth,keepaspectratio]{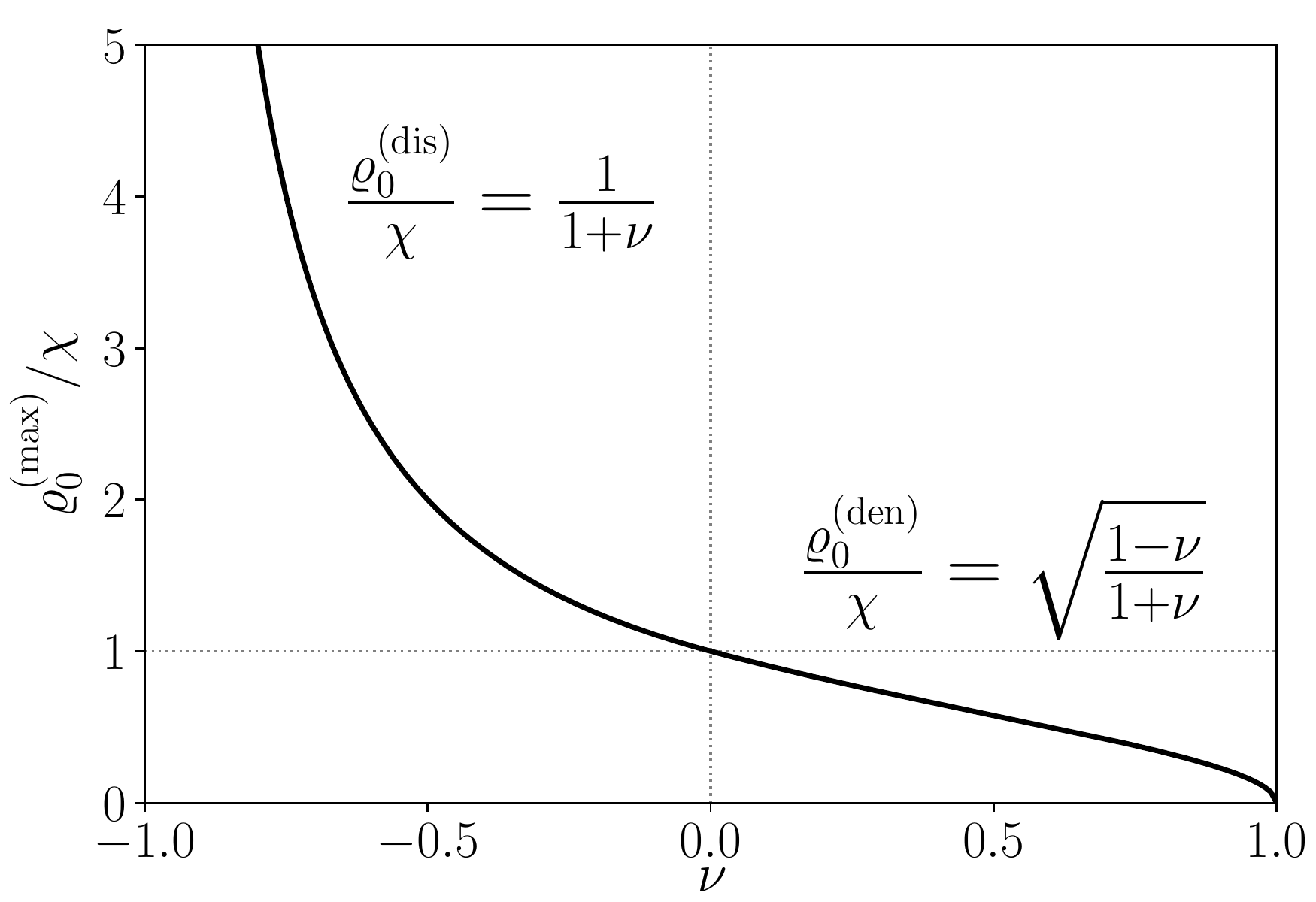}
\pull
\caption{Impact parameter of the last back-bending trajectory, dependent on the motion of the reference frame. There is a qualitative difference between the frames moving forwards (\mbox{$\nu>0$}) and those moving backwards  (\mbox{$\nu<0$}).}
\pullc
\label{figB2}
\end{figure}

\begin{figure}[t!]
\centering
\includegraphics[width=1\linewidth,keepaspectratio]{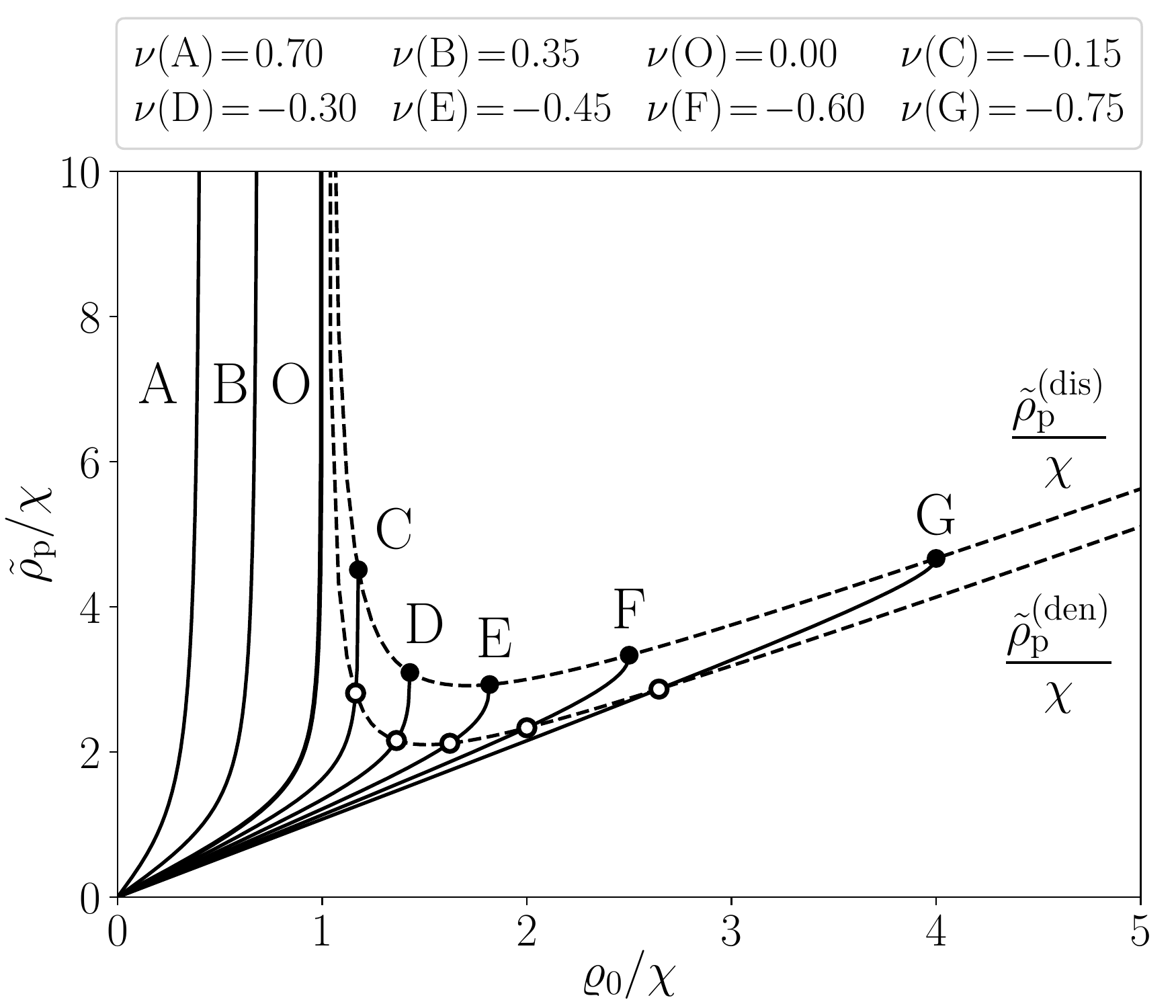}
\pull
\caption{Solutions $\tilde{\rho}_\pro$ to the extremization problem from~(\ref{extrem_rp}) for \mbox{$\eta_\tar=0.5$} and several selected values of $\nu$. The upper and lower dashed curve are given by~(\ref{exp_det}) and (\ref{exp_den}), respectively.}
\label{figB3}
\end{figure}

\noindent Figure~\ref{figB2} shows this dependence. Further expressing:
\vspace*{-1mm}
\begin{equation}
\tilde{\rho}_\pro^{(\mathrm{dis})}\equiv \tilde{\rho}_\pro\big(\R_0^{(\mathrm{dis})}\big)=\X\,\dfrac{\eta_\pro\nu-\eta_\tar}{\nu(1+\nu)},
\label{rp_det}
\vspace*{-1mm}
\end{equation}
we now see from~(\ref{r0_max}):
\vspace*{-1mm}
\begin{equation}
\tilde{\rho}_\pro^{(\mathrm{max})}=\left\{\begin{array}{cl}
\tilde{\rho}_\pro^{(\mathrm{dis})}&\ifif \; \nu\le 0\\
\infty&\ifif \; \nu\ge0
\end{array}\right. 
\vspace*{-1mm}
\end{equation}
that the geometric extent of back-bending points both qualitatively and quantitatively depends on the direction of motion of the reference frame. This is clearly visible from figure~\ref{figB3}, which shows the solution $\tilde{\rho}_\pro$ for \mbox{$\eta_\tar=0.5$} and several selected values of $\nu$. For \mbox{$\nu\ge0$} (cases $\mathrm{A}$, $\mathrm{B}$, $\mathrm{O}$) one can find back-bending points at any distance $\rho_\pro$ from the $z$-axis. However, for \mbox{$\nu<0$} (cases $\mathrm{C}$, $\mathrm{D}$, $\mathrm{E}$, $\mathrm{F}$, $\mathrm{G}$) the geometric extent of back-bending points is limited. Full circles show their end-points for a given $\nu$. Considering all such end-points (for any $\nu$) yields an upper dashed curve beyond which one cannot find a real solution $\tilde{\rho}_\pro$, regardless of $\nu$. Eliminating $\nu$ from~(\ref{r0_det}) and (\ref{rp_det}), one can even obtain an explicit expression for this curve:
\begin{equation}
\tilde{\rho}_\pro^{(\mathrm{dis})}=\frac{\R_0^{(\mathrm{dis})}\big(\R_0^{(\mathrm{dis})}-\eta_\pro\X\big)}{\R_0^{(\mathrm{dis})}-\X}.
\label{exp_det}
\end{equation}
Additionally, open circles show the points at which the solutions $\tilde{\rho}_\pro$ would diverge due to the zero in the denominator, if not for the corresponding zero in the numerator yielding the finite values from~(\ref{rp_den}). The geometric place of all such points forms a lower dashed curve, whose explicit form is found by eliminating $\nu$ from~(\ref{r0_den}) and (\ref{rp_den}):
\begin{equation}
\tilde{\rho}_\pro^{(\mathrm{den})}=\frac{\R_0^{(\mathrm{den})}\big[\big(\R_0^{(\mathrm{den})}\big)\,\!^2-\eta_\pro\X^2\big]}{\big(\R_0^{(\mathrm{den})}\big)\,\!^2-\X^2}.
\label{exp_den}
\end{equation}

\pagebreak

\vspace*{-6mm}

\begin{figure}[t!]
\centering
\includegraphics[width=1\linewidth,keepaspectratio]{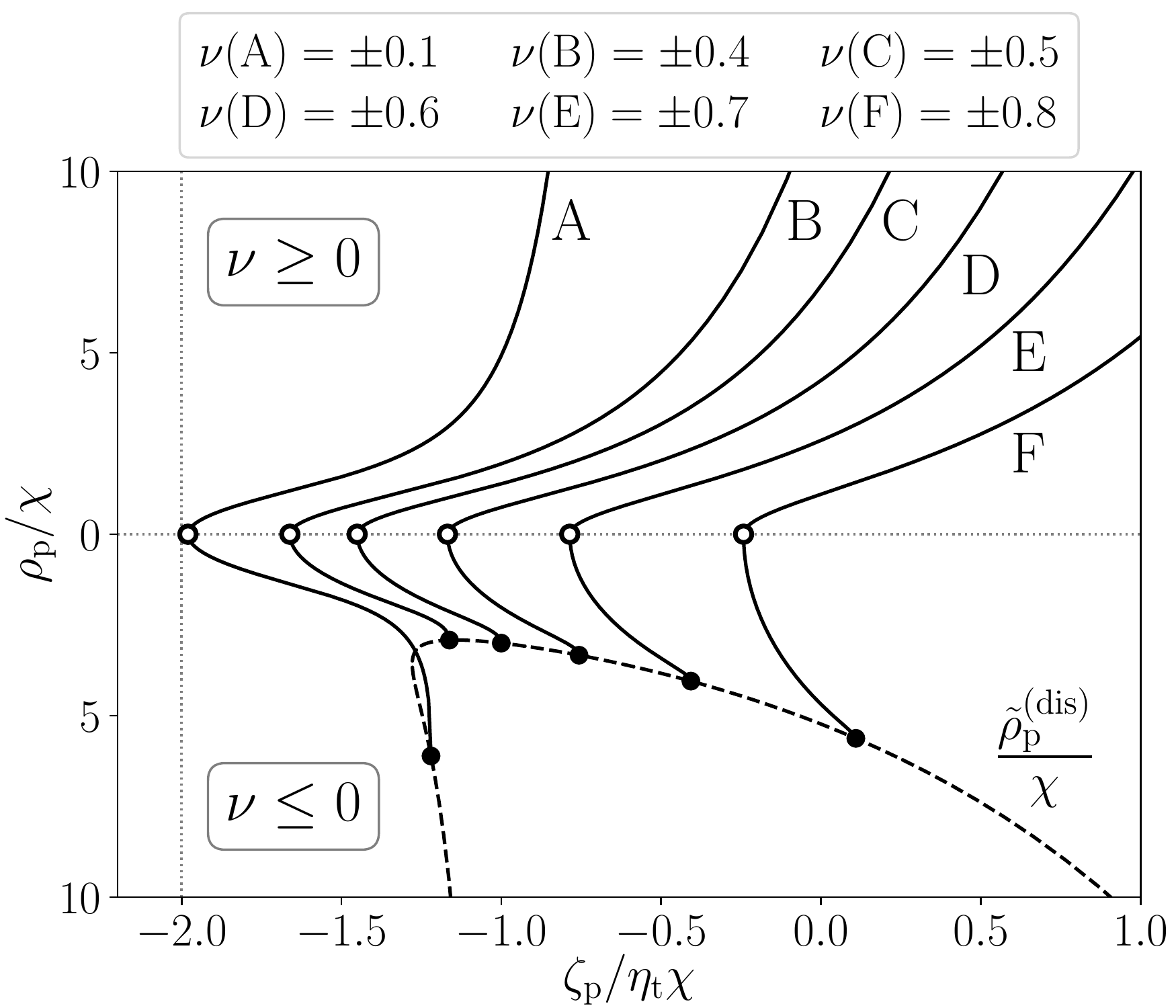}
\pull
\caption{Back-bending curves (geometric place of the trajectories' back-bending points) for \mbox{$\eta_\tar=0.5$} and several selected values of $\nu$. The lower and the upper part of the plot separately show the examples for \mbox{$\nu\le0$} and  \mbox{$\nu\ge0$}. Open circles correspond to the shadow vertex positions. Full circles are the curves' end-points, sweeping the dashed curve.}
\pullc
\label{figB4}
\end{figure}

Figure~\ref{figB4} shows the geometric place of back-bending points for \mbox{$\eta_\tar=0.5$} and several values of $\nu$, as per definition from~(\ref{sol_rp}). It should be noted that plugging the solution $\tilde{\rho}_\pro$ from~(\ref{rho_plus}) into the trajectory equation from~(\ref{zeta}) yields the dependence $\bnd_\pro(\R_0)$ on the impact parameter $\R_0$, while figure~\ref{figB4} shows the dependence $\bnd_\pro(\rho_\pro)$ on the geometric coordinate $\rho_\pro$ of the back-bending points. The upper part of the plot shows examples for \mbox{$\nu>0$}, while the lower part shows the examples for the corresponding negative values of $\nu$. The open circles show the limit of the bending-points curve, as it approaches the $z$-axis, determining the vertex position from~(\ref{vtx_eq}) (with \mbox{$\LL=\X$} used in all these plots). As in figure~\ref{figB3}, the back-bending curves for \mbox{$\nu<0$} are limited in their extent, ending at coordinates \mbox{$\big[\tilde{\rho}_\pro^{(\mathrm{dis})},\bnd_\pro\big(\tilde{\rho}_\pro^{(\mathrm{dis})}\big)\big]$}, shown by the full circles. All such end-points (for any \mbox{$-1<\nu<0$} and a given $\eta_\tar$) form a dashed curve. It would be false to say that no bending curve reaches beyond this boundary (in either the radial or the axial direction), which is clearly seen from the \mbox{$\nu=-0.1$} case.

In~\cite{zugec_supp} we have addressed an issue of confusing the scattering shadow with the geometric place of the closest approach points. From figure~\ref{figB4} it is perfectly clear that neither the back-bending curves correspond to a scattering shadow, which is a simple consequence of their precise and distinct definitions from~(\ref{zz}) and (\ref{sol_rp}). However, if one still asked for a specific and convincing counterexample, it is easily obtained in the infinitely-massive-target frame (\mbox{$\eta_\tar=1$} and \mbox{$\nu=0$}).\pagebreak

\noindent In that case the solution from~(\ref{rho_plus}) reduces to:
\vspace*{-0.5mm} 
\begin{equation}
\tilde{\rho}_\pro^{(\infty)}(\R_0)=\R_0+\frac{\X\R_0}{\sqrt{\X^2-\R_0^2}}.
\end{equation}
Since the inverse dependence $\R_0\big(\tilde{\rho}_\pro^{(\infty)}\big)$ is still long and tedious---a solution to the $4^\mathrm{th}$ degree polynomial---we keep the back-bending curve from~(\ref{sol_rp}) parametrized by~$\R_0$:
\vspace*{-1mm}
\begin{equation}
\bnd_\pro^{(\infty)}(\R_0)=-\X-\sqrt{\X^2-\R_0^2}
\label{bckbnd_inf}
\end{equation}
and compare it to the shadow equation parametrized by~$\R_0$, which is easily found by combining~(\ref{inf_r0}) and (\ref{inf_shad}):
\begin{equation}
\ZZ_\pro^{(\infty)}(\R_0)=\frac{\R_0^2}{2\X}-2\X.
\end{equation}
Having resorted to the infinitely-massive-target frame---coinciding with the fixed-target frame---we may also make a comparison with the associated closest approach curve from~(\ref{cp_zp_ro}):
\begin{equation}
\cpz_\pro^{(\infty)}(\R_0)=-\X-\frac{\X^2}{\sqrt{\X^2-\R_0^2}}.
\end{equation}
Evidently, the back-bending curve, the closest approach curve and the scattering shadow caustic are all decidedly dissimilar.

\end{document}